\documentclass[12pt]{article}
\usepackage{amsmath,amssymb,epsfig,amsfonts}
\usepackage{graphicx,subfigure, pict2e}
\usepackage[usenames,dvipsnames]{xcolor}
\usepackage[backref]{hyperref}
\usepackage{cite}
\usepackage[all]{xy}
\usepackage{verbatim} 

\makeatletter

\newcommand{\ssn}[1]{{\color{black}  #1}}

\newcommand{\be}{\begin{equation}}
\newcommand{\ee}{\end{equation}}
\newcommand{\ba}{\begin{aligned}}
\newcommand{\ea}{\end{aligned}}

\def\unit{{1\kern-.65ex {\rm l}}}
\def\1{{1\kern-.65ex {\rm l}}}

\newcount\hour \newcount\minute
\hour=\time \divide \hour by 60
\minute=\time
\def\now{%
\ifnum \hour<13
  \ifnum \hour=0 \advance \hour by 12 \number\hour:\else \number\hour:\fi%
     \ifnum \minute<10 0\fi%
     \number\minute%
\ A.M.%
\else \advance \hour by -12 \number\hour:%
  \ifnum \minute<10 0\fi%
  \number\minute%
  \ P.M.%
\fi%
}

\makeatother

\begin{document}

\baselineskip=18pt  
\numberwithin{equation}{section}  
\allowdisplaybreaks  


%
%


\thispagestyle{empty}

\vspace*{-2cm} 
\begin{flushright}
{\tt KCL-MTH-12-14}\\
\end{flushright}

\vspace*{0.8cm} 
\begin{center}
 
 
 {\LARGE The Tate Form on Steroids:\\
\bigskip
 
{Resolution and Higher Codimension Fibers}
 }

 \vspace*{1.5cm}
 Craig Lawrie and Sakura Sch\"afer-Nameki\\
 \vspace*{1cm} 
 {\it Department of Mathematics, King's College, University of London,\\
  The Strand, London WC2R 2LS, England \\}
{\tt  gmail: craig.lawrie1729, sakura.schafer.nameki}

\end{center}
\vspace*{1cm}

\noindent
F-theory on singular elliptically fibered Calabi-Yau four-folds provides a setting to geometrically study four-dimensional 
$\mathcal{N}=1$ supersymmetric gauge theories, including matter and Yukawa couplings. 
The gauge degrees of freedom arise from the codimension 1 singular loci, 
the matter and Yukawa couplings are generated at enhanced singularities in higher codimension. 
We construct the resolution of the singular Tate form for an elliptic Calabi-Yau four-fold with an ADE type singularity in codimension 1 and study the structure of the fibers in codimension 2 and 3. We determine the fibers in higher codimension which in general are of Kodaira type along minimal singular loci, and are thus consistent with the low energy gauge-theoretic intuition. Furthermore, we provide a complementary description of the fibers in higher codimension,   which will also be applicable to non-minimal singularities.  The irreducible components in the fiber in codimension 2 correspond to weights of representations of the ADE gauge group. These can split further  in codimension 3 in a way that is  consistent with the generation of Yukawa couplings. Applying this reasoning, we then venture out to study non-minimal singularities, which occur for $A$ type  along codimension 3, and for $D$ and $E$ also in codimension 2.  
The fibers in this case are non-Kodaira, however some insight into these singularities can be gained by considering the splitting of fiber components along higher codimension, which are shown to be consistent with matter and Yukawa couplings for the corresponding gauge groups.

\newpage

\tableofcontents



\section{Introduction}

It has long been known that F-theory is an ideal framework for geometric engineering of gauge theories \cite{Vafa:1996xn, Morrison:1996na, Morrison:1996pp}. F-theory compactified on singular elliptically fibered Calabi-Yau $n$-manifolds $Y_n$,
realizes supersymmetric gauge theories at low energies, which can be understood to have their origin in 7-branes wrapping cycles in the compactification manifold. Let $B_{n-1}$ denote the base of the elliptic fibration, which we shall assume to be described in terms of a Weierstrass model
\begin{equation}
y^2 = x^3 + f x + g \,,
\end{equation}
where $f$ and $g$ are sections of $K_{B_{n-1}}^{-4}$ and $K_{B_{n-1}}^{-6}$. 
The singular elliptic fibers in codimension 1 in the base were classified by Kodaira \cite{Kodaira}, and the Kodaira type of the fiber determines the gauge group of the effective theory on the 7-branes that wrap the codimension 1 locus in the base as well as the transverse flat space. From an M-theory perspective, the abelian part of the gauge bosons arise by dimensional reduction of the $C_3$-form along the $\mathbb{P}^1$s in the resolved fiber, and the remaining components arise from wrapped M2-branes.

Whenever $n\geq 3$, in addition, matter transforming in highest weight representations of the gauge group is known to come from codimension 2 loci in the base  \cite{MR1158626, Katz:1996xe, Morrison:2011mb, Esole:2011sm, MS, Grassi:2011hq}, which confirms the physical intuition that  7-branes generate matter along their intersection loci. 
Naturally, one can extend this for $n\geq 4$ and the study of codimension 3 fibers. 
Indeed, more recently, F-theory on Calabi-Yau fourfolds has played a key role in building four dimensional supersymmetric grand unified theories (GUTs) \cite{Donagi:2008ca, Beasley:2008dc, Beasley:2008kw}. 
For a complete description of a GUT model, we need in addition to gauge and matter degrees of freedom, a way to generate the Yukawa couplings of the matter fields. 7-brane intuition tells us that this should happen at the intersection of multiple branes. Geometrically, these are realized in codimension 3 in the base $B_3$. 

The generation of matter and Yukawa couplings from higher codimension loci, where the singularity type enhances, has a counterpart in the 
 effective field on the 7-branes, there matter and Yukawa can be thought to arise from Higgsing of a higher rank gauge group.  For gauge groups embedded into $E_8$, this has a precise description in terms of the Higgs bundle spectral cover \cite{Hayashi:2008ba, Donagi:2009ra, Hayashi:2009ge, Marsano:2009gv}, and more generally including additional gluing data \cite{Donagi:2011jy, Cecotti:2010bp, Donagi:2011dv, Marsano:2012bf}.
For instance for $SU(5)$ GUTs the matter and Yukawa couplings can be thought to arise by Higgsing $SU(6)$ or $SO(10)$ for matter, and $SO(12)$, or  $E_6$ for Yukawa couplings. This gauge theoretic expectation was  confirmed by studying the resolved Calabi-Yau fourfold and the fiber structure in \cite{Esole:2011sm, MS}, except for the absence of a Kodaira $E_6$ fiber in codimenion 3, which however was shown to not be an obstruction to generating the top Yukawa coupling \cite{MS}: along the putative $E_6$ codimension 3 locus the fiber component corresponding to a {\bf 10} matter representation splits further into a {\bf 10} and a {\bf 5}. 

In this context there are two natural questions that arise, which we will address in this paper: from a mathematical point of view, one can ask whether a codimension 1 singular Kodaira fiber degenerates in higher codimension to Kodaira fibers or not. The physical counterpart to this question is whether in general the geometry corroborates the gauge theoretic description of matter and Yukawas in terms of Higgsing of a higher rank gauge group. 

Key to understanding the fibers in higher codimension is the resolution of the singularities. 
 One of the main objectives of this paper is therefore to resolve elliptically fibered  Calabi-Yau fourfolds with an ADE type singularity in codimension 1. \ssn{To keep the resulting space Calabi-Yau, which is required to preserve $N=1$ supersymmetry in 4 dimensions, we impose that the resolution is crepant, i.e. the pull-back of the canonical bundle under the resolution yields the canonical bundle of the resolved geometry. In particular, if the initial geometry has trivial canonical class, this property is preserved under the resolutions.  }
 Our starting point is the Tate form \cite{Bershadsky:1996nh, Katz:2011qp} of the Kodaira singularity in codimension 1\footnote{Note that we will not assume any further structure of the sections appearing in the Tate form. Relaxing this conditions, one can of course obtain less generic fibers, e.g. models with extra sections will not be considered here. Starting with the Tate form comes with a few minor caveats, which will be discussed in the next section.}. In each case we construct the fully resolved geometry, and study the fibers in higher codimension, giving two, slightly distinct, points of view: 
 we can either consider the intersections of the fibers in higher codimension. These are summarized in table \ref{FiberSummaryTable}.
 Alternatively, we can study the splitting of the fibers along the enhanced singularities. The latter will in particular be of use for the non-minimal singularities, which we will discuss in a moment. 
  
Concretely, for codimension 2, we show that generically the intersections of the fibers along minimal singularities  are Kodaira $A$ or $D$ type. 
Likewise, following the fibers in codimension 1 to the codimension 2 locus, we show that  additional components in the fiber correspond to weights\footnote{As we will make precise in the main text, this means, that their intersections with the codimension 1 fibers, i.e. roots, are given by Dynkin labels of highest weight representations.} of highest weight representations of the gauge group. 
In codimension 3 the fibers along minimal singular loci are again generically Kodaira $A$ or $D$ type, or alternatively put, the 
fiber components become reducible, and in each case we identify these with a consistent Yukawa coupling between matter in codimension 2. 

What we said so far applies to singularities which are so-called minimal, characterized by a vanishing order of the sections $f$ and $g$ in the  Weierstrass form as well as the discriminant $\Delta$ not all exceeding $3, 5$ and $11$, respectively. In codimension 1 we  will always assume that this condition of minimality is satisfied. However, when passing to higher codimension, we will also consider non-minimal loci\footnote{We define non-minimality in this context by applying the same criterion of vanishing orders of $f$, $g$ and $\Delta$ applied to their restrictions. }. These occur for $A_n$ type singularities in codimension 3, generalizing the $E$-type Dynkin diagrams that one would obtain as fiber intersection graphs for low values of $n$. \ssn{ Along the non-minimal loci, the fibers are shown to have 
additional one-dimension higher (i.e. surface) components in the fiber, i.e. the fibration ceases to be flat, and
are most definitely not Kodaira. The surface components allow for wrapped M5 branes, which are expected to not be described by simple gauge theoretic degrees of freedom, and signal additional light states that will have to be included in the  low energy effective theory. }
Nevertheless we can study how the matter surfaces in codimension 2 split along the non-minimal loci, which is shown to be  completely consistent with the generation of Yukawa couplings (generalizing the $E_6$ Yukawa for $SU(5)$, for instance). To depict the fibers we either consider only the components that decend from the matter surfaces, yielding generalizations of  $E$-type Dynkin diagrams or we include the additional one-dimension higher components.

 In a similar way we study the non-minimal matter locus in codimension 2 for $D_n$, and show that the Cartans split into spin representations and the fundamental representation $V$. The fiber along these non-minimal loci will depend on the particular small resolution that one applies to the singular geometry. To exemplify this, we provide an alternative resolution for the non-minimal matter locus for $D_n$ in appendix \ref{app:SOReload}, where in addition the splitting includes matter in the representations $\Lambda^iV$. 
We consider these results as a starting point for studying these non-minimal singularities and their low energy effective description in more detail.

Related work for compactifications of F-theory to six dimensions has appeared  \cite{Morrison:2011mb},  where the matter representations and fibers  were studied by resolving the local singularities in codimension 2.  An alternative point of view was presented in  \cite{Grassi:2000we, Grassi:2011hq}, where a generalized version of the Tate algorithm was developed for matter. The present result seems to be in agreement with the matter found in   \cite{Grassi:2011hq} for minimal singularities.  Finally, in \cite{Candelas:2000nc} an example of a codimension 3 non-minimal singularity for an $E_7$-type singularity was considered, and studied from the point of view of singular spectral covers. This opens up the possibility of a connection with the constructions in \cite{Marsano:2012bf}. It would be interesting to apply similar methods to resolve the non-minimal loci that we encounter for $A_n$ and $D_n$.

Beyond the clarification of the codimension 2 and 3 structure of the fibers, the resolved Tate forms that we determine 
will be useful for the study of additional data that are necessary in F-theory compactification. In particular  $G$-flux $G_4\in H^{2,2} (Y, \mathbb{Z})$ can be constructed in terms of surfaces in the resolved geometry, and proper quantization can be checked using the $c_2$ of the resolved fourfold
 \cite{Marsano:2009gv, Collinucci:2010gz, MS, Krause:2011xj,  Collinucci:2012as, Marsano:2012bf, Cvetic:2012xn}. Likewise for general gauge groups, the resolved geometries in this paper can be applied to construct such $G$-flux.
 
Resolutions of singularities for some low gauge group relevant for GUT models have been studied in the literature: for $SU(5)$ in \cite{Esole:2011sm, MS }, for $E_6$ and $SO(10)$ codimension 1 singularities, the resolution of CY fourfolds has appeared in \cite{Kuntzler:2012bu,  Tatar:2012tm}, and for other  models related to the $SU(5)$ case, e.g. with additional $U(1)$ factors and connection to orientifold limits, the resolved geometries were studied in \cite{Grimm:2010ez, Krause:2011xj, Grimm:2011fx, Marsano:2012yc,Cvetic:2012xn, Mayrhofer:2012zy, Esole:2011cn, Esole:2012tf}.

The outline of the paper is as follows. 
In section \ref{sec:General} we summarize the theory used to resolve the
singular geometry and the intersection theory necessary 
to calculate the properties of the fibers in higher codimensions, in particular their splitting and charges. 
For each singularity in codimension 1, we summarized the higher codimension loci of enhanced symmetry in table \ref{BigTable}. The fibers are summarized in table \ref{FiberSummaryTable}.
In the subsequent sections we resolve the $A_n$, $D_n$ and $E_n$ singularities in turn. Each section can be read independently, although the most detailed analysis will be presented for $SU(2k+1)$. 
For each type we first fully resolve the Tate form for general $n$, then compute the intersections of exceptional divisors in codimension 1. In particular, we will consider the divisors that are obtained by fibering the irreducible components of the fibers over the codimension 1 locus, which we will denote by {\it Cartan divisors}, and their intersections reproduce the affine Dynkin diagrams of ADE type. 
For each of the codimension 2 loci in table \ref{BigTable} we determine the splitting of the codimension 1 fibers, and compute the intersections of the irreducible fiber components in codimension 2 with the Cartan divisors (labeled by the simple roots), which gives rise to weights of representation of the ADE  gauge group. The third subsection in each ADE case contains the discussion of fibers in codimension 3, and in each locus listed in table \ref{BigTable} we show what Yukawa coupling they generate. For $A_n$ and $D_n$ we furthermore discuss the fibers along the non-minimal loci in a separate subsection. 
Appendix \ref{app:Tate} lists the Tate forms, appendix \ref{app:Examples} contains low rank examples.
Finally, we include extensive appendices containing the details of the  intersection calculations and the resolved geometries.


\section{Resolution of Singularities}\label{sec:General}


\subsection{The Tate Form}

Consider a singular elliptically fibered Calabi Yau $n$-fold. 
The elliptic fibration has singularities in codimension 1, and for sufficiently large $n$ higher codimension singularities can arise, whose structure is less understood. The present case of interest, $n=4$, allows in addition codimension 2 and 3  singularities. We will study these in turn for various Tate forms, by first resolving the geometries in all codimensions and then studying the intersections along the loci of singularity enhancement. 

The starting point for our analysis is the  Weierstrass form for a  singular elliptically fibered Calabi-Yau fourfold $Y_4$ over a base  $B_3$
\be
	 y^2 = x^3 + f x + g \,.
\ee
Let $\zeta_0$ be a local coordinate on the base threefold. 
For a singularity with Kodaira fibers of a specific type along a divisor $\zeta_0=0$, 
one can globally put the Weierstrass form into the Tate form (or a suitable generalization thereof) \cite{Bershadsky:1996nh, Katz:2011qp}
\be\label{TateForm}
	T = y^2w - x^3 + b_1xyw\zeta_0^{i_1} - b_2wx^2\zeta_0^{i_2} + b_3yw^2\zeta_0^{i_3} 
			- b_4xw^2\zeta_0^{i_4} - b_6\zeta_0^{i_6}w^3 = 0 \,,
\ee
which is realized inside a $\mathbb{P}^2$ bundle
\be
X_5= \mathbb{P} (\mathcal{O}\oplus K^{-2}_{B_3} \oplus K^{-3}_{B_3} )\,.
\ee
There is a subtlety associated with this starting point. The Tate form guarantees that a singularity of a specific type is realized. However, there are some 
caveats related to bringing a given Weierstrass form globally  into the Tate form  \cite{Katz:2011qp}. In particular for the cases 
\begin{equation}
\ba
SU(m)\,,\qquad  &m= 6,7,8,9, \cr
Sp(m)\,,\qquad&  m = 3,4 \cr
SO(m) \,,\qquad&  m= 13, 14 \,,
\ea
\end{equation}
the Tate form cannot be achieved. For the $I_{2k+1}$ singularities with monodromy, which in the table in appendix \ref{app:Tate}, are denoted by $I^{ns}$ or $I^{ss}$, the Tate form cannot in general be achieved globally, however a generalized Tate form presented in  \cite{Katz:2011qp} can. For this paper we will concentrate on the simply-laced, i.e. ADE type  singularities, which in particular do not have monodromy\footnote{When refering to Kodaira fibers of type $I_n^s$, $I_n^{*s}$, $IV^{*s}$, $III^*$ and $II^*$, respectively, we sometimes will denote them by $A$, $D$ or $E$ type fibers, where it is understood, that the multiplicities are as in the Kodaira classification. 
}. The only case in which our starting point is thus not the most general one is the $SU(m)$ with $m= 6, 7, 8, 9$.  

With this minor caveat in mind, consider now the elliptically fibered CY defined by (\ref{TateForm}), which has a singularity along $\zeta_0=0$, whose type is specified by the vanishing orders $i_m$  as summarized in appendix \ref{app:Tate}.
For most purposes it will be sufficient to work in the patch $w=1$.
The various sections have the following classes
\be\label{Classes}
\begin{array}{r|l}
\hbox{Section} & \hbox{Class} \cr\hline
x & \sigma + 2 c_1 \cr
y & \sigma + 3 c_1 \cr
w & \sigma \cr
\zeta_0 & S_2 \cr
b_n&  n c_1 - i_n S_2 
\end{array}
\ee
The codimension 1 singularities are along components of the vanishing locus of the discriminant
\be
\Delta =  4f^3 + 27g^2  = O (\zeta_0^d)\,,
\ee
where $d$ is the vanishing order as specified in appendix \ref{app:Tate}. The higher codimension loci are defined below. 

The resolution of the singularity in codimension 1 in the base, i.e. along a surface $S_2$ specified by the section $\zeta_0$, is well-known to 
give rise to fibers with exceptional $\mathbb{P}^1$s, whose intersections are represented by the affine Dynkin diagrams of the ADE type of the singularity. 
From the point of view of F-theory \cite{Vafa:1996xn, Morrison:1996pp, Morrison:1996na}, this corresponds to realizing the gauge degrees of freedom on 7-branes wrapping the surface $S_2$. The abelian part of the gauge bosons has its origin in the M-theory 3-form $C_3$, decomposed into the $(1,1)$ forms $\omega^{(1,1)}$ of the exceptional $\mathbb{P}^1$s 
\begin{equation}
C_3 = A^i\wedge \omega_i^{(1,1)}\,,
\end{equation}
whereas the remaining gauge bosons arise from M2-branes wrapping the exceptional $\mathbb{P}^1$s in the fiber. We will denote the divisors obtained by fibering the exceptional $\mathbb{P}^1$s over the surface $S_2$ by Cartan divisors $D_{-\alpha_i}$, labeled by the simple roots $\alpha_i$ of the  gauge group. The section $\zeta_0$ will correspond to the affine root $-\alpha_0$. In summary: \\

{\begin{itemize}
\item[] \underline{Codimension 1}

The fibers in codimension 1 (i.e. along $\zeta_0=0$) have intersections given by the affine Dynkin diagram of the ADE type of the singularity. The Cartan divisors $D_{-\alpha_i}$ are obtained as the irreducible exceptional divisors of the resolution.  
\end{itemize}
}
\smallskip

Less explored is the structure of the fibers in higher codimension. For specific singularity types in Calabi-Yau 3- and 4-folds this was studied in 
\cite{Candelas:2000nc, Grassi:2011hq, Esole:2011sm, MS, Morrison:2011mb, Grimm:2011fx, Kuntzler:2012bu, Tatar:2012tm}. 
First consider the higher order terms in the expansion of the discriminant with respect to $\zeta_0$ 
\begin{equation}\label{DeltaExpand}
\Delta =\zeta_0^d \Delta_d +  \zeta_0^{d+i} \Delta_{d+i} +  O (\zeta_0^{d+i+1}) \,,
\end{equation}
where $\zeta_0=0$ intersected with $\Delta_d=0$ yields a higher vanishing order, and we would expect new components  to emerge in the fiber. 
To clarify what happens in the fiber,  we resolve the space in codimension 1 but also 2 and 3, where the latter are defined as
\begin{equation}\label{CodimDef}
\ba
\hbox{Codimension 2}: &\quad \zeta_0=\Delta_d=0  \cr
\hbox{Codimension 3}: &\quad \zeta_0=\Delta_{d}= \Delta_{d+i}=0 \,.
\ea
\end{equation}
The following table \ref{BigTable} lists each of the codimension 2 and 3 loci for the Tate forms in the table in appendix \ref{TateTable}. From an F-theoretic point of view, codimension 2 corresponds to the matter loci, and codimension 3 to the points, where Yukawa couplings are generated. 

\ssn{The discriminant $\Delta$ strictly speaking only identifies the singularities in codimension 1. To determine the singularities in codimension 2, one has to inspect the discriminant after resolving in codimesion 1 to determine where the geometry is still singular. This is different from our approach here: we choose to consider the higher codimension restrictions of the discriminant $\Delta$ to identify interesting loci in the geometry, and make sure that the final resolved space is smooth, i.e. has not further singularities at these loci. It would be interesting to compare the two criteria and work out their precise correlation. 

It is clear that the higher codimension restriction of the codimension 1 discriminant $\Delta$ does not imply necessarily a further degeneration of the fiber. However, as we will see in the resolved geometries, there are loci, such as the $P=Q=0$ locus for $SU(2k+1)$ discussed in (\ref{PQ2k1}), where the fiber does not degenerate further, however in terms of the splitting of the matter curves, it is consistent with generating a coupling $V \otimes {\overline{V}} \otimes {\bf 1}$. So this locus does play an important role in the effective field theory description, despite the fact that the geometry seems to not further degenerate there. 
}

In summary we find the following fiber structure in higher codimensions: along minimal singularities the resolved geometry has Kodaira fibers for $A_n$ and $D_n$ singularities in codimension 2 and 3, including multiplicities. For non-minimal loci the fibers do not intersect in the affine Dynkin diagram of a simple Lie algebra. We will study the latter following the logic put forward in \cite{MS}, by considering the splitting of the lower codimension fibers. 

\begin{itemize}
\item[] \underline{Codimension 2}

Along the codimension 2 loci the exceptional $\mathbb{P}^1$ in the fiber either remain irreducible, or split, generating new components that are labeled by weights of a highest weight representation of the gauge group, i.e. their intersections with the irreducible components of the fibers in codimension 1 are Dynkin labels of representations. Generically, this splitting gives rise to one new effective curve in the fiber. 
Along the codimension 2 loci, the Cartan divisors $D_{-\alpha_i}$ will restrict to surfaces,  the irreducible components of which will be denoted by $S_{v_i}$, where $v_i$ is either a weight of  a highest weight representation, or in case $D_{-\alpha_i}$ remains irreducible, a root. We will refer to these as  matter surfaces.

\item[] \underline{Codimension 3} 

Along the codimension 3 loci, either, the fibers remain irreducible, or a  $\mathbb{P}^1$ labeled by a weight,  becomes homologous to the sum of two others, compatible with the generation of the Yukawa coupling. Put differently, these three curves are connected by a 3-chain, which shrinks to zero size at the codimension 3 locus. The restriction of the matter surfaces $S_{v_i}$ will be denoted by the curves $\Sigma_{v_i}$.

\end{itemize}


\subsection{Higher Codimension Loci, Minimality and Flatness}\label{sec:MinFlat}

In the following table \ref{BigTable} we provide the codimension 2 and 3 loci for the Tate forms of appendix \ref{app:Tate}. 
The table contains loci, which are non-minimal. 
We marked those accordingly by putting them into brackets $[\, \cdots]$. 
A singularity is non-minimal along a codimension $1$ locus $\zeta_0=0$ if the  vanishing orders of $f$, $g$ and $\Delta$ all exceed $(3,5,11)$, i.e. it is non-minimal if 
\begin{equation}\label{NonMinDef}
f = O(\zeta_0^4) \,,\qquad
g = O(\zeta_0^6) \,,\qquad 
\Delta = O(\zeta_0^{12}) \,.
\end{equation}
The sections $f$ and $g$ of the Weierstrass form are related to the sections in the Tate form by
\begin{equation}\label{WeierTate}
\ba
f &= \frac{1}{48} \left(-\left(a_1^2+4 a_2\right){}^2+24 a_1 a_3+48 a_4\right) \cr
g&= \frac{1}{864} \left(-\left(a_1^2+4 a_2\right){}^3+36 \left(a_1 a_3+2 a_4\right) \left(a_1^2+4 a_2\right)-216 \left(a_3^2+4 a_6\right)\right) \,,
\ea
\end{equation}
where $a_n = b_n\zeta_0^{i_n}$ in (\ref{TateForm}). 

\ssn{We define non-minimality in 
 codimension 2  by 
 \begin{equation}
 f|_{\Delta_d=0} = O(\zeta_0^4)\,,\qquad  
 g|_{\Delta_d=0} = O(\zeta_0^6)\,,\qquad 
 \Delta|_{\Delta_{d}=0} = O(\zeta_0^{12}) \,,
 \end{equation}
and  in codimension 3 by 
\begin{equation}
f|_{\Delta_{d}=\Delta_{d+1}=0} = O(\zeta_0^4)\,,\qquad   
g|_{\Delta_{d}=\Delta_{d+1}=0} =  O(\zeta_0^6)\,,\qquad  
\Delta|_{\Delta_{d}=\Delta_{d+1}=0} =O(\zeta_0^{12})\,,
\end{equation}
in the notation of (\ref{CodimDef}). 
For matter this question was addressed in a Tate algorithmic way in \cite{Grassi:2011hq}.  An example for  non-minimal codimension 3 singularity and its resolution appeared in  \cite{Candelas:2000nc}. 

By explicitly resolving the non-minimal loci in higher codimension we show that these correspond in fact to loci where the fiberation ceases to be flat, and in particular generates higher dimensional components, which are not curves. This seems indicative that non-minimality implies non-flatness of the fiberation. 

}

There are a few low rank outlier cases that are minimal: 
\begin{itemize}
\item $Sp(k)$ for $k=2$: $b_1^2 + 4 b_2 =0$ and $b_1^2 + 4 b_2 =b_1 b_3 + 2 b_4=0$ are minimal
\item $Sp(k)$ for $k=3$: $b_1^2 + 4 b_2 =0$ is minimal
\item $SU(2k)$ for $k=3$: $b_1=b_2=0$ is minimal
\end{itemize}
Finally, note that the following abbreviations were used in the table:
{\footnotesize
\begin{equation}
\ba
SU(3):\qquad & P\equiv 4b_4^3 + 4b_2^3b_6 - b_2^2b_4^2 - 18b_2b_4b_6 - 27b_6^2  \cr 
G_2: \qquad &  P\equiv  4 b_6 b_2^3-b_4^2 b_2^2-18 b_4 b_6 b_2+4 b_4^3+27 b_6^2  \cr
		     & Q\equiv   2 b_3^2 b_2^3-2 b_1 b_3 b_4 b_2^2+6 b_1^2 b_6 b_2^2-b_1^2 b_4^2 b_2-9 b_3^2 b_4 b_2\cr
		 		 &\qquad    -18 b_1 b_3 b_6 b_2+12 b_1 b_3 b_4^2+27 b_3^2 b_6-9 b_1^2 b_4 b_6 \cr
SO(11) :\qquad & Q\equiv  b_1b_2b_3b_4 - b_2^2b_3^2 - b_1^2b_2b_6 + 2b_4b_6		  	 \cr
SO(13): \qquad & Q\equiv  b_6 b_1^6-b_3 b_4 b_1^5-b_4^2 b_1^4-\left(b_3^3+36 b_6 b_3\right) b_1^3 \cr
		&\qquad +6 b_4 \left(5 b_3^2-12 b_6\right) b_1^2+96 b_3 b_4^2 b_1+64 b_4^3+27 \left(b_3^2+4   b_6\right){}^2 \cr		  
SO(14): \qquad &  P\equiv  \left(b_1^3-27 b_3\right) b_3^3+b_1^2 \left(b_1^3-30 b_3\right) b_4 b_3-64 b_4^3+b_1 \left(b_1^3-96 b_3\right) b_4^2 \cr
E_8 : \qquad & Q\equiv b_3 b_4 b_1^5-b_2 b_3^2 b_1^4+b_4^2 b_1^4+b_3^3 b_1^3+8 b_2 b_3 b_4 b_1^3-8 b_2^2 b_3^2 b_1^2+8 b_2 b_4^2 b_1^2 \cr
 &\qquad-30 b_3^2 b_4 b_1^2+36 b_2 b_3^3 b_1-96 b_3 b_4^2   b_1+16 b_2^2 b_3 b_4 b_1-27 b_3^4-64 b_4^3\cr
&\qquad   -16 b_2^3 b_3^2+16 b_2^2 b_4^2+72 b_2 b_3^2 b_4	
\ea
\end{equation}
}



$${\footnotesize
	\begin{array}{|c|c|r|r|}
		\hline
		\hbox{Type} & \hbox{Group} & \hbox{Codim 2} & \hbox{Codim 3} \cr
		\hline & & &\cr
		I_2 & SU(2) &
		\begin{aligned}
			b_1^2 + 4b_2 = 0 \cr
			P \equiv b_1^2b_6 + b_2b_3^2 + 4b_2b_6 - b_4^2 - b_1b_3b_4 = 0 \cr
		\end{aligned} &
		\begin{aligned}
			b_1^2 + 4b_2 = b_1b_3 + 2b_4 = 0 \cr
			P = b_3^2 + 4b_6 = 0 \cr
			P = b_1^2 + 4b_2 = 0 \cr
			P = b_1b_3 + 2b_4 = 0 \cr
		\end{aligned} \cr
		\hline & & & \cr
		


		I_{2k}^{ns} & Sp(k) & 
		\begin{aligned}
			\left[b_1^2 + 4b_2 = 0 \right]\cr
			P \equiv b_1^2b_6 + b_2(b_3^2 + 4b_6) - b_4(b_1b_3 + b_4) = 0 \cr
		\end{aligned} &
		\begin{aligned}
			\left[b_1^2 + 4b_2 = b_1b_3 + 2b_4 = 0\right] \cr
			P = b_1b_3 + 2b_4 = 0 \cr
			P = 4b_4^2 + 4b_1b_3b_4 + b_1^2b_3^2 = 0 \cr
		\end{aligned} \cr
		\hline & & & \cr
		
		I_4^s & SU(4) & 
		\begin{aligned}
			b_1 = 0 \cr
			P \equiv b_1b_3b_4 + b_4^2 - b_1^2b_6 = 0 \cr
		\end{aligned} & 
		\begin{aligned}
			b_1 = 4b_4^2(b_2^2 - 4b_4) + b_2^2(b_3^2 + 6b_6) = 0 \cr
			P = b_2 = 0 \cr
			P = b_3^2 + 4b_6 = 0 \cr
		\end{aligned} \cr
		\hline & & & \cr

		I_{2k}^s & SU(2k) & 
		\begin{aligned}
			b_1 = 0 \cr
			P \equiv b_1b_3b_4 + b_4^2 - b_1^2b_6 = 0 \cr
		\end{aligned} & 
		\begin{aligned}
			\left[b_1 = b_2 = 0\right] \cr
			b_1 = b_4 = 0 \cr
			P = b_2 = 0 \cr
			P = b_3^2 + 4b_6 = 0 \cr
		\end{aligned} \cr
		\hline & & & \cr

		
		I_5^s & SU(5) &
		\begin{aligned}
			b_1 = 0 \cr
			P \equiv b_2b_3 + b_1(b_1b_6 - b_3b_4) = 0 \cr
		\end{aligned} &
		\begin{aligned}
			b_1 = b_2 = 0 \cr
			b_1 = b_3 = 0 \cr
			P = b_3^2 + b_4^2b_1 + 4b_1b_2b_6 = 0 \cr
		\end{aligned} \cr
		\hline & & & \cr

		I_{2k+1}^s & SU(2k+1) & 
		\begin{aligned} 
			b_1 = 0 \cr
			P \equiv b_2b_3^2 + b_1(b_1b_6 - b_3b_4) = 0 \cr
		\end{aligned} &
		\begin{aligned}
			\left[b_1 = b_2 = 0\right] \cr
			b_1 = b_3 = 0 \cr
			P = b_4^2 - 4b_2b_6 = 0 \cr
		\end{aligned} \cr

		\hline & & & \cr
	
		III & SU(2) &
		\begin{aligned}
			b_4 = 0 \cr
		\end{aligned} &
		\begin{aligned}
			b_4 = b_3^2 + 4b_6 = 0 \cr
		\end{aligned} \cr
		\hline & & & \cr

		
		IV^s & SU(3) &
		\begin{aligned}
			b_3 = 0 \cr
		\end{aligned} &
		\begin{aligned} 
			b_3 = P= 0 \cr
		\end{aligned} \cr
		\hline & & & \cr

		I_0^{*ns} & G_2 & 
		\begin{aligned}
			P = 0 \cr
		\end{aligned} &
		\begin{aligned}
			P = Q = 0 \cr
		\end{aligned} \cr
		\hline & & & \cr
		
		I_0^{*ss} & SO(7) &
		\begin{aligned}
			b_4 = 0 \cr
			P \equiv b_2^2 - 4b_4 = 0 \cr
		\end{aligned} &
		\begin{aligned}
			b_4 = b_3^2 + 4b_6 = 0 \cr
			P = (b_1b_2 - 2b_3)^2 - 16b_6 = 0 \cr
		\end{aligned} \cr
		\hline & & & \cr

		I_1^{*ns} & SO(9) &
		\begin{aligned}
			b_2 = 0 \cr
			b_3^2 + 4b_6 = 0 \cr
		\end{aligned} & 
		\begin{aligned}
			b_2 = b_3^2 + 4b_6 = 0 \cr
			b_3^2 + 4b_6 = b_1b_3 + 2b_4 = 0 \cr
		\end{aligned} \cr
		\hline & & & \cr

		I_1^{*s} & SO(10) &
		\begin{aligned}
			b_2 = 0 \cr
			b_3 = 0 \cr
		\end{aligned} &
		\begin{aligned}
			b_2 = b_3 = 0 \cr
			b_3 = 4b_2b_6 - b_4^2 = 0 \cr
		\end{aligned} \cr
		\hline& & & \cr

		I_2^{*ns} & SO(11) &
		\begin{aligned}
			b_2 = 0 \cr
			P \equiv b_4^2 - 4b_2b_6 = 0 \cr
		\end{aligned} &
		\begin{aligned}
			b_2 = b_4 = 0 \cr
			P =Q = 0 \cr
		\end{aligned} \cr
		\hline 		
	\end{array}
}$$

\begin{table}
$$
{\footnotesize	\begin{array}{|c|c|r|r|}
			\hline
		\hbox{Type} & \hbox{Group} & \hbox{Codim 2} & \hbox{Codim 3} \cr
		\hline & & & \cr
		I_2^{*s} & SO(12)^{*} &
		\begin{aligned}
			b_2 = 0 \cr
			b_4 = 0 \cr
		\end{aligned} & 
		\begin{aligned}
			\left[b_2 = b_4 = 0\right] \cr
			b_3 = b_4 = 0 \cr
		\end{aligned} \cr
		\hline & & & \cr

		I_3^{*ns} & SO(13) &
		\begin{aligned}
			\left[b_2 = 0 \right]\cr
			b_3^2 + 4b_6 = 0 \cr
		\end{aligned} &
		\begin{aligned}
			\left[b_2 = Q = 0\right] \cr
			b_3^2 + 4b_6 = b_1b_3 + 2b_4 = 0 \cr
		\end{aligned} \cr
		\hline & & & \cr

		I_3^{*s} & SO(14) &
		\begin{aligned}
			\left[b_2 = 0\right] \cr
			b_3 = 0 \cr
		\end{aligned} &
		\begin{aligned}
			\left[b_2 = P = 0\right] \cr
			b_3= b_4^2-4 b_2 b_6 = 0 \cr
		\end{aligned} \cr
		\hline & & & \cr

		I_3^{*ns} & SO(15) & 
		\begin{aligned}
			\left[b_2 = 0\right] \cr
			P \equiv b_4^2 - 4b_2b_6 = 0 \cr
		\end{aligned} &
		\begin{aligned}
			\left[b_2 = b_4 = 0\right] \cr
			\left[b_2 = b_1^4 - 64b_4 = 0\right] \cr
			P = b_2b_3^2 + b_1^2b_6 - b_1b_3b_4 = 0 \cr
		\end{aligned} \cr
		\hline & & & \cr

		I_3^{*s} & SO(16)^{*} &
		\begin{aligned}
			\left[b_2 = 0\right] \cr
			b_4 = 0 \cr
		\end{aligned} & 
		\begin{aligned}
			\left[b_2 = b_4 = 0 \right]\cr
			\left[b_2 = b_1^4 - 64b_4 = 0\right] \cr
			b_3 = b_4 = 0 \cr
		\end{aligned} \cr
		\hline & & & \cr

		I_{2k-3}^{*ns} & SO(4k+1) &
		\begin{aligned}
			\left[b_2 = 0\right] \cr
			b_3^2 + 4b_6 = 0 \cr
		\end{aligned} &
		\begin{aligned}
			\left[b_2 = b_1 = 0 \right]\cr
			\left[b_2 = b_1b_3b_4 + b_4^2 - b_1^2b_6 = 0\right] \cr
			b_3^2 + 4b_6 = b_1b_3 + 2b_4 = 0 \cr
		\end{aligned} \cr
		\hline & & & \cr

		I_{2k-3}^{*s} & SO(4k+2) &
		\begin{aligned}
			\left[b_2 = 0\right] \cr
			b_3 = 0 \cr
		\end{aligned} &
		\begin{aligned}
			\left[ b_2 = b_1 = 0\right] \cr
			\left[b_2 = b_3 = 0 \right]\cr
			\left[b_2 = b_1b_3 + b_4 = 0\right] \cr
			\left[b_2 = b_4 = 0 \right]\cr
			b_3 = 4b_2b_6 - b_4^2 = 0 \cr
		\end{aligned} \cr
		\hline & & & \cr

		I_{2k-2}^{*ns} & SO(4k+3) &
		\begin{aligned}
			\left[b_2 = 0\right] \cr
			P \equiv b_4^2 - 4b_2b_6 = 0 \cr
		\end{aligned} & 
		\begin{aligned}
			\left[b_2 = b_4 = 0\right] \cr
			\left[b_2 = b_1 = 0 \right]\cr
			P = b_2b_3^2 + b_1^2b_6 - b_1b_3b_4 = 0 \cr
		\end{aligned} \cr
		\hline & & & \cr

		I_{2k-2}^{*s} & SO(4k+4)^{*} &
		\begin{aligned}
			\left[b_2 = 0\right] \cr
			b_4 = 0 \cr
		\end{aligned} & 
		\begin{aligned}
			\left[b_2 = b_4 = 0\right] \cr
			\left[b_2 = b_1 = 0\right] \cr
			b_3 = b_4 = 0 \cr
		\end{aligned} \cr
		\hline & & & \cr

		IV^{*ns} & F_4 & 
		\begin{aligned}
			b_3^2 + 4b_6 = 0 \cr
		\end{aligned} &
		\begin{aligned}
			\left[b_3^2 + 4b_6 = b_1b_3 + 2b_4 = 0\right] \cr
		\end{aligned} \cr
		\hline & & & \cr

		IV^{*s} & E_6 &
		\begin{aligned}
			b_3 = 0 \cr
		\end{aligned} &
		\begin{aligned}
			b_3 = b_4 = 0 \cr
		\end{aligned} \cr
		\hline & & & \cr

		III^* & E_7 & 
		\begin{aligned} 
			b_4 = 0 \cr
		\end{aligned} &
		\begin{aligned}
			\left[b_4 = b_6 = 0\right] \cr
		\end{aligned} \cr
		\hline & & & \cr

		II^* & E_8 & 
		\begin{aligned}
			\left[b_6 = 0\right] \cr
		\end{aligned} &
		\begin{aligned}
			\left[b_6 = Q = 0\right] \cr
		\end{aligned} \cr
		\hline
	\end{array}
	}
$$
\caption{Summary of all codimension 2 and 3 loci for fiber types in table \ref{TateTable}. Loci that are non-minimal in the sense defined in section \ref{sec:MinFlat} are put in parenthesis.} 
\label{BigTable}
\end{table}

\newpage


\subsection{General Resolution Procedure}\label{sec:GeneralProc}

We now turn to a discussion of the general process of resolving the singular Tate forms. As explained in the introduction, we are interested in crepant resolutions. 

The Tate form (\ref{TateForm}) is singular at 
\be 
x=y=\zeta_0=0 \,,
\ee
in particular, all the derivatives vanish at this locus. 
The singularity is resolved by blowing up, which introduces a 
new exceptional $\mathbb{P}^2$, specificed by a section $\zeta =0$ and with projective coordinates $[X_1, Y_1, Z_1]$, where 
\begin{equation}
x= X_1 \zeta \,,\qquad 
y= Y_1 \zeta\,,\qquad 
\zeta_0 = Z_1 \zeta \,.
\end{equation}
We use the shorthand for this blowup 
\be\label{BUNotation}
 (x, y, \zeta_0; \zeta ) \,. 
\ee
Applying this repeatedly for all singular loci, given by the vanishing of three sections, resolves the singularity in codimension 1, for $A_n$, but for  $D_n$ and $E_n$, resolutions along loci of the type $y=\zeta=0$ at the end fully resolves the singularity in codimension 1. The canonical class under the blowup (\ref{BUNotation}) changes  by 
$K_{X} \rightarrow K_X + 2 E$, where $E$ is the class of the exceptional divisor specified by $\zeta=0$.

The resulting space (after proper transformation) however is still singular in higher codimensions. 
Small resolutions along these loci resolve the space completely. 
For instance, let $B=0$ inside $Y_4$ specify a codimension 2 singularity, and let the space be singular along $B=Y= Z=0$. A  small resolution can remedy this: introduce an exceptional $\mathbb{P}^1$ given by $\delta =0$ and having projective coordinates $[Y_1, Z_1]$, which are related to the original coordinates by $(Y,Z) =  (\delta Y_1, \delta Z_1)$. 
For this resolution we shall use the shorthand notation
\be
 (Y, Z; \delta) \,.
\ee
The canonical class changes by $K_Y \rightarrow K_Y + E$, where $E$ is the class of exceptional divisor $\delta=0$. 
Repeating this process yields eventually a fully  resolved CY fourfold. 

As observed already in \cite{Esole:2011sm} for $SU(5)$, there are various choices for the small resolutions, which are all birationally equivalent. We shall find that the general structure of the geometries after resolving the codimension 1 singularities takes the form
\be\label{BinoGeo}
y (y+ V) = \zeta_1 \cdots \zeta_\ell U \,,
\ee
where $U, V$ depend on all sections (including the exceptional sections and the $b_n$).  For $SU(5)$, $\ell=2$ and the geometry is of a so-called binomial type, and was resolved by choosing the two loci for the small resolutions to be $y= A=0$ and $(y+V) = B=0$ where $A, B = \zeta_1, \zeta_2$ or $U$. Generically $\ell >2$, and this method for small resolutions does not generalize. We will show that a lot of simplifications in this procedure arise by performing all small resolutions with $y=0$, i.e. for $SU(5)$ as we iterate in appendix \ref{app:Examples}, the two small resolutions are along $y=\zeta_1=0$ and $y=\zeta_2=0$. Generally, the small resolutions will be of the type
\be\label{SmallRes}
(y, \zeta_i; \delta_i) \,,\qquad i=1, \cdots,  \ell \,.
\ee
Note, that we could also include small resolutions along $U$ instead of one of the $\zeta_i$, however, we will consider here only the resolutions of the type  (\ref{SmallRes}). Indeed, there is an interesting web of small resolutions \cite{Esole:2011sm, LS}. The small resolutions we will consider have the property that the exceptional divisors all become irreducible, i.e. to each section in the resolved space, there is a corresponding irreducible exceptional divisor, i.e. these are Cartier.
In the following we apply this resolution to the Tate forms with ADE fibers with the aim to study the structure of the higher codimension fibers. 

For a concise notation of the geometries, it will be useful to define the following abbreviations
\be\label{ABCEDef}
\ba
	A(z) &= \prod_{i=1}^{k-1}z_{i}^{i-1} \cr
	C(z) &= \prod_{i=1}^{k-1}z_{i}^{k-(i+1)} \cr
	 B(z) &= \prod_{i=1}^{k-1} z_i \,.
\ea
\ee
Note that  e.g. $A(\zeta \delta)= \prod_{i=1}^{k-1}(\zeta_{i}\delta_i)^{i-1}$.


\begin{table}
\begin{equation}
\begin{array}{|c|c|c|}
\hline
\hbox{Group} & \hbox{Codim 2 or 3} & \hbox{Fiber} \cr\hline
\begin{picture}(60, 30)
\put(0,20){$SU(2k+1)$}
\end{picture}
 & 
\begin{picture}(70, 30)
\put(10,20){$b_1=0$}
\end{picture}
&
\begin{picture}(80, 40)
	\tiny
	\put(0, 10){\circle*{5}}
	\put(0, 30){\circle*{5}}
	\put(10, 20){\circle*{5}}
	\put(20,20){\circle*{5}}
	\put(30,20){\circle*{5}}
	\put(33, 20){$\ldots$}
	\put(50,20){\circle*{5}}
	\put(60,20){\circle*{5}}	
	\put(70,20){\circle*{5}}
	\put(80,10){\circle*{5}}
	\put(80,30){\circle*{5}}
	\put(10,20){\line(1,0){20}}
	\put(50,20){\line(1,0){20}}
	\put(10,20){\line(-1,1){10}}
	\put(10,20){\line(-1,-1){10}}
	\put(70,20){\line(1,1){10}}
	\put(70,20){\line(1,-1){10}}
	\put(-2, 1){1}
	\put(-2, 21){1}
	\put(8, 11){2}
	\put(18, 11){2}
	\put(28, 11){2}
	\put(48, 11){2}
	\put(58, 11){2}
	\put(68, 11){2}
	\put(78, 1){1}
	\put(78, 21){1}
\end{picture}

\cr\hline
 & 
\begin{picture}(70, 30)
\put(10,20){$P=0$}
\end{picture}
&
\begin{picture}(80, 40)
	\tiny
	\put(10, 10){\circle*{5}}
	\put(20,10){\circle*{5}}
	\put(30,10){\circle*{5}}
	\put(33, 10){$\ldots$}
	\put(50,10){\circle*{5}}
	\put(60,10){\circle*{5}}	
	\put(70,10){\circle*{5}}
	\put(10,10){\line(1,0){25}}
	\put(45,10){\line(1,0){25}}
	\put(10, 30){\circle*{5}}
	\put(20,30){\circle*{5}}
	\put(30,30){\circle*{5}}
	\put(33, 30){$\ldots$}
	\put(50,30){\circle*{5}}
	\put(60,30){\circle*{5}}	
	\put(70,30){\circle*{5}}
	\put(10,30){\line(1,0){20}}
	\put(45,30){\line(1,0){25}}
	\put(80,20){\circle*{5}}
	\put(0,20){\circle*{5}}
	\put(0,20){\line(1,1){10}}
	\put(0,20){\line(1,-1){10}}
	\put(70,10){\line(1,1){10}}
	\put(70,30){\line(1,-1){10}}	
	\put(8, 1){1}
	\put(18, 1){1}
	\put(28, 1){1}
	\put(48, 1){1}
	\put(58, 1){1}
	\put(68, 1){1}

	\put(8, 34){1}
	\put(18, 34){1}
	\put(28, 34){1}
	\put(48, 34){1}
	\put(58, 34){1}
	\put(68, 34){1}

	\put(-2, 11){1}
	\put(78, 11){1}
\end{picture}
\cr\hline
%
 & 
\begin{picture}(70, 30)
\put(10,20){$b_1=b_3=0$}
\end{picture}
&
\begin{picture}(90, 60)
	\tiny
	\put(0, 20){\circle*{5}}
	\put(10, 30){\circle*{5}}
	\put(10, 20){\circle*{5}}
	\put(20,20){\circle*{5}}
	\put(30,20){\circle*{5}}
	\put(32, 20){$\ldots$}
	\put(50,20){\circle*{5}}
	\put(60,20){\circle*{5}}	
	\put(70,20){\circle*{5}}
	\put(80,20){\circle*{5}}
	\put(80,30){\circle*{5}}
	\put(90,20){\circle*{5}}
	\put(0,20){\line(1,0){30}}
	\put(50,20){\line(1,0){40}}
	\put(80,20){\line(0,1){10}}
	\put(10,20){\line(0,1){10}}
	\put(-2, 11){1}
	\put(8, 11){2}
	\put(18, 11){2}
	\put(28, 11){2}
	\put(48, 11){2}
	\put(58, 11){2}
	\put(68, 11){2}
	\put(78, 11){2}
	\put(88, 11){1}
	\put(8, 34){1}
	\put(78, 34){1}

\end{picture}
\cr\hline
\hline
\begin{picture}(60, 30)
\put(0,20){$SU(2k)$}
\end{picture}
 & 
\begin{picture}(70, 30)
\put(10,20){$b_1=0$}
\end{picture}
&
\begin{picture}(80, 40)
	\tiny
	\put(0, 10){\circle*{5}}
	\put(0, 30){\circle*{5}}
	\put(10, 20){\circle*{5}}
	\put(20,20){\circle*{5}}
	\put(30,20){\circle*{5}}
	\put(33, 20){$\ldots$}
	\put(50,20){\circle*{5}}
	\put(60,20){\circle*{5}}	
	\put(70,20){\circle*{5}}
	\put(80,10){\circle*{5}}
	\put(80,30){\circle*{5}}
	\put(10,20){\line(1,0){20}}
	\put(50,20){\line(1,0){20}}
	\put(10,20){\line(-1,1){10}}
	\put(10,20){\line(-1,-1){10}}
	\put(70,20){\line(1,1){10}}
	\put(70,20){\line(1,-1){10}}
	\put(-2, 1){1}
	\put(-2, 21){1}
	\put(8, 11){2}
	\put(18, 11){2}
	\put(28, 11){2}
	\put(48, 11){2}
	\put(58, 11){2}
	\put(68, 11){2}
	\put(78, 1){1}
	\put(78, 21){1}
\end{picture}

\cr\hline
 & 
\begin{picture}(70, 30)
\put(10,20){$P=0$}
\end{picture}
&
\begin{picture}(80, 40)
	\tiny
	\put(10, 10){\circle*{5}}
	\put(20,10){\circle*{5}}
	\put(30,10){\circle*{5}}
	\put(33, 10){$\ldots$}
	\put(50,10){\circle*{5}}
	\put(60,10){\circle*{5}}	
	\put(70,10){\circle*{5}}
	\put(10,10){\line(1,0){25}}
	\put(45,10){\line(1,0){25}}
	\put(10, 30){\circle*{5}}
	\put(20,30){\circle*{5}}
	\put(30,30){\circle*{5}}
	\put(33, 30){$\ldots$}
	\put(50,30){\circle*{5}}
	\put(60,30){\circle*{5}}	
	\put(70,30){\circle*{5}}
	\put(10,30){\line(1,0){20}}
	\put(45,30){\line(1,0){25}}
	\put(0,20){\circle*{5}}
	\put(0,20){\line(1,1){10}}
	\put(0,20){\line(1,-1){10}}
	\put(70,10){\line(0,1){20}}
	\put(8, 1){1}
	\put(18, 1){1}
	\put(28, 1){1}
	\put(48, 1){1}
	\put(58, 1){1}
	\put(68, 1){1}
	
	\put(8, 34){1}
	\put(18, 34){1}
	\put(28, 34){1}
	\put(48, 34){1}
	\put(58, 34){1}
	\put(68, 34){1}

	\put(-2, 11){1}
\end{picture}
\cr\hline
 & 
\begin{picture}(70, 30)
\put(10,20){$b_1=b_4=0$}
\end{picture}
&
\begin{picture}(90, 60)
	\tiny
	\put(0, 20){\circle*{5}}
	\put(10, 20){\circle*{5}}
	\put(10, 30){\circle*{5}}
	\put(20,20){\circle*{5}}
	\put(30,20){\circle*{5}}
	\put(32, 20){$\ldots$}
	\put(50,20){\circle*{5}}
	\put(60,20){\circle*{5}}	
	\put(70,20){\circle*{5}}
	\put(80,20){\circle*{5}}
	\put(80,30){\circle*{5}}
	\put(90,20){\circle*{5}}
	\put(10, 20){\line(0,1){10}}
	\put(0,20){\line(1,0){30}}
	\put(50,20){\line(1,0){40}}
	\put(80,20){\line(0,1){10}}

	\put(-2, 11){1}
	\put(8, 11){2}
	\put(18, 11){2}
	\put(28, 11){2}
	\put(48, 11){2}
	\put(58, 11){2}
	\put(68, 11){2}
	\put(78, 11){2}
	\put(88, 11){1}
	\put(8, 34){1}
	\put(78, 34){1}
\end{picture}
\cr\hline\hline


\begin{picture}(60, 30)
\put(0,20){$SO(4k+2)$}
\end{picture}
 & 

%
\begin{picture}(70, 30)
\put(10,20){$b_3=0$}
\end{picture}
&
\begin{picture}(90, 60)
	\tiny
	\put(10, 30){\circle*{5}}
	\put(0, 20){\circle*{5}}
	\put(10, 20){\circle*{5}}
	\put(20,20){\circle*{5}}
	\put(30,20){\circle*{5}}
	\put(32, 20){$\ldots$}
	\put(50,20){\circle*{5}}
	\put(60,20){\circle*{5}}	
	\put(70,20){\circle*{5}}
	\put(80,20){\circle*{5}}
	\put(80,30){\circle*{5}}
	\put(90,20){\circle*{5}}
	\put(10,20){\line(0,1){10}}
	\put(0,20){\line(1,0){30}}
	\put(50,20){\line(1,0){40}}
	\put(80,20){\line(0,1){10}}
	\put(-2, 11){1}
	\put(8, 11){2}
	\put(18, 11){2}
	\put(28, 11){2}
	\put(48, 11){2}
	\put(58, 11){2}
	\put(68, 11){2}
	\put(78, 11){2}
	\put(88, 11){1}
	\put(8, 34){1}
	\put(78, 34){1}
\end{picture}
\cr\hline

\hline\hline
\begin{picture}(60, 30)
\put(0,20){$SO(4k+4)$}
\end{picture}

%

 & 
\begin{picture}(70, 30)
\put(10,20){$b_4=0$}
\end{picture}
&
\begin{picture}(90, 40)
	\tiny
	\put(10, 30){\circle*{5}}
	\put(0, 20){\circle*{5}}
	\put(10, 20){\circle*{5}}
	\put(20,20){\circle*{5}}
	\put(30,20){\circle*{5}}
	\put(32, 20){$\ldots$}
	\put(50,20){\circle*{5}}
	\put(60,20){\circle*{5}}	
	\put(70,20){\circle*{5}}
	\put(80,20){\circle*{5}}
	\put(80,30){\circle*{5}}
	\put(90,20){\circle*{5}}
	\put(10,20){\line(0,1){10}}
	\put(0,20){\line(1,0){30}}
	\put(50,20){\line(1,0){40}}
	\put(80,20){\line(0,1){10}}

	\put(-2, 11){1}
	\put(8, 11){2}
	\put(18, 11){2}
	\put(28, 11){2}
	\put(48, 11){2}
	\put(58, 11){2}
	\put(68, 11){2}
	\put(78, 11){2}
	\put(88, 11){1}
	\put(8, 34){1}
	\put(78, 34){1}
\end{picture}
\cr\hline
\end{array}
\end{equation}
\caption{Summary of fibers in higher codimension including multiplicities. The cases when the fibration is not flat can be found in the respective subsections. }
\label{FiberSummaryTable}
\end{table}



\subsection{Fibers in Higher Codimension}

As a summary of our results we now list the properties of the fibers in codimension 2 and 3 for $A_n$ and $D_n$, $n$ even and odd. 
We will be mainly concerned with the representations that occur in codimension 2, and the Yukawa couplings that are generated in codimension 3, as this will generalize to the non-minimal cases. 
We tabulate how the Cartan divisors split into matter representations of the gauge group, and how these further split along the Yukawa points. 
These are the data that are most relevant in order to understand the physical interpretation in terms of F-theory and 7-brane effective theories. 
The specific intersections of the fibers, which may be of interest from a geometric point of view are shown in each of the following sections \ssn{and are summarized in table \ref{FiberSummaryTable}}.
 In codimension 2 most intersections give rise to affine Dynkin diagrams, and for minimal singularities have multiplicities as required for a Kodaira fiber. 
 
\ssn{For the non-minimal cases, marked by parenthesis in table \ref{BigTable}, the fibers are  non-Kodaira throughout. In fact, we observe that  the non-minimal fibers are in fact non-flat fibrations, i.e. the dimensionality changes. Along non-minimal loci we show that there are surface components in the fiber. The presence of such fiber components is indicative that a pure field theoretic description of the low energy theory is not going to suffice.

Non-flatness of the fibration means that in the singular limit surface components that can be wrapped by M5 branes shrink to zero size, thus yielding additional degrees of freedom beyond wrapped M2 branes on the curve components of the fiber, which give rise to the gauge bosons of a SYM theory. It is thus to be expected that a field theoretic description of the effective theory will be insufficient at capturing all the degrees of freedom. It is interesting to see how to incorporate the additional M5 brane modes.  }

Nevertheless we can follow what happens to the Cartan divisors along non-minimal loci. In codimension 2, this occurs for the $SO(2n)$  groups along the matter locus $b_2=0$, where we show that the fiber splits such that the irreducible components carry the Cartan charges of spin representations. Similarly, for the $SU(n)$ groups, the codimension 3 loci, which would generalize the E-type enhancements at low $n$, yield non-minimal singularities. Nevertheless, as elaborated upon before, these can be studied in terms of  the splitting of the matter surfaces, which is consistent with the generation of certain Yukawa couplings (generalizing the top Yukawa for $SU(5)$ for instance). Matter and couplings that arise at such non-minimal loci are put in parenthesis in table \ref{BigTable}. 
\bigskip

\begin{table}
\begin{equation}
\begin{array}{|c|l|l|}
\hline
\hbox{Group} & \hbox{Codim 2/3} & \hbox{Representation/Yukawa} \cr\hline
SU(2k+1)  
&
b_1=0 
&
\Lambda^2V \,,\  \Lambda^2 \overline{V}
\cr
&
P=0 
&
V\,,\  \overline{V} 
\cr\hline
&
b_1=b_3=0
&
\bar{V} \otimes \bar{V}\otimes \Lambda^2V  \cr
&
b_1=b_2=0 
& 
\left[ \Lambda^2 \bar{V} \otimes \Lambda^2 \bar{V} \otimes \Lambda^4 V \right] \cr
&
P=Q=0 
&
V\otimes \bar{V} \otimes {\bf 1}
\cr\hline\hline

SU(2k)  
&
b_1=0 
&
\Lambda^2V \,,\  \Lambda^2 \overline{V}
\cr
&
P=0 
&
V\,,\  \overline{V} 
\cr\hline
&
b_1=b_4=0
&
\bar{V} \otimes \bar{V}\otimes \Lambda^2V  \cr
&
b_1=b_2=0 
& 
\left[ \Lambda^2 \bar{V} \otimes \Lambda^2 \bar{V} \otimes \Lambda^4 V \right] \cr
&
P=b_2=0 
&
V\otimes \bar{V} \otimes {\bf 1}
\cr\hline\hline

SO(4k+2)
&
b_3=0 
& 
V
\cr
&
b_2=0 
&
\left[ S^{\pm}  \right]
\cr
\hline
&
b_3= b_4^2 - 4 b_2 b_6=0
&
V\otimes V \otimes {\bf 1}
\cr
&
b_2= b_3=0
&
\left[ S^+\otimes S^- \otimes V\right]
\cr\hline\hline

SO(4k+4)
&
b_2=0 
&
\left[ S^+\right]
\cr
&
b_4=0 
&
V
\cr\hline
&
b_3 = b_4 = 0
&
V\otimes V \otimes {\bf 1}
\cr
& b_2= b_4=0
& \left[V\otimes V\otimes \Lambda^2V\right]
\cr\hline
\end{array}
\end{equation}
\bigskip
\caption{Matter and Yukawa couplings that arise at the codimension 2 and 3 singularities for ADE gauge groups. Entries in brackets indicate non-minimal singularities, and we will explain in what way the entries are justified in those cases. }
\end{table}


\section{Resolution of $A$ Type Singularities}

Consider the Tate forms for $A$ type singularities. 
For $SU(n)$ we need to discuss the case of $n$ even and odd separately
\begin{equation}
\ba
		n = 2k: &\qquad 0= y^2 - x^3 + b_1xy - b_2x^2\zeta_0 + b_3y\zeta_0^k 
					- b_4x\zeta_0^k - b_6\zeta_0^{2k} \cr
		n = 2k + 1: &\qquad  0= y^2 - x^3 + b_1xy - b_2x^2\zeta_0 + b_3y\zeta_0^k
					- b_4x\zeta_0^{k+1} - b_6\zeta_0^{2k+1}  \,,
\ea
\end{equation}
as the higher codimension structure is different. Note further, that for the low $k$ values, some degeneracies may occur. We provide, for some of these outlier cases in appendix \ref{app:Examples}, the explicit resolutions.



\subsection{$SU(2k+1)$}

The Tate form for $SU(2k+1)$ is 
\begin{equation}\label{TateA2k1}
0= y^2 - x^3 + b_1xy - b_2x^2\zeta_0 + b_3y\zeta_0^k	- b_4x\zeta_0^{k+1} - b_6\zeta_0^{2k+1} \,.
\end{equation}
To resolve this in codimension 1, we apply the  procedure outlined in section \ref{sec:GeneralProc}. The blowups, in the notation (\ref{BUNotation}), are 
\begin{equation}\label{SUoddBIG}
(x, y, \zeta_i; \zeta_{i+1}) \,,\qquad i=0, \cdots, k-1 \,.
\end{equation}
After proper transformation, the resulting space is smooth in codimension 1 and  takes the form
\be\label{SUCodim1Smooth}\ba
&y \left( y + b_1x + b_3 \zeta_0^kB(\zeta)C(\zeta)  \right) \cr
& \quad =   B(\zeta)\left[  \zeta_0\zeta_k b_2 x^2 	+   x^3A(\zeta)\zeta_k^k 	+ b_4x\zeta_0^{k+1} \zeta_k B(\zeta) C(\zeta) - b_6\zeta_0^{2k+1}  \zeta_k B(\zeta^2) C(\zeta^2)  \right] \,.
\ea\ee
This has the general form
\begin{equation}\label{SUGenForm}
y Y = \zeta_1 \cdots \zeta_k \, V \,.
\end{equation}
In particular, this geometry is still singular in higher codimension. We choose the following small resolutions to remedy this
\begin{equation}\label{SUoddSMALL}
(y, \zeta_i; \delta_i) \,,\qquad i= 1, \cdots , k \,.
\end{equation}
As explained earlier, there is a choice in the set of small resolutions, and their structure will be studied elsewhere \cite{LS}. Note also, that the small resolution in \cite{MS} is different from the one chosen here. This is mainly due to the fact that the one chosen here, is easily generalized to all  values of $k$,  (\ref{SUGenForm}) ensures that there is a one to one correspondence between irreducible exceptional divisors and sections. 
The fully resolved geometry for $SU(2k+1)$ is 
\be\label{TateResSUodd}
\ba
\tilde{T}_{SU(2k+1)}:&\qquad 	\cr
		y^2
		B(\delta)\delta_k &- x^3B(\zeta)A(\zeta \delta)\zeta_k^k\delta_k^{k-1} + b_1xy - b_2x^2\zeta_0B(\zeta)\zeta_k 
		+ b_3y\zeta_0^kB(\zeta\delta)C(\zeta \delta) 
			\cr &- b_4x\zeta_0^{k+1}B(\zeta^2 \delta)C(\zeta\delta)\zeta_k 
			- b_6\zeta_0^{2k+1}B(\zeta^3\delta^2)C(\zeta^2\delta^2)\zeta_k = 0 \,,
\ea
\ee
with the abbreviations (\ref{ABCEDef}). The exceptional sections are 
\begin{equation}
\ba
\zeta_i\,,&\qquad i= 0, \cdots, k \,, \cr
\delta_i\,,& \qquad i=1, \cdots, k \,.
\ea
\end{equation}
Their classes as well as the projective relations among these are listed in appendix \ref{app:SUn}.
In the following sections we will study the fibers in codimension 1, 2 and 3.


\begin{figure}
\begin{center}
\begin{picture}(280, 140)

	\put(-20,50){\circle*{10}}
	\put(-30,30){$\zeta_0$}

	\put(0,20){\circle*{10}}
	\put(40,20){\circle*{10}}
	\put(80,20){\circle*{10}}

	\put(140,20){$\ldots$}

	\put(200,20){\circle*{10}}
	\put(240,20){\circle*{10}}
	\put(280,20){\circle*{10}}

	\put(0,20){\line(1,0){110}}
	\put(170,20){\line(1,0){110}}

	\put(-5,0){$\zeta_1$}
	\put(35, 0){$\zeta_2$}
	\put(75,0){$\zeta_3$}
	\put(195,0){$\zeta_{k-2}$}
	\put(235,0){$\zeta_{k-1}$}
	\put(275,0){$\zeta_k$}


	\put(0,80){\circle*{10}}
	\put(40,80){\circle*{10}}
	\put(80,80){\circle*{10}}

	\put(140,80){$\ldots$}

	\put(200,80){\circle*{10}}
	\put(240,80){\circle*{10}}
	\put(280,80){\circle*{10}}

	\put(0,80){\line(1,0){110}}
	\put(170,80){\line(1,0){110}}

	\put(-5,95){$\delta_1$}
	\put(35, 95){$\delta_2$}
	\put(75,95){$\delta_3$}
	\put(195,95){$\delta_{k-2}$}
	\put(235,95){$\delta_{k-1}$}
	\put(275,95){$\delta_k$}

	
	\put(0,80){\line(-2,-3){20}}
	\put(0,20){\line(-2,3){20}}
	\put(280,80){\line(0,-1){60}}

\end{picture}
\end{center}
\caption{Intersection graph of the fibers in codimension 1, resulting in the affine $A_{2k}$ Dynkin diagram. The labels denote the sections which give rise to the Cartan divisors $D_{-\alpha_i}$ as is spelled out in (\ref{SUCartanDivs}).}\label{fig:DynkingSUn}
\end{figure}
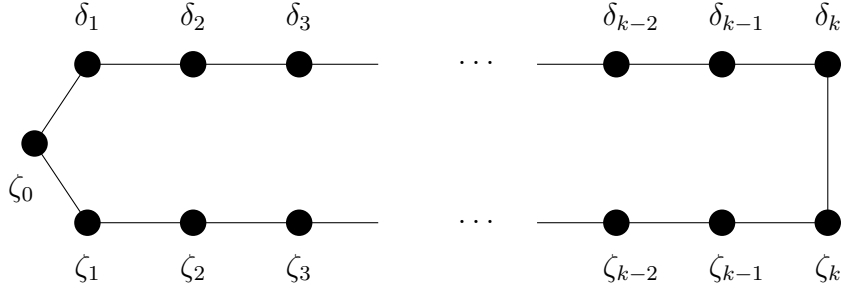

\subsubsection{Codimension 1}

The resolved geometry is given in terms of projective coordinates, which have to satisfy projective relations, which we list in appendix \ref{app:SUn}.
The small resolutions were chosen so that the vanishing locus of each section is irreducible. For each section the equation of the resolved Tate form $Y_4$ restricted to the vanishing locus of the section is listed in the following table.
\begin{equation}\label{SUCartanDivs}
	\begin{array}{l|c|l}
		\hbox{Divisor} & \hbox{Section } & \hbox{Equation in $Y_4$ }\cr\hline
		D_{-\alpha_0} & \zeta_0  &0=  y^2\delta_1 - x^3\zeta_1 + b_1xy  \cr
		D_{-\alpha_1} & \zeta_1  &0= \delta_1 + b_1x  \cr
		D_{-\alpha_i} \quad\hfill i = 2,\cdots,k-1 & \zeta_i  & 0=\delta_{i-1}\delta_i + b_1x  \cr
		D_{-\alpha_k} & \zeta_k  & 0=\delta_{k-1}\delta_k + b_1x + b_3\zeta_{k-1}\delta_{k-1}  \cr
		D_{-\alpha_{k+1}} & \delta_{k}  & 0=b_1yx - b_2x^2\zeta_k + b_3y\delta_{k-1} - b_4x\zeta_k\delta_{k-1} - b_6\zeta_k\delta_{k-1}^2  \cr
		D_{-\alpha_{2k + 1 - i}} \quad\hfill i = 2,\cdots,k-1 & \delta_i  & 0=b_1y - b_2\zeta_i\zeta_{i+1} \cr
		D_{-\alpha_{2k}} & \delta_1 & 0=b_1y - b_2\zeta_0\zeta_1\zeta_2 - \zeta_1\zeta_2^2\delta_2  \cr
	\end{array}
\end{equation}
These irreducible exceptional divisors are the  {\it Cartan divisors} $D_{-\alpha_i}$, and are naturally labeled them by the simple roots of $SU(2k+1)$, with $\alpha_0$ labeling the affine root. 
Note that the Cartan divisors in $Y_4$ can be thought of as fibering $\mathbb{P}^1$s in the resolved fiber over the surface $S_2$. Computing the intersection of the Cartan divisors with the exceptional $\mathbb{P}^1$s in the resolved fiber reproduces the (extended) Cartan matrix of the group $SU(2k+1)$, as depicted in figure \ref{fig:DynkingSUn}. To each Cartan divisor, we can associate a Cartan charge vector computed from the intersections with the $\mathbb{P}^1$s given by
\begin{equation}
	\begin{array}{l|l}
		\mbox{Cartan Divisor} & \mbox{$j$th Cartan charge} \cr\hline
		D_{-\alpha_0} & (-\alpha_0)_j= \delta_{j,1} + \delta_{j,2k} - 2\delta_{j,0} \cr
		D_{-\alpha_i} & (-\alpha_i)_j= \delta_{i,j+1} + \delta_{j,i+1} - 2\delta_{i,j} \cr
		D_{-\alpha_{2k}} & (-\alpha_{2k})_j = \delta_{j,2k-1} + \delta_{j,0} - 2\delta_{j,2k} \cr
	\end{array}
\end{equation}



\subsubsection{Codimension 2 }\label{sec:SUodd2}

Along higher codimension loci of enhanced symmetry, which we summarized in table \ref{BigTable}, the exceptional divisors will split. The particular splitting has an interesting structure, and depends on the choice of small resolutions. There are two codimension 2 matter loci for $SU(2k+1)$: $b_1=0$ and $P=0$.
The vanishing order $O(\zeta_0^{2k+1})$ of the discriminant increases along these loci as follows
\begin{equation}
\ba
b_1=0:\qquad& \Delta|_{b_1=0} = O(\zeta_0^{2k+3}) \cr
P =0:\qquad& \Delta|_{P=0} = O(\zeta_0^{2k+2}) \,,
\ea
\end{equation}
which has a clear representation theoretic interpretation: the matter that is localized corresponds to decomposition of the adjoint of $SO(4k+2)$ and $SU(2k+2)$, respectively, with respect to $SU(2k+1)$. We confirm this from the intersection diagrams in codimesion 2, which for $A_n$ singularities give rise to $D$ and $A$ type Dynkin diagrams and the fibers  have the correct multiplicities for Kodaira fibers. 

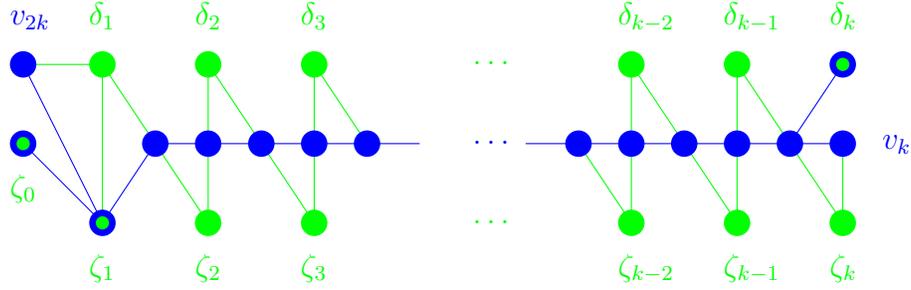
\begin{figure}
\begin{center}
\begin{picture}(280, 140)

\color{green}

	\put(0,20){\color{blue}\line(2,3){20}}
	\put(0,20){\color{blue}\line(-1,1){30}}	
	\put(0,20){\color{blue}\line(-1,2){30}}	
	\put(20,50){\color{blue}\line(1,0){100}}
	\put(160,50){\color{blue}\line(1,0){120}}
	\put(260,50){\color{blue}\line(2,3){20}}


	\put(0,80){\line(-1,0){30}}
		
	\put(0,20){\line(0,1){60}}
	\put(40,20){\line(0,1){60}}
	\put(80,20){\line(0,1){60}}
	
	\put(200,20){\line(0,1){60}}
	\put(240,20){\line(0,1){60}}
	\put(280,20){\line(0,1){30}}
	
	\put(40,20){\line(-2,3){40}}
	\put(80,20){\line(-2,3){40}}
	\put(80,80){\line(2,-3){20}}

	\put(200,20){\line(-2,3){20}}
	\put(240,20){\line(-2,3){40}}
	\put(280,20){\line(-2,3){40}}

	\put(0,20){\color{blue}\circle*{10}}
	\put(0,20){\circle*{5}}	
	\put(40,20){\circle*{10}}
	\put(80,20){\circle*{10}}

	\put(140,20){$\ldots$}

	\put(200,20){\circle*{10}}
	\put(240,20){\circle*{10}}
	\put(280,20){\circle*{10}}

	\put(-5,0){$\zeta_1$}
	\put(35, 0){$\zeta_2$}
	\put(75,0){$\zeta_3$}
	\put(195,0){$\zeta_{k-2}$}
	\put(235,0){$\zeta_{k-1}$}
	\put(275,0){$\zeta_k$}

\color{blue}
	
	\put(20,50){\circle*{10}}
	\put(40,50){\circle*{10}}
	\put(60,50){\circle*{10}}
	\put(80,50){\circle*{10}}
	\put(100,50){\circle*{10}}

	\put(140,50){$\ldots$}

	\put(180,50){\circle*{10}}
	\put(200,50){\circle*{10}}
	\put(220,50){\circle*{10}}
	\put(240,50){\circle*{10}}
	\put(260,50){\circle*{10}}
	\put(280,50){\circle*{10}}
	
	\put(295,48){$v_{k}$}

\color{green}


	\put(-30,50){\color{blue}\circle*{10}}
	\put(-30,50){\circle*{5}}
	\put(-35,30){$\zeta_0$}

{\put(-30,80){\color{blue}\circle*{10}}}
	\put(0,80){\circle*{10}}
	\put(40,80){\circle*{10}}
	\put(80,80){\circle*{10}}

	\put(140,80){$\ldots$}

	\put(200,80){\circle*{10}}
	\put(240,80){\circle*{10}}
	{\color{blue}\put(280,80){\circle*{10}}}
	\put(280,80){\circle*{5}}

	\put(-35,95){\color{blue}$v_{2k}$}
	\put(-5,95){$\delta_1$}
	\put(35, 95){$\delta_2$}
	\put(75,95){$\delta_3$}
	\put(195,95){$\delta_{k-2}$}
	\put(235,95){$\delta_{k-1}$}
	\put(275,95){$\delta_k$}

\end{picture}
\end{center}
\caption{Splitting of the Cartan divisors $D_{-\alpha_i}$ along the codimension 2 locus $b_1=0$ for $SU(2k+1)$. $\color{green}\bullet$ are  the simple roots of $A_{2k}$, green lines indicate how they split along $b_1=0$. $\color{blue}\bullet$ are the irreducible components of the fiber over the codimension 2 locus, the blue lines give their intersection graph, which reproduces the affine $D_{2k+1}$. Curves that remain irreducible are bicolored.}
\label{fig:SUnSplittingb1}
\end{figure}

Denote the matter surfaces by $S_{v}$, where $v$ is the Cartan charge obtained by intersecting the surface with the Cartan divisors. 
Along $b_1 = 0$ the matter surfaces, which are  irreducible components of the restriction of the Cartan divisors, are
\begin{equation}\label{SUoddb1Weights}
	\begin{array}{l|c|l|l}
		\mbox{Matter} & \mbox{Section} & \mbox{Equation in $Y_4|_{b_1=0}$} & j\mbox{th Cartan charge } \cr\hline
		S_{-\alpha_0} & \zeta_0 & 0= y^2\delta_1 - x^3\zeta_1 & \delta_{j,1} + \delta_{j,2k} - 2\delta_{j,0} \cr
		S_{-\alpha_1} & \zeta_1  & 0=\delta_1  & \delta_{j,2} + \delta_{j,0} - 2\delta_{j,1} \cr
		S_{v_i} \qquad i = 2,\cdots,k-1 & \zeta_i  & 0=\delta_i 
			& \delta_{j,i+1} - \delta_{j,i} + \delta_{j,2k+2-i} - \delta_{j,2k+1-i} \cr
		S_{v_{k+i}}\, \quad i = 2,\cdots,k-1 & \zeta_i  & 0=\delta_{i-1}   
			& \delta_{j,i-1} - \delta_{j,i} - \delta_{j,2k+2-i} + \delta_{j,2k+1-i} \cr
		S_{v_k} & \zeta_k & 0=b_3\zeta_{k-1} + \delta_k & -\delta_{j,k} + \delta_{j,k+2} \cr	
		S_{v_{2k}} & \delta_1  & 0=b_2\zeta_0+\zeta_2\delta_2 & \delta_{j,1} - \delta_{j,2k} \cr
		S_{-\alpha_{k+1}} & \delta_k  & 0=\delta_{k-1}b_3y - b_2x^2\zeta_k & \cr 
			& & \quad - \delta_{k-1}\zeta_k(b_4x + b_6\delta_{k-1})) 
			& \delta_{j,k} + \delta_{j,k+2} - 2\delta_{j,k+1} \cr
		S_{v_{k+1}} & \zeta_k  & 0=\delta_{k-1} & \delta_{j,k-1} - \delta_{j,k} + \delta_{j,k+1} - \delta_{j,k+2} \cr
	\end{array}
\end{equation}
Put differently, as we pass to the codimension 2 locus $b_1=0$, the following Cartan divisors become reducible and split into matter surfaces
\be\label{SUoddb1Split}
\ba
D_{-\alpha_i}  \quad &{\longrightarrow}  \quad  S_{v_i} + S_{v_{k+i}} \cr
D_{-\alpha_{2k+1-i}} \quad &{\longrightarrow}  \quad  S_{v_{i}} + S_{v_{k+i+1}}\cr
D_{-\alpha_{2k}} \quad & {\longrightarrow} \quad S_{v_1} + S_{v_{k+2}}+ S_{v_{2k}} \cr
D_{- \alpha_k} \quad &{\longrightarrow}  \quad S_{v_{k}} + S_{v_{k+1}} \,.
\ea
\ee
All other Cartan divisors remain irreducible. The splitting is depicted in figure \ref{fig:SUnSplittingb1}. 
Note that this graph also allows to read off the multiplicities: the green nodes are the initial affine $A_{2k}$ Dynkin diagram 
and the green lines indicate how the simple roots split (not their intersections). The multiplicities are in particular not always one. The intersection graph of the irreducible components labeled by weights\footnote{Fibering those curves over the codimension 2 locus yields the matter surfaces $S_v$.} is depicted in blue, and is given by the affine $D_{2k+1}$ Dynkin diagram.

The Cartan charges can be identified with Dynkin labels of weights of representations of $SU(2k+1)$. Along the codimension 2 locus $b_1=0$, the weights are in the representations
\be
{\bf  k(2k+1) }= \Lambda^2 V \qquad \hbox{and} \qquad \overline{\bf k (2k+1)}= \Lambda^2 \overline{V} \,,
\ee
where $V$ is the fundamental ${\bf 2k+1}$ of $SU(2k+1)$. Denote the highest weights by
\be
\mu_{\Lambda^2 V } = (0, 1, 0, \cdots, 0) \,,\qquad 
\mu_{\Lambda^2 \overline{V}} = (0, \cdots, 0, 1, 0) \,.
\ee
then the weights carried by the matter surfaces in codimension 2 are 
\begin{equation}
	\begin{array}{l|l}
		\mbox{Matter Surface} & \mbox{Weights} \cr\hline
		S_{-\alpha_0} &  -\alpha_0 \cr
		S_{-\alpha_1} & -\alpha_1 \cr
		S_{v_i}  &  \mu_{\Lambda^2 {V}} 
				- \left(\alpha_1 + 2 \sum_{\ell =1}^i \alpha_\ell + \sum_{\ell= i+1}^{2k+ 1-i} \alpha_\ell  \right)
				 \cr
		S_{v_{k+i}} &  \mu_{\Lambda^2 \overline{V}} 
				- \left(\alpha_{2k} + 2 \sum_{\ell =2k-i+2}^{2k-1} \alpha_{2k-\ell} + \sum_{\ell= i+1}^{2k+ 1-i} \alpha_\ell \right)				 \cr
		S_{v_{k+1}} &   \mu_{\Lambda^2 \overline{V}} 
				- \left(\alpha_{2k} + \alpha_k+ \alpha_{k+1}+ 2 \sum_{\ell = k+2}^{2k-1} \alpha_{\ell}   \right) \cr
		S_{v_k} &  \mu_{\Lambda^2 {V}} - \left(\alpha_1 + \alpha_{k+1}+ 2 \sum_{\ell =2}^k  \alpha_\ell \right)  \cr
		S_{v_{2k}} & \mu_{\Lambda^2 {V}} - \left( \sum_{\ell =2}^{2k} \alpha_\ell \right) \cr
		S_{-\alpha_{k+1}} & -\alpha_{k+1}
	\end{array}
\end{equation}

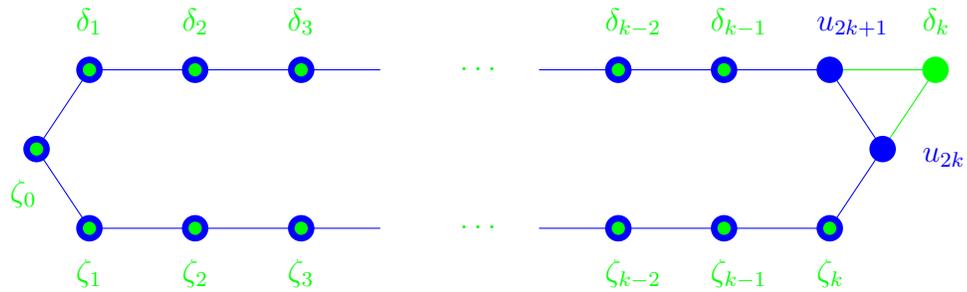
\begin{figure}
\begin{center}
\begin{picture}(280, 140)
\color{green}

	\color{blue}
	\put(0,80){\line(1,0){110}}
	\put(170,80){\line(1,0){110}}

	\put(0,20){\line(1,0){110}}
	\put(170,20){\line(1,0){110}}
		
	\put(0,80){\line(-2,-3){20}}
	\put(0,20){\line(-2,3){20}}	

	\put(280,80){\line(2,-3){20}}
	\put(280,20){\line(2,3){20}}

\color{green}

\put(320,80){\line(-1,0){40}}
\put(320,80){\line(-2,-3){20}}

\put(280,80){\color{blue}\circle*{10}}
\put(300,50){\color{blue}\circle*{10}}
\put(315,45){\color{blue}$u_{2k}$}


	\put(-20,50){\color{blue}\circle*{10}}
		\put(-20,50){\circle*{5}}
	\put(-30,30){$\zeta_0$}

	\put(0,20){\color{blue}\circle*{10}}
		\put(0,20){\circle*{5}}
	\put(40,20){\color{blue}\circle*{10}}
		\put(40,20){\circle*{5}}	
	\put(80,20){\color{blue}\circle*{10}}
		\put(80,20){\circle*{5}}

	\put(140,20){$\ldots$}

	\put(200,20){\color{blue}\circle*{10}}
		\put(200,20){\circle*{5}}
	\put(240,20){\color{blue}\circle*{10}}
		\put(240,20){\circle*{5}}
	\put(280,20){\color{blue}\circle*{10}}
		\put(280,20){\circle*{5}}

	\put(-5,0){$\zeta_1$}
	\put(35, 0){$\zeta_2$}
	\put(75,0){$\zeta_3$}
	\put(195,0){$\zeta_{k-2}$}
	\put(235,0){$\zeta_{k-1}$}
	\put(275,0){$\zeta_k$}


	\put(0,80){\color{blue}\circle*{10}}
	\put(0,80){\circle*{5}}
	\put(40,80){\color{blue}\circle*{10}}
	\put(40,80){\circle*{5}}
	\put(80,80){\color{blue}\circle*{10}}
	\put(80,80){\circle*{5}}

	\put(140,80){$\ldots$}
	
	\put(200,80){\color{blue}\circle*{10}}
	\put(200,80){\circle*{5}}
	\put(240,80){\color{blue}\circle*{10}}
	\put(240,80){\circle*{5}}

	\put(320,80){\circle*{10}}

	\put(-5,95){$\delta_1$}
	\put(35, 95){$\delta_2$}
	\put(75,95){$\delta_3$}
	\put(195,95){$\delta_{k-2}$}
	\put(235,95){$\delta_{k-1}$}
	\put(315,95){$\delta_k$}
	\put(275,95){\color{blue}$u_{2k+1}$}

\end{picture}
\end{center}
\caption{Codimension 2 splitting  for $SU(2k+1)$ along $P=0$. Bicolored nodes correspond to Cartan divisors that remain irreducible when passing to $P=0$. The only one that splits is $\delta_k$, which splits according to the green lines. The irreducible components of the fibers are blue and their intersections are given by the blue lines. Bicolored nodes remain irreducible.
}\label{fig:DynkinSplittingP}
\end{figure}


The second  codimension 2 locus, where the discriminant has increased vanishing order, is along 
\be
P = b_2b_3^2 + b_1^2b_6 - b_1b_3b_4 = 0\,.
\ee
The geometry restricted to $P=0$\footnote{At this locus we 
are explicitly not interested in the situation where $b_1 = 0$ since $b_1=P=0$  
 provides a greater enhancement of the vanishing order. For this 
reason we can scale the Tate form by $b_1^2$ without loss of generality and do the 
substitution for the $P = 0$ condition by replacing the $b_1^2 b_6$ term.}
\begin{equation}\label{SUoddYP}
	\ba
&Y_4 \cap (P=0): \cr
&\ 
		b_1^2y^2B(\delta)\delta_k 
		- b_1^2x^3B(\zeta)A(\zeta\delta)\zeta_k^k\delta_k^{k-1} + b_1^3xy 
		- b_1^2b_2x^2\zeta_0B(\zeta)\zeta_k 
		+ b_1^2b_3y\zeta_0^kB(\zeta\delta)C(\zeta\delta)
		\cr &- b_1^2b_4x\zeta_0^{k+1}B(\zeta^2\delta)C(\zeta\delta)\zeta_k
		- (b_1b_3b_4 - b_2b_3^2)\zeta_0^{2k+1}B(\zeta^3\delta^2)C(\zeta^2\delta^2)\zeta_k
		= 0 \,.
	\ea
\end{equation}
The exceptional divisors are again obtained by setting the sections $\zeta_i$ and $\delta_i$ to zero. These all remain irreducible, except for $\delta_k=0$, which restricted to the $P=0$ locus has two components
\begin{equation}\label{PSplit}
		\delta_k = 0 :\qquad  (b_1x + b_3\delta_{k-1})(b_1^2y + b_2b_3\zeta_k\delta_{k-1} - b_1\zeta_k(b_2x + b_4\delta_{k-1})) = 0  \,.
\end{equation}
 Let us first write out the irreducible matter surfaces, i.e. the irreducible components of the exceptional divisors restricted to $P=0$
\begin{equation}\label{SUoddPmatter}
	\begin{array}{l|c|l}
		\mbox{Matter Surface} & \mbox{Section}& \mbox{Equation in $Y_4|_{P=0}$} \cr\hline
		S_{-\alpha_0} & \zeta_0 &0=y^2\delta_1 - x^3\zeta_1 + b_1xy \cr
		S_{-\alpha_1} & \zeta_1 &0= \delta_1 + b_1x \cr
		S_{-\alpha_i} \quad\hfill i = 2,\cdots,k-1 &\zeta_i &0=\delta_{i-1}\delta_i + b_1x \cr
		S_{-\alpha_k} & \zeta_k &0=\delta_{k-1}\delta_k + b_1x + b_3\zeta_{k-1}\delta_{k-1} \cr
		S_{-\alpha_{2k}} & \delta_1 &0= b_1y - b_2\zeta_0\zeta_1\zeta_2 - \zeta_1\zeta_2^2\delta_2 \cr
		S_{-\alpha_{2k+1-i}} \quad\hfill i = 2,\cdots,k-1 & \delta_i  & 0=b_1y - b_2\zeta_i\zeta_{i+1} \cr
		S_{u_{2k}} & \delta_k &0= b_1x + b_3\delta_{k-1} \cr
		S_{u_{2k+1}} & \delta_k & 0=b_1^2y + b_2b_3\zeta_k\delta_{k-1} - b_1\zeta_k(b_2x + b_4\delta_{k-1}) \cr 
	\end{array}
\end{equation}
The Cartan charges, worked out from the intersection relations in appendix \ref{app:SUn}, are
\begin{equation}
	\begin{array}{l|l|l}
		\hbox{Matter Surface} & \hbox{$j$th Cartan Charge} & \hbox{Weights} \cr\hline
		S_{-\alpha_0} & \delta_{j,1} + \delta_{j,2k} - 2\delta_{j,0} & -\alpha_{0}\cr
		S_{-\alpha_1} & \delta_{j,2} + \delta_{j,0} - 2\delta_{j,1}  & - \alpha_1 \cr
		S_{-\alpha_i} \quad\hfill i = 2,\cdots,k-1 & \delta_{j,i+1} + \delta_{j,i-1} - 2\delta_{j,i} & -\alpha_i\cr
		S_{-\alpha_k} & \delta_{j,k+1} + \delta_{j,k-1} - 2\delta_{j,k} & -\alpha_{k}\cr
		S_{-\alpha_{2k}} & \delta_{j,0} + \delta_{j,2k-1} - 2\delta_{j,2k} & -\alpha_{2k} \cr
		S_{-\alpha_{k+i}} \quad\hfill i = 2,\cdots,k-1 & \delta_{j,k+i+1} + \delta_{j,k+i-1} - 2\delta_{j,k+i} & -\alpha_{k+i}\cr
		S_{u_{2k}} & \delta_{j,k} - \delta_{j,k+1} & \mu_{\overline{V}} - \sum_{\ell = k+1}^{2k} \alpha_\ell   \cr
		S_{u_{2k+1}} & \delta_{j,k+2} - \delta_{j,k+1} & \mu_{{V}} - \sum_{\ell = k+1}^{2k} \alpha_\ell  \cr
	\end{array}
\end{equation}
These Cartan charges either correspond to roots, or to weights of the (anti-)fundamental representation of $SU(2k+1)$
\be
{\bf  (2k+1) }=  V \qquad \hbox{and} \qquad \overline{\bf (2k+1)}=  \overline{V} \,,
\ee
whose highest weights we will label by
\be
\mu_{V } = (1, 0, \cdots, 0) \,,\qquad 
\mu_{\overline{V}} = (0, \cdots, 0, 1) \,.
\ee
In summary, along the matter locus $P=0$ the Cartan divisor $\delta_k=0$ splits as in (\ref{PSplit}), into two matter surfaces associated to the (anti-) fundamental representation
\begin{equation}
D_{-\alpha_{k+1}} \quad \longrightarrow \quad S_{u_{2k}} + S_{u_{2k+1}} \,.
\end{equation}
The structure of the fiber along $P=0$ is depicted in figure \ref{fig:DynkinSplittingP}. Note that in this case the fiber is of Kodaira type corresponding to an $SU(2k+2)$ affine Dynkin diagram, with the correct multiplicities. The fact that the representations, which arise here is the (anti-) fundamental one is in accord with decomposing the adjoint of $SU(2k+2)$ with respect to $SU(2k+1)$. 


\subsubsection{Codimension 3}\label{sec:SUodd3}

Finally we discuss the codimension three loci. 
These occur along
\be
b_1 = b_3 = 0\,,\qquad P=Q=0 \,,\qquad b_1 = b_2 = 0\,.
\ee
It is useful to follow the $b_1=0$ matter surfaces (i.e. codimension two fibers) through to vanishing $b_2$ or $b_3$. 
Along these loci, the vanishing order of the discriminant increases as follows
\begin{equation}
\ba
b_1=b_3 =0:\qquad& \Delta|_{b_1=b_3=0} = O(\zeta_0^{2k+4}) \cr
P=Q=0:\qquad & \Delta|_{P=Q=0} = O(\zeta_0^{3k}) \cr
b_1=b_2=0:\qquad& \Delta|_{b_1=b_2=0} = O(\zeta_0^{3k+3}) \,,
\ea
\end{equation}
where the first locus will be shown to have a simple interpretation in terms of a local enhancement to $SO(2k+4)$, which will be clear from the 
intersections of the fibers.
The case  $b_1=b_2=0$ is more subtle, as the singularity is non-minimal and the  intersections are not of a standard Dynkin type. These generalize, what in the low $k$ examples, such as $SU(5)$, corresponds to exceptional enhancements. We will discuss them in the next subsection.

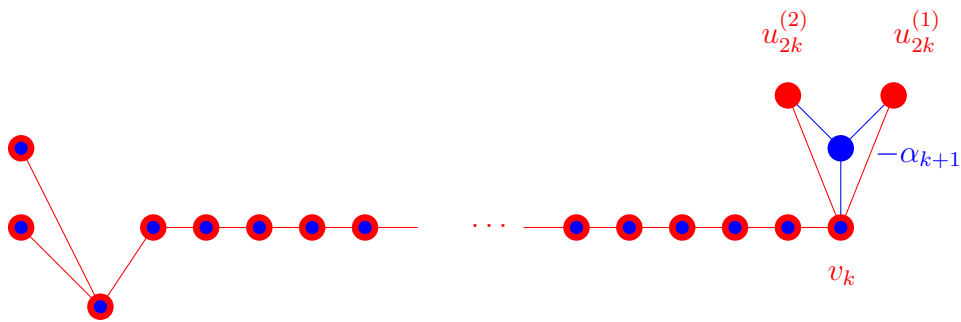
\begin{figure}
\begin{center}
\begin{picture}(280, 140)

\color{white}

	\put(0,20){\color{red}\line(2,3){20}}
	\put(0,20){\color{red}\line(-1,1){30}}	
	\put(0,20){\color{red}\line(-1,2){30}}	
	\put(20,50){\color{red}\line(1,0){100}}
	\put(160,50){\color{red}\line(1,0){120}}

	\put(280,50){\color{red}\line(-2,5){20}}
	\put(280,50){\color{red}\line(2,5){20}}

	\put(280, 80){\color{blue}\line(1,1){20}}
	\put(280, 80){\color{blue}\line(-1,1){20}}
	\put(280, 80){\color{blue}\line(0,-1){30}}


	\put(-30,50){\color{red}\circle*{10}}
	\put(-30,50){\color{blue}\circle*{5}}


	\put(0,20){\color{red}\circle*{10}}
	\put(0,20){\color{blue}\circle*{5}}


	\put(20,50){\color{red}\circle*{10}}
	\put(20,50){\color{blue}\circle*{5}}
	\put(40,50){\color{red}\circle*{10}}
	\put(40,50){\color{blue}\circle*{5}}
	\put(60,50){\color{red}\circle*{10}}
	\put(60,50){\color{blue}\circle*{5}}
	\put(80,50){\color{red}\circle*{10}}
	\put(80,50){\color{blue}\circle*{5}}
	\put(100,50){\color{red}\circle*{10}}
	\put(100,50){\color{blue}\circle*{5}}

	\put(140,50){\color{red}$\ldots$}

	\put(180,50){\color{red}\circle*{10}}
	\put(180,50){\color{blue}\circle*{5}}
	\put(200,50){\color{red}\circle*{10}}
	\put(200,50){\color{blue}\circle*{5}}
	\put(220,50){\color{red}\circle*{10}}
	\put(220,50){\color{blue}\circle*{5}}
	\put(240,50){\color{red}\circle*{10}}
	\put(240,50){\color{blue}\circle*{5}}
	\put(260,50){\color{red}\circle*{10}}
	\put(260,50){\color{blue}\circle*{5}}
	\put(280,50){\color{red}\circle*{10}}
	\put(280,50){\color{blue}\circle*{5}}
	
	\put(275,30){\color{red}$v_{k}$}


	\put(-30,80){\color{red}\circle*{10}}
	\put(-30,80){\color{blue}\circle*{5}}

	\put(300,100){\color{red}\circle*{10}}	
	\put(260,100){\color{red}\circle*{10}}	

	\put(280,80){\color{blue}\circle*{10}}

	\put(300,120){\color{red}$u_{2k}^{(1)}$}	
	\put(250,120){\color{red}$u_{2k}^{(2)}$}	
	\put(293,75){\color{blue}$-\alpha_{k+1}$}

\end{picture}
\end{center}
\caption{Codimension 3 splitting along $b_1=b_3=0$ of the $b_1=0$ codimension 2 locus for $SU(2k+1)$. The only node that splits is $\alpha_{k+1}$, which has three components $u_{2k}^{(1,2)}$ and $v_k$. The irreducible components in codimension 3 are $\color{red}\bullet$ or bi-colored, and the red lines indiate the intersections of these. Bicolored nodes remain irreducible.}
\label{fig:SUnSplittingb1b3}
\end{figure}

Consider first $b_1= b_3=0$, along which the equation for the matter surface  $S_{-\alpha_{k+1}}$ becomes reducible
\be\label{DvkSplit}
S_{-\alpha_{k+1}} \cdot [b_3] :\qquad b_1= b_3= \delta_k = \zeta_k (\delta_{k-1} (b_4 x + b_6 \delta_{k-1}) + b_2 x^2) =0 \,.
\ee
One component is precisely $\Sigma_{v_k} = S_{v_k} \cdot[b_3] $, and the other component is a quadratic whose roots will be denoted by
$\Sigma_{u_{2k}}^{(i)}$, which have $j$th  Cartan charge $(\delta_{j, k} - \delta_{j, k+1})$.
Note that these two components do not intersect, as depicted in figure \ref{fig:SUnSplittingb1b3}.
In summary, the splitting is 
\begin{equation}\label{SplittingSUodd1}
S_{-\alpha_{k+1}}  \quad \longrightarrow \quad \Sigma_{ v_k } + \Sigma_{u_{2k}^{(1)}} + \Sigma_{u_{2k}^{(2)}} \,,
\end{equation}
where we use the notation $\Sigma_{v} = S_{v} \cdot [b_3]$.
Note that the weight corresponding to $u_{2k}$ is in the anti-fundamental representation
\be
S_{u_{2k}}: \qquad  \mu_{\overline{V}} - \left(\sum_{\ell =k+1}^{2k}  \alpha_\ell \right) \,.
\ee
Put differently, (\ref{SplittingSUodd1}) implies that three matter surfaces with weights ${ v_k }$ and  two with ${u_{2k}}$ 
become homologous to each other, consistent with the group theoretic coupling 
\begin{equation}
b_1=b_3=0 :\qquad \bar{V} \otimes \bar{V}\otimes \Lambda^2V \,.
\end{equation}

Next consider the codimension 3 locus $P=Q=0$, see (\ref{BigTable}). These conditions are equivalent to 
\begin{equation}\label{PQ2k1}
P=Q=0: 
\qquad 
\ba
b_1^2 b_6-b_1 b_3 b_4+b_2 b_3^2 &=0 \cr
b_1 b_4-2 b_2 b_3 &=0  \,.
\ea
\end{equation}
We can apply the second equation to rewrite the Tate form  (\ref{SUoddYP}), where $P=0$ was already implemented. In particular, using the second equation on (\ref{PSplit}), we see that along $P=Q=0$ the two irreducible components of $\delta_k=P=0$, $S_{u_{2k}}$ and $S_{u_{2k+1}}$ intersect, and the equation for $\delta_k$ can be rewritten in an unfactored form along $P=Q=0$. This is consistent with the $S_{u_{2k}}$ and $S_{u_{2k+1}}$ being (anti-) fundamental representations, forming a coupling with a singlet along $P=Q=0$
\begin{equation}
 P=Q=0:\qquad V \otimes \overline{V} \otimes {\bf 1} \,.
\end{equation}


\begin{figure}
\begin{center}
\begin{picture}(280, 140)

\color{green}
		\put(0,80){\line(-1,0){30}}
		
	\put(0,20){\line(0,1){60}}
	\put(40,20){\line(0,1){60}}
	\put(80,20){\line(0,1){60}}
	
	\put(200,20){\line(0,1){60}}
	\put(240,20){\line(0,1){60}}
	\put(280,20){\line(0,1){30}}
	
	\put(40,20){\line(-2,3){40}}
	\put(80,20){\line(-2,3){40}}
	\put(80,80){\line(2,-3){20}}

	\put(200,20){\line(-2,3){20}}
	\put(240,20){\line(-2,3){40}}
	\put(280,20){\line(-2,3){40}}


	\put(-30,50){\color{red}\circle*{10}}
	\put(-30,50){\color{blue}\circle*{7}}
	\put(-30,50){\circle*{2}}
	\put(-35,30){$\zeta_0$}


	\put(0,20){\color{red}\circle*{10}}
	\put(0,20){\color{blue}\circle*{7}}	
	\put(0,20){\circle*{2}}

	\put(40,20){\circle*{10}}
	\put(80,20){\circle*{10}}

	\put(140,20){$\ldots$}

	\put(200,20){\circle*{10}}
	\put(240,20){\circle*{10}}
	\put(280,20){\circle*{10}}

	\put(-5,0){$\zeta_1$}
	\put(35, 0){$\zeta_2$}
	\put(75,0){$\zeta_3$}
	\put(195,0){$\zeta_{k-2}$}
	\put(235,0){$\zeta_{k-1}$}
	\put(275,0){$\zeta_k$}

	\put(0,80){\circle*{10}}
	\put(40,80){\circle*{10}}
	\put(80,80){\circle*{10}}

	\put(140,80){$\ldots$}

	\put(200,80){\circle*{10}}
	\put(240,80){\circle*{10}}

	\put(-5,95){$\delta_1$}
	\put(35, 95){$\delta_2$}
	\put(75,95){$\delta_3$}
	\put(195,95){$\delta_{k-2}$}
	\put(235,95){$\delta_{k-1}$}
	\put(275,95){$\delta_k$}


	\put(0,20){\color{red}\line(2,3){20}}
	\put(0,20){\color{red}\line(-1,1){30}}	
	\put(0,20){\color{red}\line(-1,2){30}}	
	\put(20,50){\color{red}\line(1,0){100}}
	\put(160,50){\color{red}\line(1,0){120}}

	\put(280,50){\color{red}\line(-2,5){20}}
	\put(280,50){\color{red}\line(2,5){20}}

	\put(280, 80){\color{blue}\line(1,1){20}}
	\put(280, 80){\color{blue}\line(-1,1){20}}
	\put(280, 80){\color{blue}\line(0,-1){30}}


	\put(-30,50){\color{red}\circle*{10}}
	\put(-30,50){\color{blue}\circle*{5}}


	\put(0,20){\color{red}\circle*{10}}
	\put(0,20){\color{blue}\circle*{5}}


	\put(-30,50){\color{red}\circle*{10}}
	\put(-30,50){\color{blue}\circle*{7}}
	\put(-30,50){\circle*{2}}
	\put(-35,30){$\zeta_0$}


	\put(0,20){\color{red}\circle*{10}}
	\put(0,20){\color{blue}\circle*{7}}	
	\put(0,20){\circle*{2}}


	\put(20,50){\color{red}\circle*{10}}
	\put(20,50){\color{blue}\circle*{5}}
	\put(40,50){\color{red}\circle*{10}}
	\put(40,50){\color{blue}\circle*{5}}
	\put(60,50){\color{red}\circle*{10}}
	\put(60,50){\color{blue}\circle*{5}}
	\put(80,50){\color{red}\circle*{10}}
	\put(80,50){\color{blue}\circle*{5}}
	\put(100,50){\color{red}\circle*{10}}
	\put(100,50){\color{blue}\circle*{5}}

	\put(140,50){\color{red}$\ldots$}

	\put(180,50){\color{red}\circle*{10}}
	\put(180,50){\color{blue}\circle*{5}}
	\put(200,50){\color{red}\circle*{10}}
	\put(200,50){\color{blue}\circle*{5}}
	\put(220,50){\color{red}\circle*{10}}
	\put(220,50){\color{blue}\circle*{5}}
	\put(240,50){\color{red}\circle*{10}}
	\put(240,50){\color{blue}\circle*{5}}
	\put(260,50){\color{red}\circle*{10}}
	\put(260,50){\color{blue}\circle*{5}}
	\put(280,50){\color{red}\circle*{10}}
	\put(280,50){\color{blue}\circle*{5}}
	


	\put(-30,80){\color{red}\circle*{10}}
	\put(-30,80){\color{blue}\circle*{5}}

	\put(300,100){\color{red}\circle*{10}}	
	\put(260,100){\color{red}\circle*{10}}	

	\put(280,80){\color{blue}\circle*{10}}
	\put(280,80){\color{green}\circle*{5}}

	\put(300,120){\color{red}$u_{2k}^{(1)}$}	
	\put(250,120){\color{red}$u_{2k}^{(2)}$}	
	\put(293,75){\color{blue}$-\alpha_{k+1}$}


\end{picture}
\end{center}
\caption{Summary graph: splitting along $b_1=b_3=0$ for $SU(2k+1)$. $\color{green}\bullet$ are initial Cartan divisors, green lines indicate how they split in codimesion 2 along $b_1=0$ into $\color{blue}\bullet$. Blue lines indicate the splitting of these when passing to $b_3=0$. $\color{red}\bullet$  are the irreducible components in codimension 3, and red lines indicate their intersections. }
\label{fig:SUnSplittingb1b3All}
\end{figure}
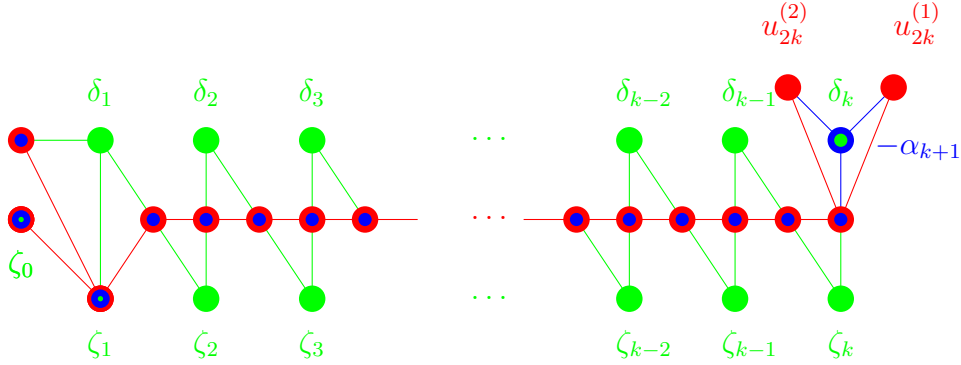


\subsubsection{Non-minimal Loci}

Finally, consider the codimension 3 locus $b_1=b_2=0$. For $k=2$ for instance this is the locus corresponding to an ``exceptional enhancement" to $E_6$. 
However, observe that, from table \ref{BigTable},  for general values of $k$ the singularity becomes non-minimal \ssn{in the sense defined in section  \ref{sec:MinFlat}}
, i.e. along $b_1=b_2=0$ the sections $f$ and $g$ defined in (\ref{WeierTate}) of the  Weierstrass form  have vanishing orders
\begin{equation}
\ba
f|_{b_1=b_2=0} &= O(\zeta_0^4) \cr
g|_{b_1=b_2=0} &= O(\zeta_0^6) \cr
\Delta|_{b_1=b_2=0} &= O(\zeta_0^{12}) \,.
\ea
\end{equation}
One way to see this explicitly is to consider the resolved Tate form (\ref{TateResSUodd}) intersected with $b_1=b_2=0$
\begin{equation}\ba
&\tilde{T}_{SU(2k+1)}|_{b_1=b_2=0}: \cr
&\delta_2 \cdots \delta_{k-1} \ \left( 
		y^2
		\delta_1\delta_k - x^3  \left(\prod_{i=3}^{k-2} \delta_i^{i-2}\right)B(\zeta)A(\zeta)\zeta_k^k\delta_k^{k-1}
		+ b_3y\zeta_0^k \delta_1 B(\zeta)C(\zeta \delta)  \right. \cr
	&\qquad \qquad  \qquad 	\left.	- b_4x\zeta_0^{k+1}\delta_1B(\zeta^2 )C(\zeta\delta)\zeta_k 
			- b_6\zeta_0^{2k+1}\delta_1B(\zeta^3\delta)C(\zeta^2\delta^2)\zeta_k \!\!\!\!\phantom{\prod_{b}^{a} }\right) = 0  \,,
\ea
\end{equation}
which makes it clear that the surfaces corresponding to 
\begin{equation}\label{NMSurface}
\aleph_i:\qquad b_1=b_2=\delta_i =0  \,,\qquad  i=2, \cdots, k-1 \,,
\end{equation}
are completely contained inside the fiber! \ssn{This violates flatness of the fibration, i.e. the dimensionality of the fiber changes. The fiber ceases to contain only curves, but at this non-minimal locus also has surface components. 
We observe this correlation of non-minimality of the singularity implying non-flatness in various instances, although we have no argument that such a connection should hold in general.
The same kind of non-minimal codimension 3 locus was analyzed in \cite{Candelas:2000nc} for $E_7$. }

Nevertheless, we dare to study these codimension 3 loci. One open question is to understand the effective theory at such points. In the present paper, our main concern is the structure of the fibers. Our philosophy was so far to follow the codimension $d$ loci to the codimension $d+1$ loci and understand how irreducible components become reducible. As we shall see, following the matter curves at $b_1=0$ to the point $b_1=b_2=0$, they split in a fashion that is consistent with the generation of Yukawa couplings for the matter representations at hand. We can then ask what the intersections of the irreducible components of the fibers are. For this we provide two ways to depict them: the first type of graph is similar to the ones we obtained from the splitting along minimal singular loci, showing the irreducible components of the fiber and their intersections -- without including the surfaces (\ref{NMSurface}), these are depicted in figures \ref{fig:SUoddSplitting} and \ref{fig:SUoddSplittingC}. For the non-minimal case it makes sense to also depict the surface components (\ref{NMSurface}), and how they intersect the remainder of the fiber, as shown in figure \ref{fig:SUoddNM}.


\begin{figure}
\begin{center}
\begin{picture}(280, 140)

\color{white}


	\put(0,20){\color{red}\line(2,3){20}}
	\put(0,20){\color{red}\line(-1,1){30}}	
	\put(20,50){\color{red}\line(0,1){70}}
	
	\put(20,50){\color{red}\line(1,0){100}}
	\put(160,50){\color{red}\line(1,0){120}}
	\put(260,50){\color{red}\line(0,1){60}}
	
	\put(-30,80){\color{blue}\line(3,-2){50}}	
	\put(-30,80){\color{blue}\line(4,3){50}}	

	\put(280,80){\color{blue}\line(-2,3){20}}
	\put(280,80){\color{blue}\line(-1,0){20}}

	\put(-30,50){\color{red}\circle*{10}}
	\put(-30,50){\color{blue}\circle*{5}}
	\put(-30,30){$\zeta_0$}


	\put(0,20){\color{red}\circle*{10}}
	\put(0,20){\color{blue}\circle*{5}}	


	
		
	
	



{\color{blue}

	\put(20,50){\color{red}\circle*{10}}	
	\put(20,50){\circle*{5}}
	\put(40,50){\color{red}\circle*{10}}	
	\put(40,50){\circle*{5}}
	\put(60,50){\color{red}\circle*{10}}	
	\put(60,50){\circle*{5}}
	\put(80,50){\color{red}\circle*{10}}	
	\put(80,50){\circle*{5}}
	\put(100,50){\color{red}\circle*{10}}		
	\put(100,50){\circle*{5}}

	\put(140,50){\color{red}$\ldots$}

	\put(180,50){\color{red}\circle*{10}}	
	\put(180,50){\circle*{5}}
	\put(200,50){\color{red}\circle*{10}}	
	\put(200,50){\circle*{5}}
	\put(220,50){\color{red}\circle*{10}}	
	\put(220,50){\circle*{5}}
	\put(240,50){\color{red}\circle*{10}}	
	\put(240,50){\circle*{5}}
	\put(260,50){\color{red}\circle*{10}}	
	\put(260,50){\circle*{5}}
	\put(280,50){\color{red}\circle*{10}}	
	\put(280,50){\circle*{5}}
}


	{\put(-30,80){\color{blue}\circle*{10}}}
	\put(-55,80){\color{blue}$v_{2k}$}
	
	{\put(20,120){\color{red}\circle*{10}}}
	\put(32,120){\color{red}$w'$}

	\put(15,30){\color{red}$v_{k+2}$}

	\put(0,80){\circle*{10}}
	\put(40,80){\circle*{10}}
	\put(80,80){\circle*{10}}

	\put(140,80){$\ldots$}

	\put(200,80){\circle*{10}}
	\put(240,80){\circle*{10}}

	\put(260,80){\color{red}\circle*{10}}
	\put(225,80){\color{red}$u_{2k}$}
	\put(260,110){\color{red}\circle*{10}}
	\put(225,110){\color{red}$u_{2k+1}$}
	
	\put(280,80){\color{blue}\circle*{10}}
	\put(295,80){\color{blue}$-\alpha_{k+1}$}


\end{picture}
\end{center}
\caption{Splitting along the $b_1=b_2=0$ codimension 3 locus for $SU(2k+1)$. $\color{blue}\bullet$  represent the fiber components along the matter locus  $b_1=0$. 
Passing to the locus $b_1=b_2=0$ two of the fiber components split, as indicated by the blue lines. $\color{red}\bullet$ labels the irreducible fiber components in codimension 3, and the red lines indicate their intersections, which are clearly not of Kodaira type. The splittings are explained in (\ref{SUYuk1}, \ref{SUYuk2}).}
\label{fig:SUoddSplitting}
\end{figure}
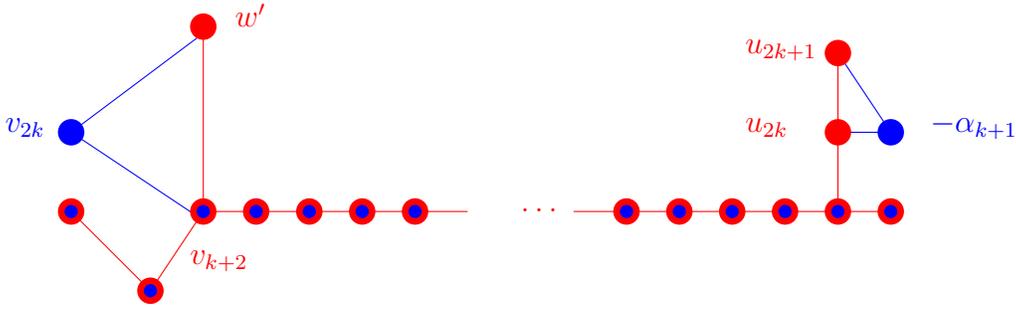



\begin{figure}
\begin{center}
\begin{picture}(280, 140)

\color{green}


	\put(0,20){\color{red}\line(2,3){20}}
	\put(0,20){\color{red}\line(-1,1){30}}	

	\put(20,50){\color{red}\line(0,1){70}}

	\put(20,50){\color{red}\line(1,0){100}}
	\put(160,50){\color{red}\line(1,0){120}}
	\put(260,50){\color{red}\line(0,1){60}}
	
	\put(-30,80){\color{blue}\line(3,-2){50}}	
	\put(-30,80){\color{blue}\line(4,3){50}}	

	\put(280,80){\color{blue}\line(-2,3){20}}
	\put(280,80){\color{blue}\line(-1,0){20}}

	\put(-30,50){\color{red}\circle*{10}}
	\put(-30,50){\color{blue}\circle*{7}}
	\put(-30,50){\circle*{2}}
	\put(-30,30){$\zeta_0$}


	\put(0,20){\color{red}\circle*{10}}
	\put(0,20){\color{blue}\circle*{7}}
	\put(0,20){\circle*{2}}	
	\put(40,20){\circle*{10}}
	\put(80,20){\circle*{10}}

	\put(140,20){$\ldots$}

	\put(200,20){\circle*{10}}
	\put(240,20){\circle*{10}}
	\put(280,20){\circle*{10}}
	
	\put(0,80){\line(-1,0){30}}
		
	\put(0,20){\line(0,1){60}}
	\put(40,20){\line(0,1){60}}
	\put(80,20){\line(0,1){60}}
	
	\put(200,20){\line(0,1){60}}
	\put(240,20){\line(0,1){60}}
	\put(280,20){\line(0,1){30}}
	
	\put(40,20){\line(-2,3){40}}
	\put(80,20){\line(-2,3){40}}
	\put(80,80){\line(2,-3){20}}

	\put(200,20){\line(-2,3){20}}
	\put(240,20){\line(-2,3){40}}
	\put(280,20){\line(-2,3){40}}

	\put(-5,0){$\zeta_1$}
	\put(35, 0){$\zeta_2$}
	\put(75,0){$\zeta_3$}
	\put(195,0){$\zeta_{k-2}$}
	\put(235,0){$\zeta_{k-1}$}
	\put(275,0){$\zeta_k$}

{\color{blue}

	\put(20,50){\color{red}\circle*{10}}	
	\put(20,50){\circle*{5}}
	\put(40,50){\color{red}\circle*{10}}	
	\put(40,50){\circle*{5}}
	\put(60,50){\color{red}\circle*{10}}	
	\put(60,50){\circle*{5}}
	\put(80,50){\color{red}\circle*{10}}	
	\put(80,50){\circle*{5}}
	\put(100,50){\color{red}\circle*{10}}		
	\put(100,50){\circle*{5}}

	\put(140,50){$\ldots$}

	\put(180,50){\color{red}\circle*{10}}	
	\put(180,50){\circle*{5}}
	\put(200,50){\color{red}\circle*{10}}	
	\put(200,50){\circle*{5}}
	\put(220,50){\color{red}\circle*{10}}	
	\put(220,50){\circle*{5}}
	\put(240,50){\color{red}\circle*{10}}	
	\put(240,50){\circle*{5}}
	\put(260,50){\color{red}\circle*{10}}	
	\put(260,50){\circle*{5}}
	\put(280,50){\color{red}\circle*{10}}	
	\put(280,50){\circle*{5}}
}


	{\put(-30,80){\color{blue}\circle*{10}}}
	{\put(20,120){\color{red}\circle*{10}}}

	\put(0,80){\circle*{10}}
	\put(40,80){\circle*{10}}
	\put(80,80){\circle*{10}}

	\put(140,80){$\ldots$}

	\put(200,80){\circle*{10}}
	\put(240,80){\circle*{10}}

	\put(260,80){\color{red}\circle*{10}}
	\put(260,110){\color{red}\circle*{10}}

	\put(280,80){\color{blue}\circle*{10}}
	\put(280,80){\circle*{5}}

	\put(-5,95){$\delta_1$}
	\put(35, 95){$\delta_2$}
	\put(75,95){$\delta_3$}
	\put(195,95){$\delta_{k-2}$}
	\put(235,95){$\delta_{k-1}$}
	\put(275,95){$\delta_k$}

\end{picture}
\end{center}
\caption{The complete splitting process for $SU(2k+1)$ to get from the irreducible fiber components labeled by roots $\color{green}\bullet$ in codimension 1, to the  irreducible fiber components in codimension 2, i.e. matter $\color{blue}\bullet$ along $b_1=0$ to the ones along codimension 3, i.e. the Yukawa $\color{red}\bullet$ locus $b_1=b_2=0$. Green and blue lines indicate how the fibers split going from codimension 1 to 2 to 3, whereas red lines indicate the intersections in codimension 3.} 
\label{fig:SUoddSplittingC}
\end{figure}
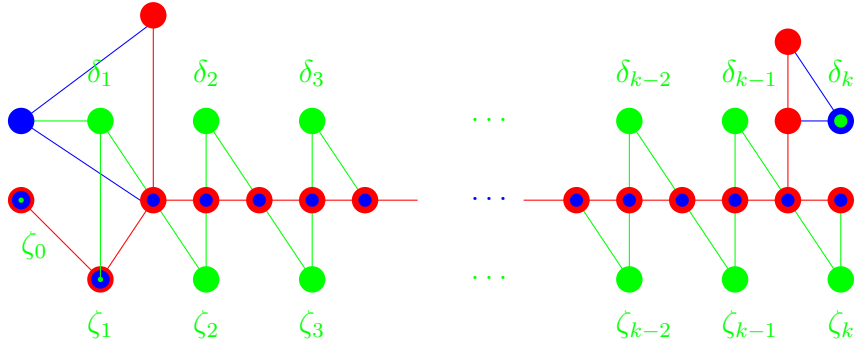



\begin{figure}\begin{center}
\begin{picture}(400, 140)

\put(50,10){\color{blue}\polygon*(10,10)(40,40)(70,10)}
\put(110,10){\color{blue}\polygon*(10,10)(40,40)(70,10)}
\put(170,10){\color{blue}\polygon*(10,10)(40,40)(70,10)}

\put(250, 30){\color{blue}$\cdots$}

\put(260,10){\color{blue}\polygon*(10,10)(40,40)(70,10)}
\put(320,10){\color{blue}\polygon*(10,10)(40,40)(70,10)}

\put(50,10){\linethickness{2pt}\color{Red}\polyline(10,10)(40,40)(70,10)(100,40)(130,10)}
\put(170,10){\linethickness{2pt}\color{Red}\polyline(10,10)(40,40)(70,10)}
\put(260,10){\linethickness{2pt}\color{Red}\polyline(10,10)(40,40)(70,10)(100,40)(130,10)}

\put(10,10){\linethickness{2pt}\color{Red}
\polyline(20,40)(50,10)
\Line(-20,40)(20,40)
\Line(0, 40)(0, 60)
\Line(-20,10)(50,10)
}

\put(375,35){\linethickness{2pt}\color{Red}\line(1,1){20}}
\put(380,70){\linethickness{2pt}\color{Red}\line(1,-1){30}}
\put(390,20){\linethickness{2pt}\color{Red}\line(1,0){30}}

\put(85, 30){\color{white}$\delta_2$}
\put(145, 30){\color{white}$\delta_3$}
\put(205, 30){\color{white}$\delta_4$}
\put(290, 30){\color{white}$\delta_{k-2}$}
\put(350, 30){\color{white}$\delta_{k-1}$}

\put(5, 75){\color{Red}$\zeta_0$}

\end{picture}
\end{center}
\caption{Fiber along the codimension 3 locus $b_1=b_2=0$. Here, the red line is the dual graph to the intersection graph in figure \ref{fig:SUoddSplitting}. The blue triangles are the 1 dimension higher (i.e. surface) components $\aleph_i$ given by the vanishing of $\delta_i$  as in (\ref{NMSurface}) that enter the fiber along this non-minimal locus, thus rendering the fibration non-flat.  The two edges that are part of the red intersection graph arise from the intersection of the components $\aleph_i$ with $\zeta_i$ and $\zeta_{i+1}$.}
\label{fig:SUoddNM}
\end{figure}
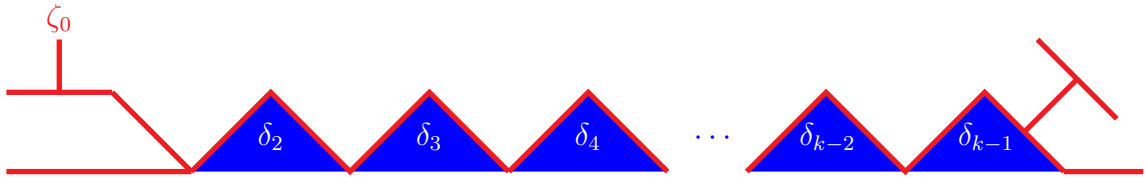

As discussed, let us consider the splitting of the $b_1=0$ matter surfaces along the locus $b_1=b_2=0$. 
There are two matter surfaces that split
\be
\ba
S_{v_{2k}}  \cdot [b_2]  &:\qquad \delta_1  = \delta_2\zeta_2=0  \cr
S_{ -\alpha_{k+1}} \cdot [b_2] &: \qquad \delta_k = \delta_{k-1} (b_3 y - \zeta_k (b_4 x + b_6 \delta_{k-1})) =0 \,.
\ea
\ee 
For the first one it is clear that one of the components is the curve obtained by restricting $S_{v_{k+2}}$, $S_{ v_{k+2}} \cdot [b_2] \equiv \Sigma_{v_{k+2}}$, so that 
\be\label{SUYuk1}
S_{v_{2k}} \quad \longrightarrow \quad \Sigma_{ v_{k+2}} + \Sigma_{w'} \,,
\ee
where the Cartan charge of the additional component is 
\be
(w')_j = \delta_{j, 2} - \delta_{j, 2k -1 } \,,
\ee
which is a weight in the representation
$\Lambda^4 V$. The corresponding Yukawa coupling corresponds to
\begin{equation}
b_1 = b_2= 0:\qquad \Lambda^2 \overline{V} \otimes \Lambda^2 \overline{V} \otimes \Lambda^4 V \,.
\end{equation}

Finally, $S_{-\alpha_{k+1}}$, which as we discussed descends to  codimension 2 from a root, 
splits as
\be\label{SUYuk2}
S_{-\alpha_{k+1}} \quad \longrightarrow \quad \Sigma_{u_{2k}} + \Sigma_{u_{2k+1}} \,,
\ee
where the charges are computed from the intersection relations in the appendix
\be
\ba
u_{2k}&=  -(-\delta_{j, k} + \delta_{j, k+1}) = \mu_{\overline{V}} - \left( \sum_{\ell =k+1}^{2k} \alpha_{\ell}\right)\cr
{u_{2k+1}} &= +  (-\delta_{j, k+1}+ \delta_{j, k+2})  = \mu_{{V}} - \left( \sum_{\ell =1}^{k} \alpha_{\ell}\right) \,.
\ea
\ee
This corresponds to a root splitting into a fundamental and anti-fundamental representation. 
The splittings are depicted in figure \ref{fig:SUoddSplitting}. 

As we explained earlier, these diagrams are in no way expected to be a full description of the fiber. Their intersection topology is not of Dynkin 
type and seem to be a generalization of exceptional Dynkin diagrams to higher rank. The algebras associated to this are not finite dimensional Lie algebras. We do not expect a simple Higgs mechanism, like in the minimal case, to describe the effective theory. As a final point, it may be useful to depict the intersections including the additional surface components (\ref{NMSurface}), which can be seen in figure \ref{fig:SUoddNM}.

We should add a remark about $SU(5)$, which is discussed in detail in appendix \ref{app:Examples}, 
which like many low $k$ cases is not quite following the splitting pattern as given in figure  \ref{fig:SUoddSplitting}.
In particular, the fiber component $\delta_1=\delta_2=0$ joins with the components that split off from $\delta_{k=2}=0$, and give rise to the intersections of 
an $E_6$ diagram. Note however, that no affine $E_6$ fiber is observed, as already well known from \cite{Esole:2011sm, MS}. This Yukawa is  consistent with $w'$ in (\ref{SUYuk1}) being a weight of $\Lambda^4V \cong V$ in the case of $SU(5)$.



\subsection{$SU(2k)$}

The procedure for $SU(2k)$ is quite similar to the $SU(2k+1)$, however, as the fibers split slightly differently, we will give a brief discussion of the 
codimension 1, 2, and 3 structure. 
The Tate form for $SU(2k)$ is
\begin{equation}
0= y^2 - x^3 + b_1xy - b_2x^2\zeta_0 + b_3y\zeta_0^k 
					- b_4x\zeta_0^k - b_6\zeta_0^{2k} \,.
\end{equation}
The ordered set of resolutions required to resolve this geometry in all codimensions, is
\begin{equation}\label{SUevenBIGSMALL}
	\begin{aligned}
		(x, y, \zeta_i; \zeta_{i+1}) &\qquad i = 0,\dots,k-1 \cr
		(y, \zeta_i; \delta_i) &\qquad i = 1,\dots,k-1 \,,
	\end{aligned}
\end{equation}
Note that after the blowups the geometry takes again the form (\ref{SUGenForm})
like for $SU(2k+1)$, and we applied the same class of small resolutions. 
The fully resolved geometry, after proper transformation is then
\begin{equation}\label{SUCodim1SmoothEVEN}
	\begin{aligned}
		y^2B(\delta) &- x^3B(\zeta)A(\zeta\delta)\zeta_k^k + b_1xy - b_2x^2\zeta_0B(\zeta)\zeta_k 
			+ b_3y\zeta_0^kB(\zeta\delta)C(\zeta\delta) \cr &- b_4x\zeta_0^kB(\zeta)C(\zeta\delta) 
			- b_6\zeta_0^{2k}B(\zeta^2\delta)C(\zeta^2\delta^2) = 0 \,.
	\end{aligned}
\end{equation}
The projectivity relations are again included in appendix \ref{app:SUn}.


\subsubsection{Codimension 1 }

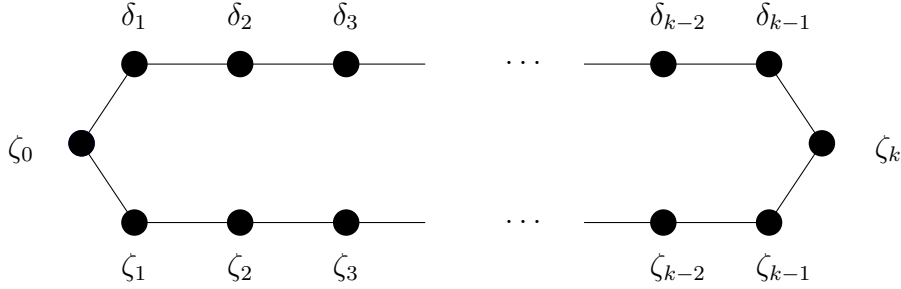
\begin{figure}
\begin{center}
\begin{picture}(280, 160)

	\put(-20,50){\color{blue}\circle*{10}}
	\put(-20,50){\circle*{10}}
	\put(-48,45){$\zeta_0$}

	\put(0,20){\circle*{10}}
	\put(40,20){\circle*{10}}
	\put(80,20){\circle*{10}}

	\put(140,20){$\ldots$}

	\put(200,20){\circle*{10}}
	\put(240,20){\circle*{10}}
	\put(260,50){\circle*{10}}

	\put(0,20){\line(1,0){110}}
	\put(170,20){\line(1,0){70}}

	\put(-5,0){$\zeta_1$}
	\put(35, 0){$\zeta_2$}
	\put(75,0){$\zeta_3$}
	\put(195,0){$\zeta_{k-2}$}
	\put(235,0){$\zeta_{k-1}$}
	
	\put(280,45){$\zeta_k$}


	\put(0,80){\circle*{10}}
	\put(40,80){\circle*{10}}
	\put(80,80){\circle*{10}}

	\put(140,80){$\ldots$}

	\put(200,80){\circle*{10}}
	\put(240,80){\circle*{10}}

	\put(0,80){\line(1,0){110}}
	\put(170,80){\line(1,0){70}}

	\put(-5,95){$\delta_1$}
	\put(35, 95){$\delta_2$}
	\put(75,95){$\delta_3$}
	\put(195,95){$\delta_{k-2}$}
	\put(235,95){$\delta_{k-1}$}

	
	\put(0,80){\line(-2,-3){20}}
	\put(0,20){\line(-2,3){20}}
	\put(240,80){\line(2,-3){20}}
	\put(240,20){\line(2,3){20}}

\end{picture}
\end{center}
\caption{Intersection graph of the fibers in codimension 1, resulting in the affine $A_{2k-1}$ Dynkin diagram. The labels are explained in (\ref{CartansSUeven}).}\label{fig:DynkinSUeven}
\end{figure}

The irreducible exceptional divisors are 
\begin{equation}\label{CartansSUeven}
	\begin{array}{l|c|l}
	\hbox{Divisor} & \hbox{Section} & \hbox{Equation in $Y_4$} \cr\hline
		D_{-\alpha_0} & \zeta_0   &  0=y^2\delta_1 - x^3\zeta_1 + b_1xy  \cr
		D_{-\alpha_1} & \zeta_1   & 0=\delta_1 + b_1x  \cr
		D_{-\alpha_i} \quad\hfill i = 2,\cdots,k-1  & \zeta_i  & 0=\delta_{i-1}\delta_i + b_1x  \cr
		D_{-\alpha_k} & \zeta_k  &  0=y^2\delta_{k-1} + b_1xy + b_3y\zeta_{k-1}\delta_{k-1} - b_4x\zeta_{k-1} 
										- b_6\zeta_{k-1}^2\delta_{k-1}  \cr
		D_{-\alpha_{k+1}} & \delta_{k-1}  & 0= b_1y - b_2\zeta_k\zeta_{k-1} - b_4\zeta_{k-1}\delta_{k-2}  \cr
		D_{-\alpha_{2k-i}} \quad\hfill i = 2,\cdots,k-2 & \delta_i  & 0=b_1y - b_2\zeta_i\zeta_{i+1}  \cr
		D_{-\alpha_{2k-1}} & \delta_1  &  0=b_1y - b_2\zeta_0\zeta_1\zeta_2 - \zeta_1\zeta_2^2\delta_2 
	\end{array}
\end{equation}
The intersections follow from the relations in appendix \ref{app:SUn}, and reproduce the affine $SU(2k)$ Dynkin diagram, where $\zeta_0$ corresponds to the divisors $D_{-\alpha_0}$. The intersection of the sections $\zeta_i$ and $\delta_i$ are as in figure \ref{fig:DynkinSUeven}.

\subsubsection{Codimension 2 }

The codimension 2 enhancements are along $b_1 = 0$ 
or  $P = b_1b_3b_4 + b_4^2 - b_1^2b_6 = 0$.  The vanishing of the discriminant, which is generically $O(\zeta_0^{2k})$, increases as follows
\begin{equation}
\ba
b_1=0:\qquad& \Delta|_{b_1=0} = O(\zeta_0^{2k+2}) \cr
P =0:\qquad& \Delta|_{P=0} = O(\zeta_0^{2k+1}) \,,
\ea
\end{equation}
corresponding to a local enhancement to $SO(4k)$ and $SU(2k+1)$, as will be demonstrated from the intersection graph of the fiber in codimesion 2. 



\begin{figure}
\begin{center}
\begin{picture}(280, 140)

\color{green}

	\put(0,20){\color{blue}\line(2,3){20}}
	\put(0,20){\color{blue}\line(-1,1){30}}	
	\put(0,20){\color{blue}\line(-1,2){30}}	
	\put(20,50){\color{blue}\line(1,0){100}}
	\put(160,50){\color{blue}\line(1,0){120}}

	\put(240,80){\line(1,0){40}}
	\put(240,50){\color{blue}\line(5,4){40}}


	\put(-30,50){\color{blue}\circle*{10}}
	\put(-30,50){\circle*{5}}
	\put(-48,45){$\zeta_0$}


	\put(0,20){\color{blue}\circle*{10}}
	\put(0,20){\circle*{5}}	
	\put(40,20){\circle*{10}}
	\put(80,20){\circle*{10}}

	\put(140,20){$\ldots$}

	\put(200,20){\circle*{10}}
	\put(240,20){\circle*{10}}

	\put(0,80){\line(-1,0){30}}
		
	\put(0,20){\line(0,1){60}}
	\put(40,20){\line(0,1){60}}
	\put(80,20){\line(0,1){60}}
	
	\put(200,20){\line(0,1){60}}
	\put(240,20){\line(0,1){60}}

	\put(40,20){\line(-2,3){40}}
	\put(80,20){\line(-2,3){40}}
	\put(80,80){\line(2,-3){20}}

	\put(200,20){\line(-2,3){20}}
	\put(240,20){\line(-2,3){40}}

	\put(-5,0){$\zeta_1$}
	\put(35, 0){$\zeta_2$}
	\put(75,0){$\zeta_3$}
	\put(195,0){$\zeta_{k-2}$}
	\put(235,0){$\zeta_{k-1}$}


	\put(20,50){\color{blue}\circle*{10}}
	\put(40,50){\color{blue}\circle*{10}}
	\put(60,50){\color{blue}\circle*{10}}
	\put(80,50){\color{blue}\circle*{10}}
	\put(100,50){\color{blue}\circle*{10}}

	\put(140,50){\color{blue}$\ldots$}

	\put(180,50){\color{blue}\circle*{10}}
	\put(200,50){\color{blue}\circle*{10}}
	\put(220,50){\color{blue}\circle*{10}}
	\put(240,50){\color{blue}\circle*{10}}

	\put(280,50){\color{blue}\circle*{10}}
	\put(280,50){\circle*{5}}
	\put(290,45){$\zeta_k$}


	\put(-30,80){\color{blue}\circle*{10}}
	\put(0,80){\circle*{10}}
	\put(40,80){\circle*{10}}
	\put(80,80){\circle*{10}}

	\put(140,80){$\ldots$}

	\put(200,80){\circle*{10}}
	\put(240,80){\circle*{10}}
	\put(280,80){\color{blue}\circle*{10}}

	\put(-35,95){\color{blue}$v_{2k}$}
	\put(-5,95){$\delta_1$}
	\put(35, 95){$\delta_2$}
	\put(75,95){$\delta_3$}
	\put(195,95){$\delta_{k-2}$}
	\put(235,95){$\delta_{k-1}$}
	\put(275,95){\color{blue}$v_{k+1}$}

\end{picture}
\end{center}
\caption{Splitting of the Cartan divisors $D_{-\alpha_i}$ along the codimension 2 locus $b_1=0$ for $SU(2k)$. $\color{green}\bullet$ are  the simple roots of $A_{2k-1}$, green lines indicate how they split along $b_1=0$. $\color{blue}\bullet$ are the irreducible components of the fiber over the codimension 2 locus, the blue lines give their intersection graph. Curves that remain irreducible are bicolored. The intersections and multiplicities are those of a $D_{2k}$ Kodaira fiber.}
\label{fig:SUevenSplittingb1}
\end{figure}
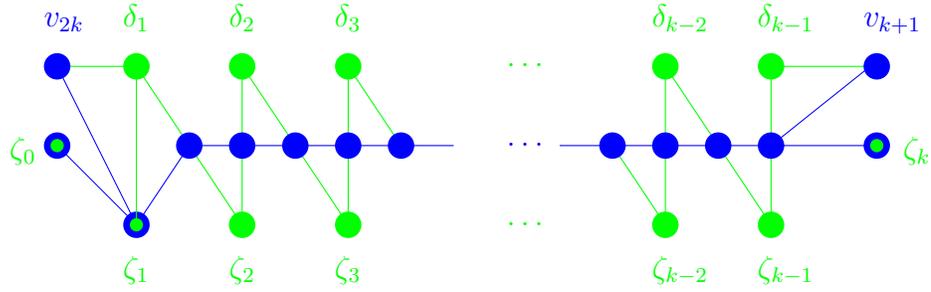



\begin{figure}
\begin{center}
\begin{picture}(280, 140)
\color{green}

	\color{blue}
	\put(0,80){\line(1,0){110}}
	\put(170,80){\line(1,0){110}}

	\put(0,20){\line(1,0){110}}
	\put(170,20){\line(1,0){110}}
		
	\put(0,80){\line(-2,-3){20}}
	\put(0,20){\line(-2,3){20}}	

\color{green}

	\put(280,80){\line(2,-3){20}}
	\put(280,20){\line(2,3){20}}


\put(280,20){\color{blue}\line(0,1){60}}

\put(240,80){\circle*{5}}
\put(280,80){\color{blue}\circle*{10}}

\put(300,50){\circle*{10}}
\put(315,45){$\zeta_k$}

\put(280,20){\color{blue}\circle*{10}}


	\put(-20,50){\color{blue}\circle*{10}}
		\put(-20,50){\circle*{5}}
	\put(-48,45){$\zeta_0$}

	\put(0,20){\color{blue}\circle*{10}}
		\put(0,20){\circle*{5}}
	\put(40,20){\color{blue}\circle*{10}}
		\put(40,20){\circle*{5}}	
	\put(80,20){\color{blue}\circle*{10}}
		\put(80,20){\circle*{5}}

	\put(140,20){\color{blue}$\ldots$}

	\put(200,20){\color{blue}\circle*{10}}
		\put(200,20){\circle*{5}}
	\put(240,20){\color{blue}\circle*{10}}
		\put(240,20){\circle*{5}}

	\put(-5,0){$\zeta_1$}
	\put(35, 0){$\zeta_2$}
	\put(75,0){$\zeta_3$}
	\put(195,0){$\zeta_{k-2}$}
	\put(235,0){$\zeta_{k-1}$}

	\put(275,0){\color{blue}$u_{2k}$}


	\put(0,80){\color{blue}\circle*{10}}
	\put(0,80){\circle*{5}}
	\put(40,80){\color{blue}\circle*{10}}
	\put(40,80){\circle*{5}}
	\put(80,80){\color{blue}\circle*{10}}
	\put(80,80){\circle*{5}}

	\put(140,80){\color{blue}$\ldots$}
	
	\put(200,80){\color{blue}\circle*{10}}
	\put(200,80){\circle*{5}}
	\put(240,80){\color{blue}\circle*{10}}
	\put(240,80){\circle*{5}}


	\put(-5,95){$\delta_1$}
	\put(35, 95){$\delta_2$}
	\put(75,95){$\delta_3$}
	\put(195,95){$\delta_{k-2}$}
	\put(235,95){$\delta_{k-1}$}

	\put(275,95){\color{blue}$u_{k}$}

\end{picture}
\end{center}
\caption{Intersection graph of the fibers for $SU(2k)$ in codimension 2 along $P=0$. The root ${\alpha_k}$ splits into 
weights $u_k$ and $u_{2k}$. The intersections of the irreducible components $\color{blue}\bullet$  in codimension 2 give the affine $A_{2k}$ Dynkin diagram. }\label{fig:MatterPSUeven}
\end{figure}
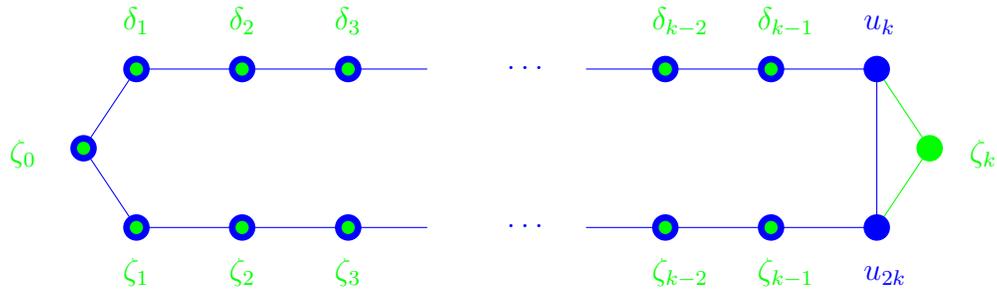


The irreducible divisors for the $b_1=0$ matter locus are
\begin{equation}
	\begin{array}{l|c|l|l}
		\mbox{Matter} & \mbox{Section} & \mbox{Equation in }Y_4|_{b_1=0} & j\text{th Cartan charge} \cr\hline
		S_{-\alpha_0} & \zeta_0 & 0=y^2\delta_1 - x^3\zeta_1 & \delta_{j,1} + \delta_{j,2k-1} - 2\delta_{j,0} \cr 
		S_{-\alpha_1} & \zeta_1& 0=\delta_1 & \delta_{j,0} + \delta_{j,2} - 2\delta_{j,1} \cr
		S_{v_i} \quad\hfill i = 2,\cdots,k-1 & \zeta_i & 0=\delta_i  & \delta_{j,i+1} - \delta_{j,i} + \delta_{j,2k+1-i} - \delta_{j,2k-i} \cr
		S_{-\alpha_k} & \zeta_k &0=\delta_{k-1}y^2 - b_4x\zeta_{k-1} &  \delta_{j,k+1} + \delta_{j,k-1} - 2\delta_{j,k}\cr 
		& & \qquad+ \delta_{k-1}(b_3y\zeta_{k-1} - b_6\zeta_{k-1}^2) 
			& \cr
		S_{v_{k+1}} & \delta_{k-1} &0= b_2\zeta_k+b_4\delta_{k-2} & \delta_{j,k-1} - \delta_{j,k+1} \cr
		S_{v_{k+i}} \quad\hfill i = 2,\cdots,k-1 & \zeta_i & 0=\delta_{i-1} & \delta_{j,i-1} - \delta_{j,i} + \delta_{j,2k-i} - \delta_{2k+1-i} \cr
		S_{v_{2k}} & \delta_1 & 0=b_2\zeta_0+\zeta_2\delta_2 & \delta_{j,1} - \delta_{j,2k-1} \cr 
	\end{array}
\end{equation}
The weights that appear are either of the $\Lambda^2 V$ or $\Lambda^2\overline{V}$, where $V={\bf 2k}$ is the fundamental representation. 
In summary, along $b_1=0$ the following Cartans split 
\begin{equation}\ba
D_{-\alpha_i} &\quad \longrightarrow \quad  S_{v_i} + S_{v_{k+i}}\cr
D_{-\alpha_{2k-i}} & \quad \longrightarrow \quad S_{v_i} + S_{v_{k+i+1}} \cr
D_{-\alpha_{k+1}} & \quad \longrightarrow \quad S_{v_{k-1}} + S_{v_{k+1}} \cr
D_{-\alpha_{2k-1}} & \quad \longrightarrow \quad S_{v_1} + S_{v_{k+2}} + S_{v_{2k}} \,.
\ea\end{equation}

The other codimension 2 locus at which the discriminant enhances is
\be
P = b_1b_3b_4 + b_4^2 - b_1^2b_6 = 0 \,,
\ee
without $b_1$ vanishing.  Again, we can,
without loss of generality, scale the Tate form by $b_1^2$ and perform the 
$P = 0$ substitution, resulting in 
\begin{equation}
	\begin{aligned}
		b_1^2y^2B(\delta) &- b_1^2x^3B(\zeta)A(\zeta\delta)\zeta_k^k + b_1^3xy 
		- b_1^2b_2x^2\zeta_0B(\zeta)\zeta_k 
		+ b_1^2b_3y\zeta_0^kB(\zeta\delta)C(\zeta\delta) 
		\cr&- b_1^2b_4x\zeta_0^kB(\zeta)C(\zeta\delta) 
		- (b_1b_3b_4 + b_4^2)\zeta_0^{2k}B(\zeta^2\delta)C(\zeta^2\delta^2) = 0 \,.
	\end{aligned}
\end{equation}
The irreducible exceptional divisors along $P=0$ are 
\begin{equation}
	\begin{array}{l|c|l|l}
		\mbox{Matter} & \mbox{Section} & \mbox{Equation in $Y_4|_{P=0}$}& j\mbox{th Cartan charge} \cr\hline
		S_{-\alpha_0} & \zeta_0 &0= y^2\delta_1 - x^3\zeta_1 + b_1xy & \delta_{j,1} + \delta_{j,2k-1} - 2\delta_{j,0} \cr 
		S_{-\alpha_1} & \zeta_1 &0= \delta_1 + b_1x & \delta_{j,0} + \delta_{j,2} - 2\delta_{j,1} \cr
		S_{-\alpha_i}, \quad \hfill i=2,\cdots,k-1 & \zeta_i &0= \delta_{i-1}\delta_i + b_1x & \delta_{j,i-1} + \delta_{j,i+1} - 2\delta_{j,i} \cr
		S_{u_k} & \zeta_k &0= b_1y - b_4\zeta_{k-1} & -\delta_{j,k} + \delta_{j,k+1} \cr
		S_{-\alpha_{2k-1}} & \delta_1 &0= b_1y - b_2\zeta_0\zeta_1\zeta_2 - \zeta_1\zeta_2^2\delta_2 
			& \delta_{j,0} + \delta_{j,2k-2} - 2\delta_{j,2k-1} \cr 
		S_{-\alpha_{2k-i}}, \quad\hfill i=2,\cdots,k-2 & \delta_i &0= b_1y - b_2\zeta_i\zeta_{i+1} & \delta_{j,2k-i-1} + \delta_{j,2k-i+1} - 2\delta_{j,2k-i} \cr
		S_{-\alpha_{k+1}} & \delta_{k-1} &0= b_1y - b_2\zeta_{k-1}\zeta_k & \cr & & \quad - b_4\zeta_{k-1}\delta_{k-2} 
			& \delta_{j,k} + \delta_{j,k+2} - 2\delta_{j,k+1} \cr
		S_{u_{2k}} & \zeta_k &0= b_1^2x + b_1y\delta_{k-1} & \cr & & \quad + \delta_{k-1}(b_4\zeta_{k-1} +  b_1b_3\zeta_{k-1}) & \delta_{j,k-1} - \delta_{j,k} \cr
	\end{array}
\end{equation}
Using the calculations in the appendix \ref{app:SUn} the Cartan charges of the components that split along the locus $P=0$ are found to be of the fundamental representation $V={\bf 2k}$.
In particular, the Cartan divisor associated to the vanishing of the section $\zeta_k$ becomes reducible and splits 
\begin{equation}
D_{-\alpha_k} \quad \longrightarrow\quad S_{u_k} + S_{u_{2k}} \,.
\end{equation}


\subsubsection{Codimension 3 }

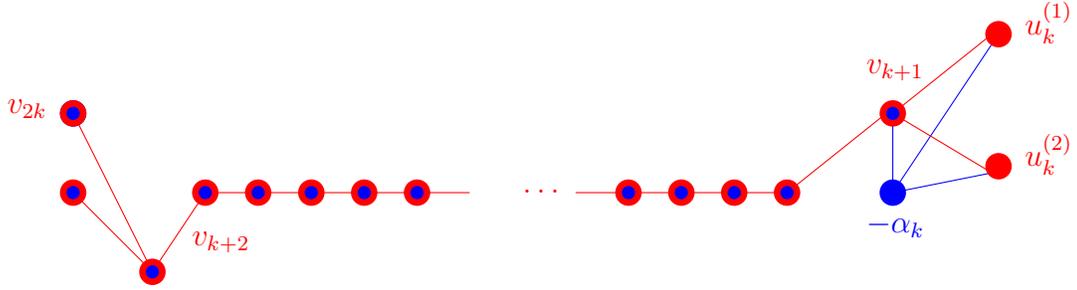
\begin{figure}
\begin{center}
\begin{picture}(280, 160)

\color{white}

	\put(0,20){\color{red}\line(-1,2){30}}	
	\put(0,20){\color{red}\line(2,3){20}}
	\put(0,20){\color{red}\line(-1,1){30}}	

	\put(20,50){\color{red}\line(1,0){100}}
	\put(160,50){\color{red}\line(1,0){80}}

	\put(280,80){\color{blue}\line(0,-1){30}}
	\put(280,50){\color{blue}\line(5,1){40}}
	\put(280,50){\color{blue}\line(2,3){40}}
	
	\put(240,50){\color{red}\line(5,4){40}}
	
	\put(280,80){\color{red}\line(5,4){40}}
	\put(280,80){\color{red}\line(5,-3){40}}

	\put(-30,50){\color{red}\circle*{10}}
	\put(-30,50){\color{blue}\circle*{5}}
	\put(-30,30){$\zeta_0$}


	{\put(-30,80){\color{blue}\circle*{10}}}
	\put(-55,80){\color{red}$v_{2k}$}
	\put(15,30){\color{red}$v_{k+2}$}

	\put(-30,80){\color{red}\circle*{10}}
	\put(-30,80){\color{blue}\circle*{5}}

	\put(320,110){\color{red}\circle*{10}}
	\put(320,60){\color{red}\circle*{10}}
	\put(330,110){\color{red}$u_{k}^{(1)}$}
	\put(330,60){\color{red}$u_k^{(2)}$}

	\put(280,80){\color{red}\circle*{10}}
	\put(280,80){\color{blue}\circle*{5}}

	\put(270,95){\color{red}$v_{k+1}$}


	\put(0,20){\color{red}\circle*{10}}
	\put(0,20){\color{blue}\circle*{5}}


	\put(20,50){\color{red}\circle*{10}}
	\put(20,50){\color{blue}\circle*{5}}

	\put(40,50){\color{red}\circle*{10}}
	\put(40,50){\color{blue}\circle*{5}}

	\put(60,50){\color{red}\circle*{10}}
	\put(60,50){\color{blue}\circle*{5}}

	\put(80,50){\color{red}\circle*{10}}
	\put(80,50){\color{blue}\circle*{5}}

	\put(100,50){\color{red}\circle*{10}}
	\put(100,50){\color{blue}\circle*{5}}

	\put(140,50){\color{red}$\ldots$}

	\put(180,50){\color{red}\circle*{10}}
	\put(180,50){\color{blue}\circle*{5}}

	\put(200,50){\color{red}\circle*{10}}
	\put(200,50){\color{blue}\circle*{5}}

	\put(220,50){\color{red}\circle*{10}}
	\put(220,50){\color{blue}\circle*{5}}

	\put(240,50){\color{red}\circle*{10}}
	\put(240,50){\color{blue}\circle*{5}}

	\put(280,50){\color{blue}\circle*{10}}
	\put(270,35){\color{blue}$-\alpha_k$}


\end{picture}
\end{center}
\caption{The fibers along the codimension 3 locus $b_1=b_4=0$ for $SU(2k)$. The irreducible components in codimension 3 are depicted in terms of red and bicolored dots. 
Red lines give their intersections. The blue dots are the codimension 2 $b_1=0$ fibers, most of which remain irreducible (specified by the bicoloring), except for $-\alpha_{k}$ and the blue line indicates how $-\alpha_{k}$ splits. The intersections are of $D$ type Kodaira. The multiplicities are read off from the splitting from one codimension lower.}
\label{fig:SUevenSplittingb1b4}
\end{figure}

Along the codimension 3 loci, as in the $SU(2k+1)$ case, only a few of the matter surfaces split further. The vanishing order of the discriminant increases to
\begin{equation}
\ba
b_1=b_4 =0:\qquad& \Delta|_{b_1=b_3=0} = O(\zeta_0^{2k+3}) \cr
P= b_2=0:\qquad & \Delta|_{P=b_2=0} = O(\zeta_0^{3k}) \cr
b_1=b_2=0:\qquad& \Delta|_{b_1=b_2=0} = O(\zeta_0^{3k}) \,.
\ea
\end{equation}
Along $b_1=b_4=0$ note that $P=0$ and indeed, we observe the following matter splitting
\begin{equation}
S_{-\alpha_k} \quad \longrightarrow \quad \Sigma_{v_{k+1}} + \Sigma_{u_k^{(1)}}+ \Sigma_{u_k^{(2)}} \,,
\end{equation}
where $u_k^{(i)}$ label the two curves, which are obtained as the two solutions of $\zeta_k=b_1=b_4=0$, which both have 
charge $u_k$, and intersect along $\zeta_{k-1}=0$. This is consistent with the Yukawa coupling
\begin{equation}
b_1= b_4=0:\qquad  \bar{V} \otimes \bar{V} \otimes \Lambda^2 V  \,.
\end{equation}
Figure \ref{fig:SUevenSplittingb1b4} shows the intersection graph of the fiber along this codimension 3 locus, which is of  affine $D$ type, 
and  the multiplicities which can be read off from the splitting of the nodes in codimension 2, are consistent with a  Kodaira fiber. 
Similarly, $P= b_2=0 $ yields the coupling 
\begin{equation}
P=b_2 =0 :\qquad V \otimes \overline{V} \otimes {\bf 1} \,.
\end{equation}

\begin{figure}
\begin{center}
\begin{picture}(280, 140)

\color{white}

	\put(0,80){\line(-1,0){30}}
		
	\put(0,20){\line(0,1){60}}
	\put(40,20){\line(0,1){60}}
	\put(80,20){\line(0,1){60}}
	
	\put(200,20){\line(0,1){60}}
	\put(240,20){\line(0,1){60}}

	\put(40,20){\line(-2,3){40}}
	\put(80,20){\line(-2,3){40}}
	\put(80,80){\line(2,-3){20}}

	\put(200,20){\line(-2,3){20}}
	\put(240,20){\line(-2,3){40}}

	\put(0,20){\color{blue}\line(2,3){20}}
	\put(0,20){\color{blue}\line(-1,1){30}}	
	\put(20,50){\color{red}\line(0,1){70}}

	\put(240,80){\line(1,0){40}}


	\put(0,20){\color{red}\line(2,3){20}}
	\put(0,20){\color{red}\line(-1,1){30}}	

	\put(240,50){\color{red}\line(5,4){40}}
	\put(20,50){\color{red}\line(1,0){100}}
	\put(160,50){\color{red}\line(1,0){120}}
	
	\put(-30,80){\color{blue}\line(3,-2){50}}	
	\put(-30,80){\color{blue}\line(4,3){50}}	


	\put(-30,50){\color{red}\circle*{10}}
	\put(-30,50){\color{blue}\circle*{5}}
	\put(-30,30){$\zeta_0$}


	{\put(-30,80){\color{blue}\circle*{10}}}
	\put(-55,80){\color{blue}$v_{2k}$}
	
	{\put(20,120){\color{red}\circle*{10}}}
	\put(32,120){\color{red}$w'$}

	\put(15,30){\color{red}$v_{k+2}$}


	\put(0,20){\color{red}\circle*{10}}
	\put(0,20){\color{blue}\circle*{5}}	

	\put(40,20){\circle*{10}}
	\put(80,20){\circle*{10}}

	\put(140,20){$\ldots$}

	\put(200,20){\circle*{10}}
	\put(240,20){\circle*{10}}

	\put(-5,0){$\zeta_1$}
	\put(35, 0){$\zeta_2$}
	\put(75,0){$\zeta_3$}
	\put(195,0){$\zeta_{k-2}$}
	\put(235,0){$\zeta_{k-1}$}


	\put(20,50){\color{red}\circle*{10}}
	\put(20,50){\color{blue}\circle*{5}}

	\put(40,50){\color{red}\circle*{10}}
	\put(40,50){\color{blue}\circle*{5}}

	\put(60,50){\color{red}\circle*{10}}
	\put(60,50){\color{blue}\circle*{5}}

	\put(80,50){\color{red}\circle*{10}}
	\put(80,50){\color{blue}\circle*{5}}

	\put(100,50){\color{red}\circle*{10}}
	\put(100,50){\color{blue}\circle*{5}}

	\put(140,50){\color{blue}$\ldots$}

	\put(180,50){\color{red}\circle*{10}}
	\put(180,50){\color{blue}\circle*{5}}

	\put(200,50){\color{red}\circle*{10}}
	\put(200,50){\color{blue}\circle*{5}}

	\put(220,50){\color{red}\circle*{10}}
	\put(220,50){\color{blue}\circle*{5}}

	\put(240,50){\color{red}\circle*{10}}
	\put(240,50){\color{blue}\circle*{5}}

	\put(280,50){\color{red}\circle*{10}}
	\put(280,50){\color{blue}\circle*{5}}
	\put(290,45){$\zeta_k$}


	\put(-30,80){\color{blue}\circle*{10}}
	\put(0,80){\circle*{10}}
	\put(40,80){\circle*{10}}
	\put(80,80){\circle*{10}}

	\put(140,80){$\ldots$}

	\put(200,80){\circle*{10}}
	\put(240,80){\circle*{10}}
	
	\put(280,80){\color{red}\circle*{10}}
	\put(280,80){\color{blue}\circle*{5}}

	\put(-5,95){$\delta_1$}
	\put(35, 95){$\delta_2$}
	\put(75,95){$\delta_3$}
	\put(195,95){$\delta_{k-2}$}
	\put(235,95){$\delta_{k-1}$}
	\put(275,95){\color{blue}$v_{k+1}$}

\end{picture}
\end{center}
\caption{Codimension 3 locus $b_1=b_2=0$ for $SU(2k)$. The irreducible components in codimension 3 are depicted in terms of red and bicolored dots. 
Red lines give their intersections. The blue dots are the codimension 2 $b_1=0$ fibers, most of which remain irreducible (specified by the bicoloring) and the blue line indicates how $v_{2k}$ splits. 
}
\label{fig:SUevenSplittingb1b2}
\end{figure}
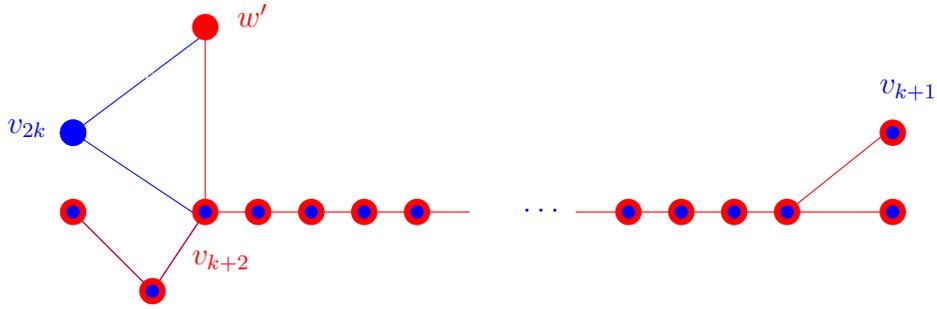

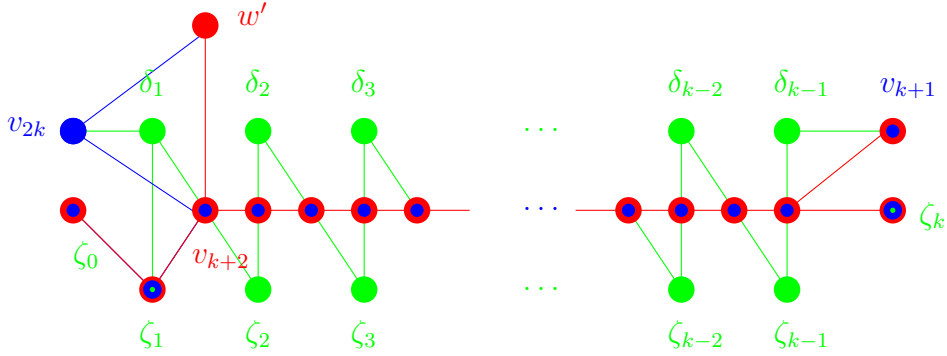
\begin{figure}
\begin{center}
\begin{picture}(280, 160)

\color{green}

	\put(0,80){\line(-1,0){30}}
		
	\put(0,20){\line(0,1){60}}
	\put(40,20){\line(0,1){60}}
	\put(80,20){\line(0,1){60}}
	
	\put(200,20){\line(0,1){60}}
	\put(240,20){\line(0,1){60}}

	\put(40,20){\line(-2,3){40}}
	\put(80,20){\line(-2,3){40}}
	\put(80,80){\line(2,-3){20}}

	\put(200,20){\line(-2,3){20}}
	\put(240,20){\line(-2,3){40}}

	\put(0,20){\color{blue}\line(2,3){20}}
	\put(0,20){\color{blue}\line(-1,1){30}}	
	\put(20,50){\color{red}\line(0,1){70}}

	\put(240,80){\line(1,0){40}}


	\put(0,20){\color{red}\line(2,3){20}}
	\put(0,20){\color{red}\line(-1,1){30}}	

	\put(240,50){\color{red}\line(5,4){40}}
	\put(20,50){\color{red}\line(1,0){100}}
	\put(160,50){\color{red}\line(1,0){120}}
	
	\put(-30,80){\color{blue}\line(3,-2){50}}	
	\put(-30,80){\color{blue}\line(4,3){50}}	


	\put(-30,50){\color{red}\circle*{10}}
	\put(-30,50){\color{blue}\circle*{5}}
	\put(-30,30){$\zeta_0$}


	{\put(-30,80){\color{blue}\circle*{10}}}
	\put(-55,80){\color{blue}$v_{2k}$}
	
	{\put(20,120){\color{red}\circle*{10}}}
	\put(32,120){\color{red}$w'$}

	\put(15,30){\color{red}$v_{k+2}$}


	\put(0,20){\color{red}\circle*{10}}
	\put(0,20){\color{blue}\circle*{7}}	
	\put(0,20){\circle*{2}}

	\put(40,20){\circle*{10}}
	\put(80,20){\circle*{10}}

	\put(140,20){$\ldots$}

	\put(200,20){\circle*{10}}
	\put(240,20){\circle*{10}}

	\put(-5,0){$\zeta_1$}
	\put(35, 0){$\zeta_2$}
	\put(75,0){$\zeta_3$}
	\put(195,0){$\zeta_{k-2}$}
	\put(235,0){$\zeta_{k-1}$}


	\put(20,50){\color{red}\circle*{10}}
	\put(20,50){\color{blue}\circle*{5}}

	\put(40,50){\color{red}\circle*{10}}
	\put(40,50){\color{blue}\circle*{5}}

	\put(60,50){\color{red}\circle*{10}}
	\put(60,50){\color{blue}\circle*{5}}

	\put(80,50){\color{red}\circle*{10}}
	\put(80,50){\color{blue}\circle*{5}}

	\put(100,50){\color{red}\circle*{10}}
	\put(100,50){\color{blue}\circle*{5}}

	\put(140,50){\color{blue}$\ldots$}

	\put(180,50){\color{red}\circle*{10}}
	\put(180,50){\color{blue}\circle*{5}}

	\put(200,50){\color{red}\circle*{10}}
	\put(200,50){\color{blue}\circle*{5}}

	\put(220,50){\color{red}\circle*{10}}
	\put(220,50){\color{blue}\circle*{5}}

	\put(240,50){\color{red}\circle*{10}}
	\put(240,50){\color{blue}\circle*{5}}

	\put(280,50){\color{red}\circle*{10}}
	\put(280,50){\color{blue}\circle*{7}}
	\put(280,50){\circle*{2}}
	\put(290,45){$\zeta_k$}


	\put(-30,80){\color{blue}\circle*{10}}
	\put(0,80){\circle*{10}}
	\put(40,80){\circle*{10}}
	\put(80,80){\circle*{10}}

	\put(140,80){$\ldots$}

	\put(200,80){\circle*{10}}
	\put(240,80){\circle*{10}}
	
	\put(280,80){\color{red}\circle*{10}}
	\put(280,80){\color{blue}\circle*{5}}

	\put(-5,95){$\delta_1$}
	\put(35, 95){$\delta_2$}
	\put(75,95){$\delta_3$}
	\put(195,95){$\delta_{k-2}$}
	\put(235,95){$\delta_{k-1}$}
	\put(275,95){\color{blue}$v_{k+1}$}
	
\end{picture}
\end{center}
\caption{Summary of splittings for codimension 3 locus $b_1=b_2=0$ for $SU(2k)$. Green dots are the fiber components in codimension 1, blue dots in codimension 2, and red in codimension 3. Red lines are the intersections in codimension 3, whereas blue and green lines indicate how the codimension $d$ splits into the codimension $d+1$ fibers. In particular tricolored nodes remain irreducible in all codimension. }
\label{fig:SUevenSplittingb1b22}
\end{figure}

Finally, there is the non-minimal locus $b_1=b_2=0$ which is very similar to the $SU(2k+1)$ case. Again, new surface components enter the fiber, like the $\aleph_i$  defined in (\ref{NMSurface})\ssn{, and the fibration ceases to be flat}. We again consider the splitting of the matter surfaces that we discussed along $b_1=0$. 
The matter surface that splits is $S_{v_{2k}}$, which corresponds to the $b_2 \zeta_0 + \zeta_2 \delta_2$ component of  $\delta_1=b_1=0$ inside $Y_4$. Along $b_2=0$ this splits further into two components 
\begin{equation}
S_{v_{2k}} \cdot [b_2] :\qquad \delta_1 =\delta_2 \zeta_2=0 \,.
\end{equation}
The Cartan charges are readily determined from the appendix \ref{app:SUn} and show that along $b_1=b_2=0$ we generate a new component in the fiber, given by the curve
\begin{equation}
\Sigma_{w'} : \qquad b_1= b_2= \delta_1 =\delta_2 =0 \,,
\end{equation}
the Cartan charges of which are
\begin{equation}
({w'})_j=  \delta_{j, 2} - \delta_{2k-1, j} \,,
\end{equation}
and is the Dynkin label of a weight of the $\Lambda^4V$ representation. 
The splitting of $S_{v_{2k}}$ along $b_2=0$ is then
\begin{equation}
S_{v_{2k}} \quad \longrightarrow \quad \Sigma_{w'} + \Sigma_{v_{k+2}} \,,
\end{equation}
where $ \Sigma_{v_{k+2}}=  S_{v_{k+2}} \cdot [b_2]$. The intersections of the irreducible components in the fiber are depicted in figures 
\ref{fig:SUevenSplittingb1b2} and \ref{fig:SUevenSplittingb1b22}. This is consistent with the Yukawa coupling 
\begin{equation}
b_1= b_2=0:\qquad \Lambda^2 \bar{V} \otimes \Lambda^2 \bar{V} \otimes \Lambda^4V  \,,
\end{equation}
keeping of course in mind the non-minimality of this locus. 


\newpage
\section{Resolution of $D$ Type Singularities}

Similarly to the $A_n$ case the $D_n$ splits  into 
odd and even $n$, and we shall study these in turn. The main difference to the $SU(n)$ case is that for $SO(2\ell)$ there are non-minimal loci in codimension 2.


\subsection{$SO(4k+2)$}
\label{sec:SOodd}

The Tate form for an $SO(4k+2)$ singularity above $\zeta_0 = 0$ is
\begin{equation}\label{SOoddTate}
        y^2 - x^3 + b_1xy\zeta_0 -b_2x^2\zeta_0 + b_3y\zeta_0^k 
        - b_4x\zeta_0^{k+1} - b_6\zeta_0^{2k+1} = 0 \,.
\end{equation}
The singularity in the fiber can be resolved in all codimensions using the  
notation of (\ref{BUNotation}) by the following  resolutions
\begin{equation}\label{SOoddBlowups}
        \begin{array}{rl}
                (x, y, \zeta_i; \zeta_{i+1}) & \quad i = 0,\cdots,k-1 \cr
                (y, \zeta_i;\delta_i) & \quad i = 1,\cdots,k \cr
                (\zeta_i, \delta_i; \kappa_i) & \quad i = 1,\cdots,k-1 \cr
                (\zeta_{i+1}, \delta_i; \epsilon_i)& \quad i=1, \cdots ,k-1 \,. 
        \end{array}
\end{equation} 
Note that after the first two sets of resolutions, i.e. $(x, y, \zeta_i; \zeta_{i+1})$ and $(y, \zeta_i;\delta_i) $, the geometry takes the form
\begin{equation}
B(\delta)\delta_k  U = B(\zeta) \zeta_k V\,,
\end{equation}
which again makes the different choices of small resolutions explicit. We will, as for $SU(n)$, pick one such small resolution and leave the study of the network of flops to future work. 
The geometry obtained
after applying all the resolutions (\ref{SOoddBlowups}) on the space defined
by (\ref{SOoddTate}) is
\begin{equation}\label{TateResSOodd}
        \begin{aligned}
		&y^2{B(\delta)\delta_k}
                - x^3 \zeta_k^k\delta_k^{k-1}
		B(\zeta{\epsilon})A(\zeta\delta\kappa^2 \epsilon^2)
		+ b_1xy\zeta_0{B(\zeta\delta\kappa\epsilon)\zeta_k\delta_k}
                \cr &
		- b_2x^2\zeta_0B(\zeta){\zeta_k}
                + b_3y\zeta_0^kB(\zeta\delta\kappa) C(\zeta\delta\kappa^2\epsilon^2)
                \cr &
                - b_4x\zeta_0^{k+1}\zeta_k \, B(\zeta^2\delta\kappa^2 \epsilon)
                        C(\zeta\delta\kappa^2\epsilon^2)
                - b_6\zeta_0^{2k+1}\zeta_k \, B(\zeta^3\delta^2\kappa^4 \epsilon^2)
                        C(\zeta^2 \delta^2\kappa^4\epsilon^4) =0\,,
        \end{aligned}
\end{equation}
along with the projective relations between the sections that are
documented in appendix \ref{app:SOodd}. We use the convinient
shorthand (\ref{ABCEDef}).

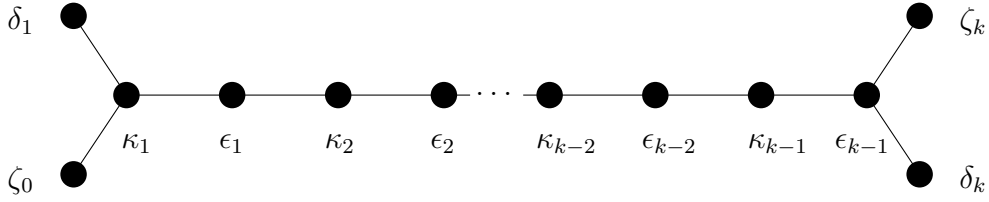
\begin{figure}
\begin{center}
\begin{picture}(280, 120)

	\put(-20,70){\circle*{10}}
	\put(-45,65){$\delta_1$}

	\put(-20,10){\circle*{10}}
	\put(-45,5){$\zeta_0$}


	\put(300,70){\circle*{10}}
	\put(315,65){$\zeta_k$}

	\put(300,10){\circle*{10}}
	\put(315,5){$\delta_k$}

	\put(0,40){\circle*{10}}
	\put(40,40){\circle*{10}}
	\put(80,40){\circle*{10}}
	\put(120,40){\circle*{10}}

	\put(132,40){$\ldots$}

	\put(160,40){\circle*{10}}
	\put(200,40){\circle*{10}}
	\put(240,40){\circle*{10}}
	\put(280,40){\circle*{10}}
	
	\put(0,40){\line(1,0){130}}
	\put(150,40){\line(1,0){130}}

	\put(-2,20){$\kappa_1$}
	\put(35, 20){$\epsilon_1$}
	\put(75,20){$\kappa_2$}
	\put(115,20){$\epsilon_2$}
	\put(155,20){$\kappa_{k-2}$}
	\put(195,20){$\epsilon_{k-2}$}
	\put(235,20){$\kappa_{k-1}$}
	\put(268,20){$\epsilon_{k-1}$}

	
	\put(0,40){\line(-2,-3){20}}
	\put(0,40){\line(-2,3){20}}
	\put(280,40){\line(2,-3){20}}
	\put(280,40){\line(2,3){20}}

\end{picture}
\end{center}
\caption{Intersection graph of the fibers in codimension 1, resulting in the affine $D_{2k+1}$ Dynkin diagram.}\label{fig:DynkinSOodd}
\end{figure}


\subsubsection{Codimension 1 }

The projective relations in appendix \ref{app:SOodd} allow us
to write the equation in $Y_4$ corresponding to the vanishing
of the exceptional sections in a particularly simple form. The irreducible exceptional divisors are
\begin{equation}
        \begin{array}{l|c|l}\label{SOoddDivs} 
                \mbox{Divisor } & \mbox{Section } & \mbox{Equation in } Y_4  \cr\hline
                D_{-\alpha_0} & \zeta_0 & 0 = y^2\delta_1 - x^3\zeta_1  \cr
                D_{-\alpha_1} & \delta_1  & 0 = \delta_2 \kappa_2^2 \epsilon_1 + b_2\zeta_0  \cr
                D_{-\alpha_2} & \kappa_1 & 0 = \delta_1 - \zeta_1 \zeta_2 (b_2 \zeta_0 + \zeta_2 \epsilon_1) \cr
                D_{-\alpha_{2i+1}} \, \quad i = 1,\cdots,k-2 & \epsilon_i  & 0 = \delta_i - b_2 \zeta_{i+1} \cr
                D_{-\alpha_{2i}} \qquad i = 2,\cdots,k-1 & \kappa_i & 0 = \delta_{i-1}\delta_i - b_2 \zeta_i \zeta_{i+1}  \cr 
                D_{-\alpha_{2k-1}} & \epsilon_{k-1} & 0 = \delta_{k-1} (b_3 \kappa_{k-1} + \delta_k) - b_2 \zeta_k \cr
                D_{-\alpha_{2k}} & \zeta_k & 0 = \delta_k + b_3\zeta_{k-1} \kappa_{k-1} \cr
                D_{-\alpha_{2k+1}} & \delta_k & 0 = b_3 \delta_{k-1} y-\zeta_k b_2 x^2-\delta_{k-1} \epsilon_{k-1}\zeta_k (b_4 x+b_6 \delta_{k-1} \epsilon_{k-1})) 
        \end{array}
\end{equation}
The $D_{-\alpha_i}$s are the {Cartan divisors}, the label $\alpha_i$
refers to the affine simple $SO(4k+2)$ roots. Computing the intersections
of these Cartan divisors with each other will allow us to construct a
Cartan charge vector. The procedure for calculating the intersections
of these divisors is explained in appendix \ref{app:SOodd}. To each
Cartan divisor we construct the $j$th element of the Cartan charge
vector by intersecting that Cartan divisor with $D_{-\alpha_j}$. The
Cartan charge vectors are
\begin{equation}
        \begin{array}{l|l}
                \text{Cartan divisor } & j\text{th Cartan charge} \cr\hline
                D_{-\alpha_0} & \delta_{j,2} - 2\delta_{j,0} \cr
                D_{-\alpha_1} & \delta_{j,2} - 2\delta_{j,1} \cr
                D_{-\alpha_2} & \delta_{j,0} + \delta_{j,1} + \delta_{j,3} - 2\delta_{j,2} \cr
                D_{-\alpha_{2i+1}} & \delta_{j,2i} + \delta_{j,2i+2} - 2\delta_{j,2i+1} \cr
                D_{-\alpha_{2i}} & \delta_{j,2i-1} + \delta_{j,2i+1} - 2\delta_{j,2i} \cr
                D_{-\alpha_{2k-1}} & \delta_{j,2k-2} + \delta_{j,2k} + \delta_{j,2k+1} - 2\delta_{j,2k-1} \cr
                D_{-\alpha_{2k}} & \delta_{j,2k-1} - 2\delta_{j,2k} \cr
                D_{-\alpha_{2k+1}} & \delta_{j,2k-1} - 2\delta_{j,2k+1} \cr
        \end{array}
\end{equation}
which are the (affine) simple roots $\alpha_i$ of  $SO(4k+2)$  in the
canonical ordering.  Figure \ref{fig:DynkinSOodd} summarizes the intersection graph with the labels refering to the sections.


\subsubsection{Codimension 2}

Along the higher codimension loci, as listed in table \ref{BigTable},
the equation in $Y_4$ for the vanishing of the exceptional
sections can become reducible. 
Along the two matter loci $b_2=0$ and $b_3=0$ the vanishing order of the discriminant, which is $O(\zeta_0^{2k+3})$ increases to
\begin{equation}
\ba
b_3=0:\qquad& \Delta|_{b_3=0} = O(\zeta_0^{2k+4})\cr
b_2=0:\qquad& \Delta|_{b_2=0} = O(\zeta_0^{2k+6})  \,.
\ea
\end{equation}
  For low rank cases, such as $SO(10)$, $b_2=0$ corresponds to an exceptional enhancement, e.g. $E_6$, in particular this gives rise to spin  representations. For higher values of $k$, $b_2=0$ is non-minimal. 
The matter surfaces at a
particular codimension 2 locus will be the irreducible
components of (\ref{SOoddDivs}) at that locus. The specifics of
the intersections of these with the Cartan divisors are again given in appendix \ref{app:SOodd}.

\begin{figure}
\begin{center}
\begin{picture}(280, 140)
\color{green}

\color{blue}	
	\put(0,40){\line(-2,-3){20}}
	\put(0,40){\line(-2,3){20}}
	\put(280,40){\line(2,3){20}}
	\put(0,40){\line(1,0){130}}
	\put(150,40){\line(1,0){130}}
	\put(300,70){\line(1,0){40}}
	\put(300,70){\line(4,-3){40}}
	
\color{green}
	\put(300,10){\line(0,1){60}}
	\put(300,10){\line(2,3){40}}
	\put(300,10){\line(4,3){40}}

 	\put(-20,70){\color{blue}\circle*{10}}
	\put(-20,70){\circle*{5}}
	\put(-45,65){$\delta_1$}

 	\put(-20,10){\color{blue}\circle*{10}}
	\put(-20,10){\circle*{5}}
	\put(-45,5){$\zeta_0$}

 	\put(340,70){\color{blue}\circle*{10}}
	\put(340,40){\color{blue}\circle*{10}}

 	\put(300,70){\color{blue}\circle*{10}}
	\put(300,70){\circle*{5}}

	\put(295,85){$\zeta_k$}
	\put(355,75){\color{blue}$u_{2k+1}^{(2)}$}
	\put(355,35){\color{blue}$u_{2k+1}^{(1)}$}

	\put(300,10){\circle*{10}}
	\put(315,5){$\delta_k$}

	\put(0,40){\color{blue}\circle*{10}}
	\put(0,40){\circle*{5}}

	\put(40,40){\color{blue}\circle*{10}}	
	\put(40,40){\circle*{5}}
 	\put(80,40){\color{blue}\circle*{10}}
	\put(80,40){\circle*{5}}
 	\put(120,40){\color{blue}\circle*{10}}
	\put(120,40){\circle*{5}}
	
	\put(132,40){\color{blue}$\ldots$}

 	\put(160,40){\color{blue}\circle*{10}}
	\put(160,40){\circle*{5}}
 	\put(200,40){\color{blue}\circle*{10}}
	\put(200,40){\circle*{5}}
 	\put(240,40){\color{blue}\circle*{10}}
	\put(240,40){\circle*{5}}
 	\put(280,40){\color{blue}\circle*{10}}	
	\put(280,40){\circle*{5}}

	\put(-2,20){$\kappa_1$}
	\put(35, 20){$\epsilon_1$}
	\put(75,20){$\kappa_2$}
	\put(115,20){$\epsilon_2$}
	\put(155,20){$\kappa_{k-2}$}
	\put(195,20){$\epsilon_{k-2}$}
	\put(235,20){$\kappa_{k-1}$}
	\put(268,20){$\epsilon_{k-1}$}

\color{blue}

\end{picture}
\end{center}
\caption{
Intersection graph of the fibers in codimension 2 along $b_3=0$ for $SO(4k+2)$. $\color{blue}\bullet$ (and bicolored nodes) are the irreducible components in codimension 2, green lines indicate the splitting, in particular $\delta_k=0$ splits off $u_{2k+1}^{(i)}$. Blue lines are the intersections of the irreducible components in codimension 2. From the latter we see that the intersection graph is an affine $D$ type Dynkin diagram, and the multiplicities are those of a Kodaira fiber. 
 }\label{fig:Matterb3SOodd}
\end{figure}
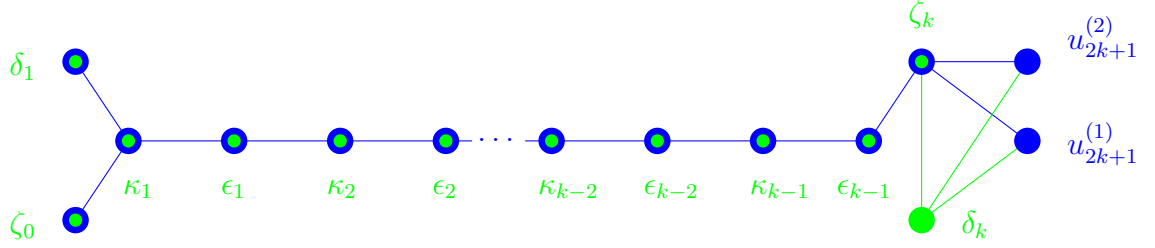


Along the codimension 2 locus $b_3 = 0$ the irreducible components of the restrictions of the Cartan divisors, i.e. matter surfaces, and the corresponding Cartan charges are
\begin{equation}
        \begin{array}{l|c|l|l}\label{SOevenb2}
                \text{Matter surface } & \text{Section}  & \hbox{Equation in $Y_{b_3=0}$} & j\text{th Cartan charge} \cr\hline
                  S_{-\alpha_0} & \zeta_0 & 0 = y^2\delta_1 - x^3\zeta_1 & \delta_{j,2} - 2\delta_{j,0} \cr
                S_{-\alpha_1} & \delta_1  & 0 = \delta_2 \kappa_2^2 \epsilon_1 + b_2\zeta_0   & \delta_{j,2} - 2\delta_{j,1}\cr
                S_{-\alpha_2} & \kappa_1 & 0 = \delta_1 - \zeta_1 \zeta_2 (b_2 \zeta_0 + \zeta_2 \epsilon_1) & \delta_{j,0} + \delta_{j,1} + \delta_{j,3} - 2\delta_{j,2}\cr
                S_{-\alpha_{2i+1}} \, \hfill   i = 1,\cdots,k-2 & \epsilon_i  & 0 = \delta_i - b_2 \zeta_{i+1} & \delta_{j,2i} + \delta_{j,2i+2} - 2\delta_{j,2i+1}\cr
                S_{-\alpha_{2i}}  \hfill   i = 2,\cdots,k-1 & \kappa_i & 0 = \delta_{i-1}\delta_i - b_2 \zeta_i \zeta_{i+1} 
                							& \delta_{j,2i-1} + \delta_{j,2i+1} - 2\delta_{j,2i}  \cr 
                S_{-\alpha_{2k-1}} & \epsilon_{k-1} & 0 = \delta_{k-1} \delta_k - b_2 \zeta_k 
                							& \delta_{j,2k-2} + \delta_{j,2k} + \delta_{j,2k+1} - 2\delta_{j,2k-1}\cr
                S_{-\alpha_{2k}} & \zeta_k & 0 = \delta_k  & \delta_{j,2k-1} - 2\delta_{j,2k} \cr
                S_{-\alpha_{2k+1}} & \delta_k & 0 = \zeta_k b_2 x^2 & \delta_{j,2k-1} - 2\delta_{j,2k+1} \cr
                				&&+\delta_{k-1} \epsilon_{k-1}\zeta_k (b_4 x+b_6 \delta_{k-1} \epsilon_{k-1}) &
               \cr
                  S_{u_{2k+1}^{(i)}}\qquad i=1,2& \delta_k &  0 = b_2 x^2  & \delta_{j,2k} -\delta_{j,2k+1} \cr
                && +\delta_{k-1} \epsilon_{k-1} (b_4 x+b_6 \delta_{k-1} \epsilon_{k-1})&
        \end{array}
\end{equation}
Note that $u_{2k+1}^{(1,2)}$ labels the divisors corresponding to the roots of the quadratic polynomial in the last line, which are generically distinct curves, as its discriminant does not generically vanish along $b_3=0$.
The  only non-trivial splitting occurs for $D_{-\alpha_{2k+1}}$,
as is depicted in figure \ref{fig:Matterb3SOodd}
\begin{equation}
        \ba
                D_{-\alpha_{2k+1}}  &\quad\longrightarrow\quad S_{-\alpha_{2k}} + S_{u_{2k+1}^{(1)}} + S_{u_{2k+1}^{(2)}} \,.
        \ea
\end{equation}
The charges are related to the highest weight $(\mu_{V})_j = \delta_{j,1}$ of the fundamental representation by
\begin{equation}
S_{u_{2k+1}^{(i)}} :\qquad  \mu_{V} - \left(\sum_{{i=1}}^{2k-1}\alpha_i + \alpha_{2k+1}\right)  \,.
\end{equation}
We discuss the non-minimal matter locus $b_2=0$ in section \ref{sec:SONM}. 


\subsubsection{Codimension 3 }

For a generic $SO(4k+2)$ the codimension 3 loci of symmetry enhancement are along
\begin{equation}
      \ba
                     b_3 = b_4^2 - 4b_2b_6 &= 0\cr
                b_2 = b_{1,3,4} & = 0 \cr
                b_2 = b_1b_3 + b_4 &= 0       \,.
        \ea
\end{equation}
Only along $b_2=b_3=0$  do  the matter surfaces split further, however, this is a non-minimal locus, which will be discussed in the next subsection. 

The only minimal singularity enhancement occurs along $b_3= b_4^2 - 4 b_2 b_6=0$. Consider the matter surfaces $S_{u_{2k+1}^{(i)}}$. We can assume, as this is codimension 3, that  without loss of generality,  $b_2$ does not vanish in addition. The discriminant of the quadratic equation defining  $S_{u_{2k+1}^{(i)}}$ is $b_4^2 - 4 b_2 b_6=0$, so that along this codimension 3 locus, the equation for these matter surfaces reduces to
\begin{equation}
 (2 b_2 x+b_4 \delta_{k-1} \epsilon_{k-1})^2 =0\,.
\end{equation}
I.e. the two distinct matter surfaces become degenerate, corresponding to a Yukawa coupling of the two ${4k+2}$ dimensional  fundamental $V$ representations in $S_{u_{2k+1}^{(i)}}$ 
\begin{equation}
 b_3 = b_4^2 - 4b_2b_6 = 0:\qquad V \otimes V \otimes {\bf 1} \,.
\end{equation}

\subsubsection{Non-minimal Loci}
\label{sec:SONM}

Finally consider the matter locus $b_2=0$, which is non-minimal. 
\ssn{Again, we observe that additional surface components appear in the fiber, which therefore ceases to be flat. This is easily seen as the resolved geometry (\ref{TateResSOodd}) along $b_2=0$ has the form}
\begin{equation}
\delta_2 \cdots \delta_{k-1} \left( \cdots \right)  =0 \,,
\end{equation}
which shows that the following loci are completely contained inside the fiber
\begin{equation}\label{AlephSO}
\aleph_{i}:\qquad b_2=\delta_i=0\,,\qquad i=2, \cdots, k-1 \,.
\end{equation}
As in the case for the non-minimal codimension 3 locus for $A_n$, we do not capitulate, but see what we can learn from the splitting of the Cartan divisors in codimension 1 along this locus. 
There are two observations which we will arrive at: 
first of all, the Cartan divisors split into surfaces which carry Cartan charges corresponding to spin representations, and in addition, 
depending on the specific small resolution, in addition there are matter surfaces in $\Lambda^{i} V$, for $V$ the fundamental representation. 
We consider here the resolution (\ref{SOoddBlowups}) and give another example resolution of the $D_{2k+1}$ singularities in appendix \ref{app:SOReload}, where the matter surfaces arising along this non-minimal locus are different.

\begin{figure}
\begin{center}
\begin{picture}(350, 140)
\color{green}

	
	\put(0,40){\color{blue}\line(-2,-3){20}}

	\put(0,40){\color{blue}\line(1,0){40}}
	\put(40,40){\color{blue}\line(2,3){20}}
	
	\put(80,40){\color{green}\line(2,3){20}}
	\put(80,40){\color{green}\line(-2,3){20}}	
	\put(60,70){\color{blue}\line(1,0){40}}	
	
	\put(120,40){\color{blue}\line(2,3){20}}
	\put(120,40){\color{blue}\line(-2,3){20}}		
	\put(160,40){\color{green}\line(-2,3){20}}
	\put(140,70){\color{blue}\line(1,0){20}}
	\put(160,40){\color{green}\line(2,3){10}}

	\put(200,40){\color{green}\line(-2,3){10}}	
	\put(200,40){\color{green}\line(2,3){20}}
	\put(220,70){\color{blue}\line(-1,0){20}}
	\put(240,40){\color{blue}\line(2,3){20}}
	\put(240,40){\color{blue}\line(-2,3){20}}	
	\put(260,70){\color{blue}\line(1,0){20}}
	
	\put(280,40){\color{green}\line(-2,3){20}}
	\put(280,40){\color{green}\line(2,3){20}}
	\put(300,70){\color{blue}\line(-1,0){20}}
	\put(320,40){\color{green}\line(0,1){30}}
	\put(300,70){\color{blue}\line(1,0){80}}
	\put(320,40){\color{green}\line(2,3){20}}

	\put(380,10){\color{green}\line(-1,0){40}}
	\put(380,10){\color{green}\line(-5,4){40}}
	\put(340,40){\color{blue}\line(0,-1){30}}
	\put(340,40){\color{blue}\line(-2,3){20}}
	
	\put(60,70){\color{blue}\line(0,1){30}}
	\put(-20,70){\color{green}\line(1,0){80}}
	\put(-22,72){\color{green}\line(1,0){80}}

	\put(-20,70){\color{green}\line(5,2){80}}
	
	\put(-20,70){\circle*{10}}
	\put(-45,65){$\delta_1$}

	\put(-20,10){\color{blue}\circle*{10}}
	\put(-20,10){\circle*{5}}
	\put(-45,5){$\zeta_0$}

	\put(60,100){\color{blue}\line(3,-2){40}}

	\put(-25,70){\color{green}\line(5,-2){70}}





	\put(60, 100){\color{blue}\circle*{10}}

	\put(60,70){\color{blue}\circle*{10}}
	\put(100,70){\color{blue}\circle*{10}}
	\put(140,70){\color{blue}\circle*{10}}
	\put(220,70){\color{blue}\circle*{10}}
	\put(260,70){\color{blue}\circle*{10}}
	\put(300,70){\color{blue}\circle*{10}}
	\put(320,70){\color{blue}\circle*{10}}
	\put(340,70){\color{blue}\circle*{10}}
	
	\put(340,40){\color{blue}\circle*{10}}
	\put(340,10){\color{blue}\circle*{10}}

	\put(0,40){\color{blue}\circle*{10}}
	\put(0,40){\circle*{5}}
	\put(40,40){\color{blue}\circle*{10}}
	\put(40,40){\circle*{5}}	
	\put(80,40){\circle*{10}}
	
	\put(120,40){\color{blue}\circle*{10}}
	\put(120,40){\circle*{5}}
	\put(160,40){\circle*{10}}

	\put(175,40){$\ldots$}

	\put(200,40){\circle*{10}}
	\put(240,40){\color{blue}\circle*{10}}	
	\put(240,40){\circle*{5}}
	\put(280,40){\circle*{10}}
	\put(320,40){\circle*{10}}

	\put(-2,20){$\kappa_1$}
	\put(35, 20){$\epsilon_1$}
	\put(75,20){$\kappa_2$}
	\put(115,20){$\epsilon_2$}
	\put(155,20){$\kappa_{3}$}
	\put(195,20){$\kappa_{k-2}$}
	\put(235,20){$\epsilon_{k-2}$}
	\put(275,20){$\kappa_{k-1}$}
	\put(310,20){$\epsilon_{k-1}$}


	\put(380,70){\color{blue}\circle*{10}}
	\put(380,70){\circle*{5}}
	\put(395,65){$\zeta_k$}

	\put(380,10){\circle*{10}}
	\put(395,5){$\delta_k$}
	
\end{picture}
\end{center}
\caption{Intersection graph of the fibers in codimension 2 along $b_2=0$ for $SO(4k+2)$. $\color{blue}\bullet$} are the irreducible components in codimension 2, bicolored nodes remain irreducible when passing to  codimension 2. Green lines indicate the splitting, blue lines the intersections.   \label{fig:Matterb2SOodd} 
\end{figure}
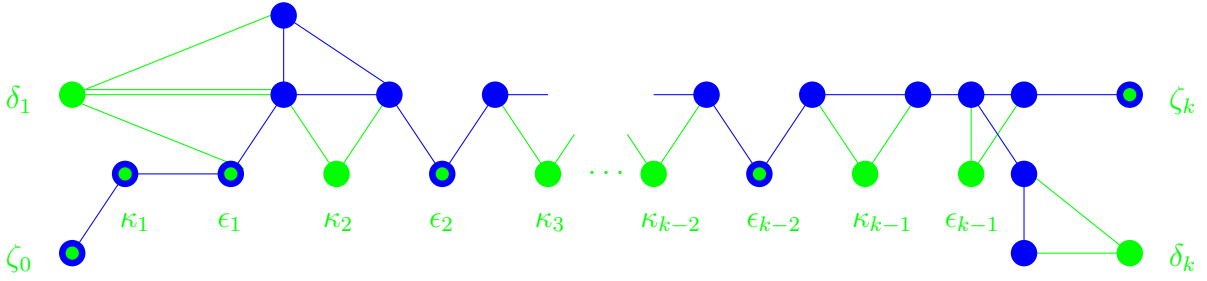



\begin{figure}\begin{center}
\begin{picture}(400, 140)

\put(50,10){\color{blue}\polygon*(10,10)(0,40)(50,40)(40,10)}
\put(50,10){\linethickness{2pt}\color{Red}\polygon(10,10)(0,40)(50,40)(40,10)}
\put(100,10){\color{blue}\polygon*(10,10)(0,40)(50,40)(40,10)}
\put(150,10){\color{blue}\polygon*(10,10)(0,40)(50,40)(40,10)}

\put(250,10){\color{blue}\polygon*(10,10)(0,40)(50,40)(40,10)}
\put(300,10){\color{blue}\polygon*(10,10)(0,40)(50,40)(40,10)}

\put(220, 30){\color{blue}$\cdots$}

\put(100,10){\linethickness{2pt}\color{Red}\polyline(0,40)(10,10)(40,10)(50,40)}
\put(150,10){\linethickness{2pt}\color{Red}\polyline(0,40)(10,10)(40,10)(50,40)}
\put(250,10){\linethickness{2pt}\color{Red}\polyline(0,40)(10,10)(40,10)(50,40)}
\put(300,10){\linethickness{2pt}\color{Red}\polyline(0,40)(10,10)(40,10)}

\put(10,10){\linethickness{2pt}\color{Red}
\Line(46,25)(10,25)
\Line(30,55)(30,25)
\Line(30,40)(10,40)
}

\put(300,10){\linethickness{2pt}\color{Red}
\Line(20,10)(20,-20)
\Line(20,10)(20,-20)
\Line(0,-10)(20,-10)

\Line(30,10)(30,-20)
\Line(30,10)(30,-20)
\Line(50,-10)(30,-10)

}

\put(70,30){\color{white}$\delta_{2}$}
\put(120,30){\color{white}$\delta_{3}$}
\put(170,30){\color{white}$\delta_{4}$}
\put(265,30){\color{white}$\delta_{k-2}$}
\put(315,30){\color{white}$\delta_{k-1}$}

\end{picture}
\end{center}

\caption{Fiber along the non-minimal codimension 2 locus $b_2=0$. The red line is the dual graph to the (blue line) intersection graph in figure \ref{fig:Matterb2SOodd}. The blue semi-hexagons are the 1 dimension higher components $\aleph_i$ given by the vanishing of $\delta_i$  as in (\ref{AlephSO}) that enter the fiber along $b_2=0$. The fiberation is not flat. 
 The edges that are part of the red intersection graph arise from the intersection of the components $\aleph_i$ with $\kappa_i$, $\epsilon_{i}$ and $\kappa_{i+1}$.}
\label{fig:SOoddNM}
\end{figure}
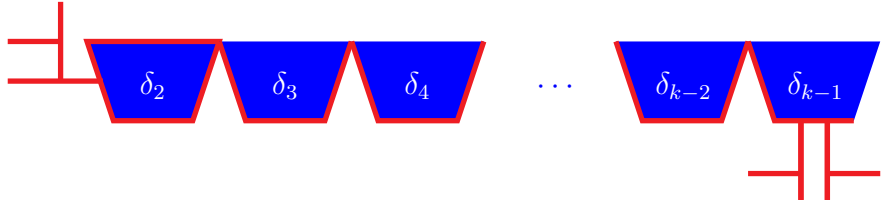

Consider the Cartan divisors (\ref{SOoddDivs}) along $b_2=0$. These split 
into irreducible matter surfaces, which have  Cartan charges given in the following table
\begin{equation}
        \begin{array}{l|c|l|l|l}
                \text{Matter surface } & \text{Section} & \hbox{Equation in $Y|_{b_2=0}$} & j\text{th Cartan charge} & \hbox{Rep}\cr\hline
                S_{-\alpha_0} & \zeta_0 & 0 = y^2\delta_1 - x^3\zeta_1 	&  \delta_{j,2} - 2\delta_{j,0} & \hbox{Adj} \cr
                S_{v_0} & \delta_1  & 0 = \delta_2 				& \delta_{j, 4} & \Lambda^4 V \cr
                S_{v_1} & \delta_1  & 0 = \kappa_2  			& -\delta_{j, 1} + \delta_{j, 3} - \delta_{j, 4} & \hbox{Adj}\cr
                %
                S_{-\alpha_2} & \kappa_1 & 0 = \delta_1 - \zeta_1 \zeta_2^2\epsilon_1 	& \delta_{j,0} + \delta_{j,1} + \delta_{j,3} - 2\delta_{j,2} & \hbox{Adj} \cr
                S_{-\alpha_{2i+1}} \hfill\quad i = 1,\cdots,k-2 & \epsilon_i  & 0 = \delta_i	& \delta_{j,2i} + \delta_{j,2i+2} - 2\delta_{j,2i+1}& \hbox{Adj}  \cr
                                S_{-\alpha_{2k}} & \zeta_k & 0 = \delta_k + b_3\zeta_{k-1} \kappa_{k-1}	&  \delta_{j,2k-1} - 2\delta_{j,2k} & \hbox{Adj} \cr
                S_{v_{3i}} \hfill\quad i = 2,\cdots,k-1 & \kappa_i & 0 = \delta_{i-1}	& \delta_{j, 2i-1} -\delta_{j,2i}  & V\cr 
                S_{v_{3i+1}} \hfill\quad i = 2,\cdots,k-1 & \kappa_i & 0 = \delta_i  	&  \delta_{j, 2i+1} -\delta_{j,2i} & V\cr 
                S_{s_{3k-1}} & \epsilon_{k-1} & 0 = \delta_{k-1}  			& \delta_{j, 2k-2} - \delta_{j, 2k-1} + \delta_{2k+1}&S^{-}\cr
                S_{s_{3k}} & \epsilon_{k-1} & 0 =  b_3 \kappa_{k-1} + \delta_k &-\delta_{j, 2k-1} + \delta_{2k}& S^{+}\cr
                S_{s_{3k+1}} & \delta_k & 0 = b_3  y  					&- \delta_{j, 2k+1} & S^{-} \cr 
      			&& \            - \epsilon_{k-1}\zeta_k (b_4 x+b_6 \delta_{k-1} \epsilon_{k-1})) && \cr
                S_{s_{3k+2}} & \delta_k & 0 = \delta_{k-1} & \delta_{j, 2k-1} - \delta_{j, 2k+1} & S^+ \cr
                \cr
        \end{array} 
\end{equation}
The last column indicates the representation. In summary, when passing to the codimension 2 locus $b_2=0$ the following Cartan divisors split into matter surfaces
\begin{equation}
  \ba
  		D_{-\alpha_1} & \quad \longrightarrow\quad S_{-\alpha_3} +2 \times  S_{v_0} + S_{v_1} \cr
		D_{-\alpha_{2i}} & \quad \longrightarrow\quad S_{v_{3i}} + S_{v_{3i+1}}\cr
		D_{-\alpha_{2k-1}} & \quad \longrightarrow\quad S_{s_{3k-1}} + S_{s_{3k}} \cr
		D_{-\alpha_{2k+1}} & \quad \longrightarrow\quad S_{s_{3k+1}} + S_{s_{3k+3}}\,.
\ea
\end{equation}

The Cartan charge vector can be associated with Dynkin labels of weights of certain
representations of ${SO(4k+2)}$. Denote the highest weights  of the spin representations by
\begin{equation}\ba
        (\mu_{S^+})_j &= \delta_{2k+1,j} \cr
        (\mu_{S^-})_j &= \delta_{2k,j} \,. 
\ea\end{equation}
Each matter surface whose Cartan charge vector is not a
simple root is listed below, along with the Cartan charge
vector.
\begin{equation}
        \begin{array}{l|l}
                \text{Matter surface } & \text{Weight } \cr\hline
S_{s_{3k-1}}   & \mu_{S^-} - \left(\alpha_{2k-1} + \alpha_{2k} \right)\cr              
S_{s_{3k}} 	& \mu_{S^+} - \left(\sum_{i=1}^{2k-1} i \alpha_i  + (k-1)\alpha_{2k} + k \alpha_{2k+1}   \right)  \cr
S_{s_{3k+1}}   & \mu_{S^-} - \left(\sum_{i=1}^{2k-1} i \alpha_i  + k\alpha_{2k} + k \alpha_{2k+1} \right)\cr
S_{s_{3k+2}}  & \mu_{S^+} - \alpha_{2k+1}         \cr
        \end{array}
\end{equation}
Along this non-minimal locus, it is clear that we do not expect a Dynkin diagram like intersection of the irreducible fiber components. 
As for codimension 3 for $SU(n)$ we can draw the intersection graph of the irreducible fiber components 
above $b_2=0$, which  are depicted in figure \ref{fig:Matterb2SOodd}. 
The graph is not a standard Dynkin diagram, however, for low $k$ values it reduces to $E$ type Dynkin diagrams. 
Alternatively we can include the additional components of the fiber $\aleph_i$ and how they intersect with the remaining components, as in figure \ref{fig:SOoddNM}. 

The codimension 3 non-minimal locus, which results in further splitting of the $b_2=0$ matter is along $b_2= b_3=0$, where
 $S_{v_{3k+1}}$  splits as
\begin{equation} 
\delta_k=0 \,,\qquad  \epsilon_{k-1}\zeta_k (b_4 x+b_6 \delta_{k-1} \epsilon_{k-1})) =0\,,
\end{equation}
which translates into
 \begin{equation}\label{SOoddcodim3}
        \ba
		S_{s_{3k+1}} &\quad\longrightarrow\quad \Sigma_{s_{3k}} + \Sigma_{u_{2k+1}}  + \Sigma_{-\alpha_{2k}}\,,
        \ea
\end{equation}
which is consistent with a coupling 
\begin{equation}
b_2= b_3=0 :\qquad S^+\otimes S^-\otimes V \,.
\end{equation}
In summary, considering the non-minimal loci results in higher dimensional components in the fiber.  However, we can nevertheless study some properties of these loci in codimension 2 and 3 by following the splitting of the Cartan divisors that are present in codimension 1. Interestingly, they generate representations, and couplings, that one would not expect from an ordinary field theoretic Higgsing of a higher rank $A$ or $D$ type gauge group.


\subsection{$SO(4k+4)$}

An elliptically fibered Calabi-Yau fourfold with an $SO(4k+4)$ singularity along $\zeta_0=0$ can be put into Tate form
\begin{equation}
        y^2 - x^3 + b_1xy\zeta_0 - b_2x^2\zeta_0 + b_3y\zeta_0^{k+1} - b_4x\zeta_0^{k+1} - b_6\zeta_0^{2k+1}  =0\,,
\end{equation}
along with the additional condition that $b_2x^2 + b_4x + b_6$ factors.
We shall choose to satisfy this extra condition by defining our
$SO(4k+4)$ singular Calabi-Yau by
\begin{equation}
        y^2 - x^3 + b_1xy\zeta_0 - b_2x^2\zeta_0 + b_3y\zeta_0^{k+1} - b_4x\zeta_0^{k+1} =0 \,.
\end{equation}
The discriminant of this Tate form is
\begin{equation}\label{DEDiscrim}
        \Delta(T_{D_{\text{even}}}) = 256b_2^2b_4^2\zeta_0^{2k+4} 
        + O(\zeta_0^{2k+5}) \,,
\end{equation}
which has the correct order noted in appendix \ref{app:Tate}. In particular, we have not
inadvertantly increased the order into a different type of
singularity by setting $b_6 = 0$. This choice for solving the additional factorization condition 
is not the most general such solution, as already noted in \cite{Bershadsky:1996nh}. 
This space can be resolved by
performing the following blowups, expressed in the notation of
(\ref{BUNotation})
\begin{equation}
        \ba
                (x, y, \zeta_i; \zeta_{i+1}) &\qquad  i = 0,\cdots,k \cr
                (y, \zeta_i, ; \delta_i) &\qquad  i = 1,\cdots,k \cr
                (\zeta_i, \delta_i; \kappa_i) & \qquad i=1, \cdots, k\cr
                (\zeta_{i+1}, \delta_i; \epsilon_i)&  \qquad i=1, \cdots, k-1\,.
        \ea
\end{equation}
The geometry obtained after proper transformation, corresponding to a Calabi-Yau with a resolved
$SO(4k+4)$ singularity, is then 
\begin{equation}
        \begin{aligned}
                &y^2B(\delta)\delta_k
                - x^3 \zeta_k^k\zeta_{k+1}^{k+1}\delta_k^{k-1}\kappa_k^{2k-2}
                B(\zeta\epsilon)A(\zeta\delta\kappa^2 \epsilon^2)
                + b_1xy\zeta_0\zeta_{k+1}B(\zeta\delta\kappa\epsilon)
                \cr &
                - b_2x^2\zeta_0B(\zeta)\zeta_k\zeta_{k+1}
                + b_3y\zeta_0^{k+1} \zeta_k B(\zeta^2\delta^2\kappa^2\epsilon^2) C(\zeta\delta\kappa^2\epsilon^2)
                \cr &
                - b_4x\zeta_0^{k+1}\zeta_k \, B(\zeta^2\delta\kappa^2 \epsilon)
                        C(\zeta\delta\kappa^2\epsilon^2)
               =0\,,
        \end{aligned}
\end{equation}
along with the projectivity conditions between the sections given in
appendix \ref{app:SOeven}. 


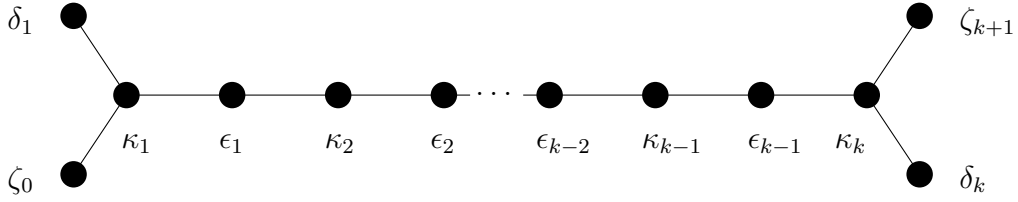
\begin{figure}
\begin{center}
\begin{picture}(280, 120)

	\put(-20,70){\circle*{10}}
	\put(-45,65){$\delta_1$}

	\put(-20,10){\circle*{10}}
	\put(-45,5){$\zeta_0$}


	\put(300,70){\circle*{10}}
	\put(315,65){$\zeta_{k+1}$}

	\put(300,10){\circle*{10}}
	\put(315,5){$\delta_k$}

	\put(0,40){\circle*{10}}
	\put(40,40){\circle*{10}}
	\put(80,40){\circle*{10}}
	\put(120,40){\circle*{10}}

	\put(132,40){$\ldots$}

	\put(160,40){\circle*{10}}
	\put(200,40){\circle*{10}}
	\put(240,40){\circle*{10}}
	\put(280,40){\circle*{10}}
	
	\put(0,40){\line(1,0){130}}
	\put(150,40){\line(1,0){130}}

	\put(-2,20){$\kappa_1$}
	\put(35, 20){$\epsilon_1$}
	\put(75,20){$\kappa_2$}
	\put(115,20){$\epsilon_2$}
	\put(155,20){$\epsilon_{k-2}$}
	\put(195,20){$\kappa_{k-1}$}
	\put(235,20){$\epsilon_{k-1}$}
	\put(268,20){$\kappa_{k}$}

	
	\put(0,40){\line(-2,-3){20}}
	\put(0,40){\line(-2,3){20}}
	\put(280,40){\line(2,-3){20}}
	\put(280,40){\line(2,3){20}}

\end{picture}
\end{center}
\caption{Intersection graph of the fibers in codimension 1, resulting in the affine $D_{2k+2}$ Dynkin diagram.}\label{fig:DynkinSOeven}
\end{figure}


\subsubsection{Codimension 1 }

The irreducible exceptional divisors for $SO(4k+4)$ are obtained as
\begin{equation}\label{DECartans} 
        \begin{array}{l|c|l|l}
                \text{Divisor } & \text{Section } & \text{Equation in } Y_4 & j\text{th Cartan charge} \cr\hline
                D_{-\alpha_0} & \zeta_0 & 0 = y^2\delta_1 - x^3\zeta_1 & \delta_{j,2} - 2\delta_{j,0} \cr
                D_{-\alpha_1} & \delta_1 & 0 = b_2 \zeta_0+\delta_2 \kappa_2^2 \epsilon_1 & \delta_{j,2} - 2\delta_{j,1} \cr
                D_{-\alpha_2} & \kappa_1 & 0 = \delta_1 - \zeta_1\zeta_2 (b_2 \zeta_0 + \zeta_2 \epsilon_1) 
                					& \delta_{j,0} + \delta_{j,1} + \delta_{j,3} - 2\delta_{j,2} \cr
                D_{-\alpha_{2i+1}} \hfill\quad i=1, \cdots, k-1& \epsilon_i & 0 = \delta_i - b_2 \zeta_{i+1} 
                						& \delta_{j,2i} + \delta_{j,2i+2} - 2\delta_{j,2i+1} \cr
                D_{-\alpha_{2i}} \hfill\quad i=2, \cdots, k-1 & \kappa_i& 0=\delta_{i-1}\delta_i - b_2 \zeta_{i} \zeta_{i+1}&\delta_{j,2i-1} + \delta_{j,2i+1} - 2\delta_{j,2i} \cr
                D_{-\alpha_{2k}} & \kappa_k & 0 = \delta_{k-1} (\delta_k-b_4 \zeta_{k} \epsilon_{k-1})
                        & \delta_{j,2k-1} + \delta_{j,2k+1} + \delta_{j,2k+2} - 2\delta_{j,2k} \cr
                        &&\quad -b_2 \zeta_k \zeta_{k+1} &\cr
                D_{-\alpha_{2k+1}} & \delta_k & 0 = b_2\zeta_{k+1} + b_4\delta_{k-1} & \delta_{j,2k} - 2\delta_{j,2k+1} \cr
                D_{-\alpha_{2k+2}} & \zeta_{k+1} & 0 = \delta_k y (b_3 \zeta_k \kappa_k+y) & \delta_{j,2k} - 2\delta_{j,2k+2} \cr
                &&\quad -b_4 \zeta_k x &\cr
        \end{array}
\end{equation}
The intersections are reproduced  in figure \ref{fig:DynkinSOeven}, 
where the simple roots of the standard ${SO(4k+4)}$ Lie group are the
Cartan charge vector associated to the labelled section above each node.

\subsubsection{Codimension 2 }

As can be seen from (\ref{DEDiscrim}) the codimension 2 loci
are $b_2 = 0$ or  $b_4 = 0$, along which  the Cartan
divisors descend to matter surfaces, with some of the divisors
becoming reducible. Each matter surface will also have a Cartan charge vector,
obtained by intersecting that matter surface with the Cartan
divisors, in the manner of appendix \ref{app:SOeven}.

When $b_4 = 0$ the only Cartan divisor that becomes reducible is $\zeta_{k+1}=0$, which yields three components, one of which is $S_{-\alpha_{2k+1}}$ and the remaining ones are 
\begin{equation}
        \begin{array}{l|l|l|l|l}
                \text{Matter surface } & \text{Section} & \text{Equation in } Y_4|_{b_4 = 0} & j\text{th Cartan charge} &\hbox{Rep}\cr\hline
                S_{u_1^{(1)}} & \zeta_{k+1} & 0 = y & \delta_{j,2k+1} - \delta_{j,2k+2} & V\cr
                 S_{u_1^{(2)}} & \zeta_{k+1} & 0 = (y + b_3\zeta_k\kappa_{k})& \delta_{j,2k+1} - \delta_{j,2k+2} & V \cr
        \end{array}
\end{equation}
In summary along $b_4=0$ the splitting is
\begin{equation}
        D_{-\alpha_{2k+2}} \quad\longrightarrow\quad S_{-\alpha_{2k+1}} +  S_{u_1^{(1)}}+  S_{u_1^{(2)}}  \,.
\end{equation}
The Cartan charge vector associated to each matter surface is a weight
in some representation of $SO(4k+4)$. The representations each is associated to
are as follows
\begin{equation}
        \begin{array}{l|l}
                \text{Matter surface } & \text{Weight } \cr\hline
                S_{u_1} & \mu_{V} - \left(\sum_{{i=1}}^{2k-1}\alpha_i + \alpha_{2k+1}\right)  \cr
		S_{s_{2k}} & \mu_{S^+} - \alpha_{2k} - \alpha_{2k+2} \cr
		 S_{s_{2k+1}} & \mu_{S^+} - \left(\sum_{i=1}^{2k} i \alpha_i + k \alpha_{2k+1} + k \alpha_{2k+2} \right)\cr
         \end{array}
\end{equation}

\begin{figure}
\begin{center}
\begin{picture}(280, 140)
\color{green}

\color{blue}	
	\put(0,40){\line(-2,-3){20}}
	\put(0,40){\line(-2,3){20}}
	\put(0,40){\line(1,0){130}}
	\put(150,40){\line(1,0){130}}
	\put(300,10){\line(1,0){40}}
	\put(300,10){\line(4,3){40}}
	\put(300,10){\line(-2,3){20}}
	
\color{green}
	\put(300,70){\line(0,-1){60}}
	\put(300,70){\line(2,-3){40}}
	\put(300,70){\line(4,-3){40}}

 	\put(-20,70){\color{blue}\circle*{10}}
	\put(-20,70){\circle*{5}}
	\put(-45,65){$\delta_1$}

 	\put(-20,10){\color{blue}\circle*{10}}
	\put(-20,10){\circle*{5}}
	\put(-45,5){$\zeta_0$}

 	\put(340,10){\color{blue}\circle*{10}}
	
 	\put(300,70){\color{green}\circle*{10}}
 	\put(340,40){\color{blue}\circle*{10}}
	
	\put(295,85){$\zeta_{k+1}$}
	\put(350,40){\color{blue}$u_{2k+1}^{(2)}$}
	\put(350,10){\color{blue}$u_{2k+1}^{(1)}$}
	
	\put(300,10){\color{blue}\circle*{10}}
	\put(300,10){\circle*{5}}
	
	\put(295,-10){$\delta_k$}

	\put(0,40){\color{blue}\circle*{10}}
	\put(0,40){\circle*{5}}

	\put(40,40){\color{blue}\circle*{10}}	
	\put(40,40){\circle*{5}}
 	\put(80,40){\color{blue}\circle*{10}}
	\put(80,40){\circle*{5}}
 	\put(120,40){\color{blue}\circle*{10}}
	\put(120,40){\circle*{5}}
	
	\put(132,40){\color{blue}$\ldots$}

 	\put(160,40){\color{blue}\circle*{10}}
	\put(160,40){\circle*{5}}
 	\put(200,40){\color{blue}\circle*{10}}
	\put(200,40){\circle*{5}}
 	\put(240,40){\color{blue}\circle*{10}}
	\put(240,40){\circle*{5}}
 	\put(280,40){\color{blue}\circle*{10}}	
	\put(280,40){\circle*{5}}

	\put(-2,20){$\kappa_1$}
	\put(35, 20){$\epsilon_1$}
	\put(75,20){$\kappa_2$}
	\put(115,20){$\epsilon_2$}
	\put(155,20){$\epsilon_{k-2}$}
	\put(195,20){$\kappa_{k-2}$}
	\put(235,20){$\epsilon_{k-1}$}
	\put(268,20){$\kappa_{k}$}

\color{blue}

\end{picture}
\end{center}
\caption{
Intersection graph of the fibers in codimension 2 along $b_4=0$ for $SO(4k+4)$. $\color{blue}\bullet$ (and bicolored nodes) are the irreducible components in codimension 2, green lines indicate the splitting. 
 }\label{fig:Matterb3SOeven}
\end{figure}
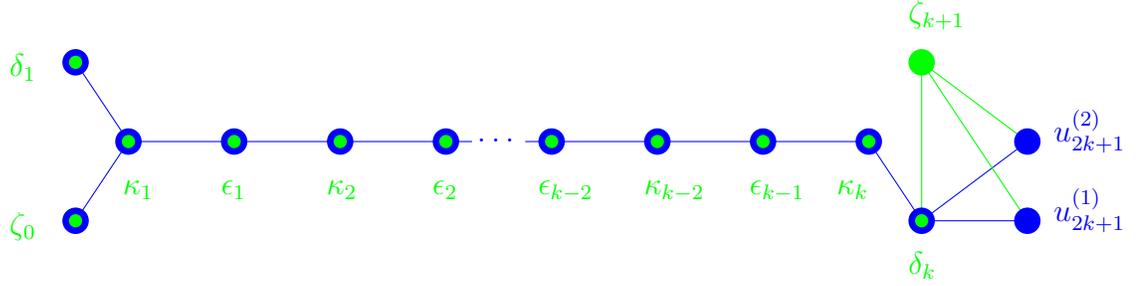

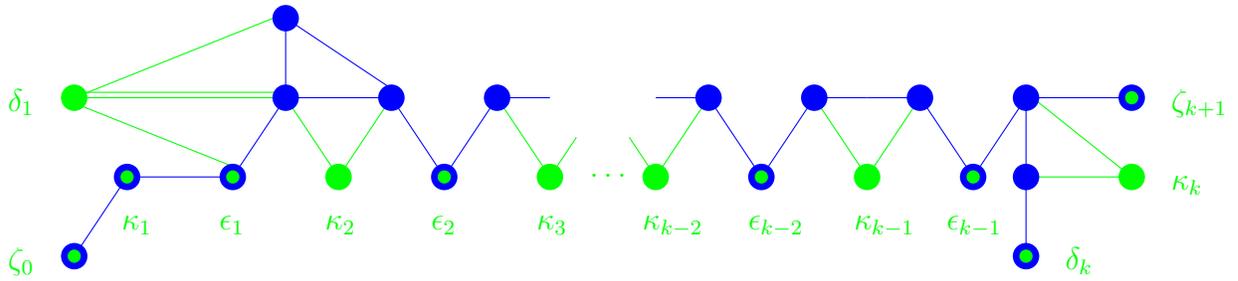
\begin{figure}
\begin{center}
\begin{picture}(350, 140)
\color{green}

	
	\put(0,40){\color{blue}\line(-2,-3){20}}

	\put(0,40){\color{blue}\line(1,0){40}}
	\put(40,40){\color{blue}\line(2,3){20}}
	
	\put(80,40){\color{green}\line(2,3){20}}
	\put(80,40){\color{green}\line(-2,3){20}}	
	\put(60,70){\color{blue}\line(1,0){40}}	
	
	\put(120,40){\color{blue}\line(2,3){20}}
	\put(120,40){\color{blue}\line(-2,3){20}}		
	\put(160,40){\color{green}\line(-2,3){20}}
	\put(140,70){\color{blue}\line(1,0){20}}
	\put(160,40){\color{green}\line(2,3){10}}

	\put(200,40){\color{green}\line(-2,3){10}}	
	\put(200,40){\color{green}\line(2,3){20}}
	\put(220,70){\color{blue}\line(-1,0){20}}
	\put(240,40){\color{blue}\line(2,3){20}}
	\put(240,40){\color{blue}\line(-2,3){20}}	
	\put(260,70){\color{blue}\line(1,0){20}}
	
	\put(280,40){\color{green}\line(-2,3){20}}
	\put(280,40){\color{green}\line(2,3){20}}
	\put(300,70){\color{blue}\line(-1,0){20}}
	\put(340,70){\color{blue}\line(1,0){40}}
	\put(320,40){\color{blue}\line(2,3){20}}
	\put(320,40){\color{blue}\line(-2,3){20}}

	\put(380,40){\color{green}\line(-1,0){40}}
	\put(380,40){\color{green}\line(-5,4){40}}
	\put(340,70){\color{blue}\line(0,-1){60}}
	
	\put(60,70){\color{blue}\line(0,1){30}}
	\put(-20,70){\color{green}\line(1,0){80}}
	\put(-22,72){\color{green}\line(1,0){80}}

	\put(-20,70){\color{green}\line(5,2){80}}
	
	\put(-20,70){\circle*{10}}
	\put(-45,65){$\delta_1$}

	\put(-20,10){\color{blue}\circle*{10}}
	\put(-20,10){\circle*{5}}
	\put(-45,5){$\zeta_0$}

	\put(-25,70){\color{green}\line(5,-2){65}}

	\put(60,100){\color{blue}\line(3,-2){40}}


	\put(60, 100){\color{blue}\circle*{10}}

	\put(60,70){\color{blue}\circle*{10}}
	\put(100,70){\color{blue}\circle*{10}}
	\put(140,70){\color{blue}\circle*{10}}
	\put(220,70){\color{blue}\circle*{10}}
	\put(260,70){\color{blue}\circle*{10}}
	\put(300,70){\color{blue}\circle*{10}}
	\put(340,70){\color{blue}\circle*{10}}
	
	\put(340,40){\color{blue}\circle*{10}}
	\put(340,10){\color{blue}\circle*{10}}
	\put(340,10){\circle*{5}}
	\put(355,5){$\delta_k$}

	\put(0,40){\color{blue}\circle*{10}}
	\put(0,40){\circle*{5}}
	\put(40,40){\color{blue}\circle*{10}}
	\put(40,40){\circle*{5}}	
	\put(80,40){\circle*{10}}
	
	\put(120,40){\color{blue}\circle*{10}}
	\put(120,40){\circle*{5}}
	\put(160,40){\circle*{10}}

	\put(175,40){$\ldots$}

	\put(200,40){\circle*{10}}
	\put(240,40){\color{blue}\circle*{10}}	
	\put(240,40){\circle*{5}}
	\put(280,40){\circle*{10}}
	\put(320,40){\color{blue}\circle*{10}}
	\put(320,40){\circle*{5}}

	\put(-2,20){$\kappa_1$}
	\put(35, 20){$\epsilon_1$}
	\put(75,20){$\kappa_2$}
	\put(115,20){$\epsilon_2$}
	\put(155,20){$\kappa_{3}$}
	\put(195,20){$\kappa_{k-2}$}
	\put(235,20){$\epsilon_{k-2}$}
	\put(275,20){$\kappa_{k-1}$}
	\put(310,20){$\epsilon_{k-1}$}


	\put(380,70){\color{blue}\circle*{10}}
	\put(380,70){\circle*{5}}
	\put(395,65){$\zeta_{k+1}$}

	\put(380,40){\circle*{10}}
	\put(395,35){$\kappa_k$}
	
\end{picture}
\end{center}
\caption{Intersection graph of the fibers in codimension 2 along $b_2=0$ for $SO(4k+4)$. $\color{blue}\bullet$} are the irreducible components in codimension 2, bicolored nodes remain irreducible when passing to  codimension 2. Green lines indicate the splitting, blue lines the intersections. From the latter we see that the intersection graph is not an ADE Dynkin diagram.    \label{fig:DynkinSOevenb2} 
\end{figure}

At $b_2 = 0$ the singularity is, like for $SO(4k+2)$, non-minimal, and the fibration has surface components. We determine the irreducible matter surfaces from the restriction of the Cartan divisors as
\begin{equation}
        \begin{array}{l|c|l|l|l}
                \text{Matter surface } & \text{Section} & \text{Equation in } Y_4|_{b_2 = 0} & j\text{th Cartan charge}& \hbox{Rep} \cr\hline
                S_{-\alpha_0} & \zeta_0 & 0 = y^2\delta_1 - x^3\zeta_1 & \delta_{j,2} - 2\delta_{j,0}  & \hbox{Adj}\cr
                S_{v_0} & \delta_1 & 0 = \delta_2 & \delta_{j,4} & \Lambda^4 V \cr
                S_{v_6}& \kappa_2 & 0 =  \delta_1  & -\delta_{j, 1} + \delta_{j, 3} - \delta_{j, 4} & \hbox{Adj}\cr
		S_{v_7}	& \kappa_2 & 0= \delta_2 & \delta_{j, 1} - \delta_{j, 4}+ \delta_{j, 5}& \hbox{Adj} \cr
                S_{-\alpha_2} & \kappa_1 & 0 = \delta_1 - \zeta_1\zeta_2^2 \epsilon_1
                					& \delta_{j,0} + \delta_{j,1} + \delta_{j,3} - 2\delta_{j,2}& \hbox{Adj}\cr
                S_{-\alpha_{2i+1}} \hfill\quad i=1, \cdots, k-1& \epsilon_i & 0 = \delta_i 
                						& \delta_{j,2i} + \delta_{j,2i+2} - 2\delta_{j,2i+1} & \hbox{Adj}\cr
                S_{-\alpha_{2k+1}} & \delta_k & 0 =  b_4\delta_{k-1} & \delta_{j,2k} - 2\delta_{j,2k+1}& \hbox{Adj}\cr
                S_{-\alpha_{2k+2}} & \zeta_{k+1} & 0 = \delta_k y (b_3 \zeta_k \kappa_k+y) & \delta_{j,2k} - 2\delta_{j,2k+2} & \hbox{Adj}\cr
                &&\quad -b_4 \zeta_k x &\cr
                S_{v_{3i}} \hfill\quad i=3, \cdots, k-1 & \kappa_i& 0=\delta_{i-1} 
                					&   \delta_{j, 2i-1} - \delta_{j, 2i} & V\cr
 		S_{v_{3i+1}} \hfill\quad i=3, \cdots, k-1 & \kappa_i& 0=\delta_i 
						& \delta_{j, 2i+1} - \delta_{j, 2i} & V \cr
                S_{s_{2k}} & \kappa_k & 0 = \delta_{k-1} 
                        & \delta_{j, 2k-1} - \delta_{j, 2k} + \delta_{j, 2k+1}  & S^+\cr
                 S_{s_{2k+1}} & \kappa_k & 0 =  \delta_k-b_4 \zeta_{k} \epsilon_{k-1}
                        &\delta_{j, 2k+2} -\delta_{j, 2k} & S^+\cr
        \end{array}
\end{equation}
where the Cartan divisors that split into matter surfaces nontrivially along $b_2=0$ are
\begin{equation}
        \begin{aligned}
                D_{-\alpha_1} &\quad\longrightarrow\quad S_{-\alpha_3} + 2\times S_{v_6} + S_{v_0} \cr
                D_{-\alpha_{2i}} &\quad\longrightarrow\quad S_{v_{3i}}  + S_{v_{3i+1}} \cr
                D_{-\alpha_{2k}} &\quad\longrightarrow\quad S_{s_{2k}} +S_{s_{2k+1}} \,.
        \end{aligned}
\end{equation}
The fibers for both codimension 2 loci are depicted in figures \ref{fig:Matterb3SOeven} and \ref{fig:DynkinSOevenb2}.

\subsubsection{Codimension 3 }

$SO(4k+4)$ has 3 codimension 3 enhancements of the discriminant. These are
\begin{equation}
        \begin{aligned}
                        b_4 &= b_3 = 0 \cr
                b_2 &= b_{1, 4} = 0 \,.
                        \end{aligned}
\end{equation}
The first locus is the only one that is minimal. The two distinct matter surfaces $S_{u_1^{(i)}}$ become degenerate at this locus, both are given by the equation $b_3=b_4= \zeta_{k+1}=y=0$, so that this codimension 3 locus allows the generation of the Yukawa coupling
\begin{equation}
b_3= b_4=0:\qquad V\otimes V\otimes {\bf 1} \,.
\end{equation}
Along the non-minimal locus $b_2=b_4=0$, the $b_2=0$ matter surface splits further as 
\begin{equation}
S_{-\alpha_{2k+2}} \quad\longrightarrow\quad \Sigma_{-\alpha_{2k+1}}  + \Sigma_{u^{(1)}_1} + \Sigma_{u^{(2)}_1}\,.
\end{equation}
Along the other locus no further splittings occur. 


\section{Resolution of $E$ Type  Singularities }

Finally, there are three exceptional cases left, which we will discuss in turn, including their higher codimension structure. 


\subsection{$E_6$}

For $E_6$ all singularities up and including codimension 3 are minimal. 
The Tate form for $E_6$ is
\be
y^2 -x^3-b_6 \zeta _0^5-b_2 \zeta _0^2 x^2-b_4 \zeta _0^3 x+b_1 \zeta _0 x y+b_3 \zeta _0^2 y =0 \,,
\ee
and  applying the following resolutions to the  geometry  resolves it fully
\be
\ba
&(x, y, \zeta_0; \zeta_1)          \cr
&(x, y, \zeta_1; \zeta_2)  \cr
&(y, \zeta_1;\delta_1) \cr
& (y, \zeta_2;\delta_2) \cr
&(\delta _1,\zeta _1;\kappa _1)\cr
& (\delta _1,\zeta _2;\epsilon _1) \cr
&(\delta _1,\delta _2;\epsilon_2)\cr
&(\delta _1,\epsilon _1;\epsilon _3)   \,.
\ea\ee
After taking the proper transform yields
\be\ba
&\delta _1 \delta _2  \epsilon _2 y^2+\delta _1 \zeta_1 \kappa _1 y \zeta _0 \left(b_3 \zeta _0+b_1 \delta _2 \zeta _2 x \epsilon _1 \epsilon _2 \epsilon _3\right)
\cr
&=\zeta _1 \zeta _2 \epsilon _1 \left(b_6 \delta _1^2 \zeta _1^2 \kappa _1^4 \epsilon _1 \epsilon _2 \epsilon _3^3 \zeta _0^5+\delta _1 \zeta _1 \kappa _1^2 x
   \epsilon _3 \zeta _0^2 \left(b_4 \zeta _0+b_2 \delta _2 \zeta _2 x \epsilon _1 \epsilon _2 \epsilon _3\right)+\delta _2 \zeta _2 x^3\right)
   \,.
\ea\ee
The irreducible exceptional divisors, which intersect in an affine $E_6$ Dynkin diagram, are 
\be
\begin{array}{l|l|l}
\hbox{Divisor} &\hbox{Section} & \hbox{Equation in } {Y} \cr\hline
D_{-\alpha_6}& \zeta_{0} & 0 =\delta _1 y^2-\zeta _1 x^3  \cr
D_{-\alpha_1}& \zeta_{2} & 0= b_3 \zeta _1 \kappa _1+\delta _2\cr
D_{-\alpha_0} &{\delta}_{2} & 0=b_3 y-\zeta _2 \epsilon _1 \left(b_4 x+b_6 \epsilon _1 \epsilon _2\right)\cr
D_{-\alpha_5} &{\kappa}_{1} & 0=\delta _1-\zeta _1 \zeta _2^2 \epsilon _1\cr
D_{-\alpha_3} &\epsilon_1&  0=b_3 \kappa _1+\delta _2 \epsilon _2\cr
D_{-\alpha_2} &\epsilon_{2} & 0=b_3 \delta _1 y-\epsilon _1 \left(b_4 \delta _1 \epsilon _3+\delta _2\right)\cr
D_{-\alpha_4} &\epsilon_{3} & 0=\delta _1 \left(b_3 \kappa _1+\epsilon _2\right)-\epsilon _1
\end{array}
\ee
The higher codimension structure was discussed in \cite{Kuntzler:2012bu} with the result that 
 along $b_3=0$ the following Cartan divisors split
 \begin{equation}
  \ba
  D_{-\alpha_3} &\quad \longrightarrow \quad  S_{(-1,1, 1, -1, 0, 0, 0)}+ S_{(1,0,-1,-1,1,0,0)}\cr
  D_{-\alpha_2} &\quad \longrightarrow \quad S_{(0, 0,-1,1,0,0,0)}+S_{(1,0,-1,-1,1,0,0)} \cr
  D_{-\alpha_0} &\quad \longrightarrow \quad S_{-\alpha_1} + S_{(-1, 1, 1, -1, 0, 0, 0)} + S_{(-1,1,0,0,0,0,0)} \,,\cr
  \ea
\end{equation}
which are all weights of  the ${\bf 27}$ representation.
At the codimension 3 locus $b_3=b_4=0$ the matter surface $S_{(-1,1,0,0,0,0,0)}$, given by $\delta_2= b_4x + b_6 \epsilon_1\epsilon_2=0$  becomes reducible 
\begin{equation}
S_{(-1,1,0,0,0,0,0)} \quad \longrightarrow \quad \Sigma_{(-1,1,1,-1,0,0,0)} + \Sigma_{(1,0,-1,-1,1,0,0)} \,.
\end{equation}
This is consistent with the ${\bf 27}^3$ Yukawa coupling in $E_6$. 


\subsection{$E_7$}

For $E_7$ there is one codimension 2 locus of symmetry enhancement, and one codimension 3 one, which is 
non-minimal, and has been discussed in \cite{Candelas:2000nc}. 
The Tate form for an $E_7$ singularity at $\zeta_0=0$ is 
\begin{equation}
	y^2 - x^3 + b_1xy\zeta_0 - b_2x^2\zeta_0^2 + b_3y\zeta_0^3 - b_4x\zeta_0^3 - b_6\zeta_0^5 \,.
\end{equation}
To resolve the space, we  apply the following blowups 
\begin{equation}
\ba
&		(x, y, \zeta_0; \zeta_1) \cr
&		(x, y, \zeta_1; \zeta_2) \cr
&		(y, \zeta_1; \delta_1) \cr
&		(y, \zeta_2; \delta_2) \cr
&		(\zeta_2, \delta_1; \epsilon_1) \cr
&		(\zeta_1, \delta_1;\epsilon_2)\cr
&		(\zeta_2, \delta_2; \epsilon_3) \cr
&		(\delta_1, \delta_2;\epsilon_4)\cr
&		(\delta_2, \epsilon_1;\epsilon_5)\cr
&		(\epsilon_1, \epsilon_4; \epsilon_6) \,,
\ea
\end{equation}
where the notation is that defined above (\ref{BUNotation}).  Note that the last three small resolutions ensure, in particular, that the exceptional divisors are all irreducible. 
The exceptional sections are 
\be\label{E7Cartans}
\{ \zeta_0\,,\  \epsilon_2  \,,\ \epsilon_4   \,,\ \delta_1   \,,\ \epsilon_6   \,,\ \epsilon_5   \,,\  \epsilon_3  \,,\ \delta_2   \} \,,
\ee
in terms of which the resolved geometry is 
\begin{equation}
\ba
 &\delta _1 \delta _2 \zeta _1 \zeta _2 y   \epsilon _1 \epsilon _2 \epsilon _3 \epsilon _4 \epsilon _5 \epsilon _6 \zeta _0
   \left(b_3 \delta _1 \zeta _1 \epsilon _1 \epsilon _2^2 \epsilon _4 \epsilon _5
   \epsilon _6^2 \zeta _0^2+b_1 x\right)
   +\delta _1 \delta _2 y^2 \epsilon _4=
   \cr
&\zeta _1 \zeta _2 \epsilon _1 \left(\delta _2 \zeta _2 x^2 \epsilon _3^2 \epsilon _5
   \left(b_2 \delta _1 \zeta _1 \epsilon _1 \epsilon _2^2 \epsilon _4 \epsilon _5
   \epsilon _6^2 \zeta _0^2+x\right)+\delta _1 \zeta _1 \epsilon _2^2 \zeta _0^3
   \left(b_6 \delta _1 \zeta _1 \epsilon _1 \epsilon _2^2 \epsilon _4 \epsilon _5
   \epsilon _6^2 \zeta _0^2+b_4 x\right)\right)
   \ea
\end{equation}
along with the projective relations given in  appendix \ref{app:E7}. 
The exceptional divisors in (\ref{E7Cartans}) intersect  in an affine $E_7$ Dynkin diagram
\begin{equation}
	\begin{array}{c|c|l|l}
		\mbox{Cartan Divisor} & \mbox{Section } & \mbox{Equation in $Y_4$} & \mbox{Cartan charges } \cr\hline
		D_{-\alpha_0} & \zeta _{0} & 0 = \delta _1 y^2-\zeta _1 x^3  & (-2, 1, 0, 0, 0, 0, 0, 0) \cr
		D_{-\alpha_1} & \epsilon_2 & 0 = \delta _1-\zeta _1 \epsilon _1 & (1, -2, 0, 1, 0, 0, 0, 0) \cr
		D_{-\alpha_2} & \epsilon_4  & 0 = -b_4 \delta _1-\delta _2 \epsilon _5&(0, 0, -2, 0, 1, 0, 0, 0) \cr
		D_{-\alpha_3} &  \delta_1 & 0 =\epsilon _1  & (0, 1, 0, -2, 1, 0, 0, 0 ) \cr
		D_{-\alpha_4} &  \epsilon_6 & 0 =
		-b_4 \delta _1 \epsilon _1+\delta _1 \epsilon _4-\epsilon _5\epsilon _1& (0, 0, 1, 1, -2, 1, 0, 0) \cr
		D_{-\alpha_5} &\epsilon_5 & 0 = \delta _2 \epsilon _4-b_4 \epsilon _1  & (0, 0, 0, 0, 1, -2, 1, 0) \cr
		D_{-\alpha_6} & \epsilon_3   & 0 = \delta _2-\zeta _2 \epsilon _1 \left(b_4 x+b_6 \epsilon _1
   \epsilon _5\right) & (0, 0, 0, 0, 0, 1, -2, 1) \cr
		D_{-\alpha_7} & \delta_2  & 0 = b_4 x+ b_6 \epsilon _4 \epsilon _5  & ( 0, 0, 0, 0, 0, 0, 1, -2) \cr
	\end{array}
\end{equation}
At the codimension 2 enhacement locus $b_4=0$, it is clear that $D_{-\alpha_5}$ and $D_{-\alpha_5}$ becomes reducible. 
The splitting is as follows
\begin{equation}\ba
D_{-\alpha_5} &\quad \longrightarrow \quad S_{(0,0,-1,0,1,-1,0,1)} + S_{(0,0,-1,0,0,1,0,-1)} \cr
D_{-\alpha_6} &\quad \longrightarrow \quad S_{(0,0,-1,0,1,-1,0,1)} + S_{(0,0,1,0,0,-1,1,-1)} \,,
\ea
\end{equation}
where the weights of the matter surfaces $S_{v}$ correspond to the ${\bf 56}$ representation. 
Finally, the codimension 3 locus at $b_4= b_6=0$ is non-minimal. Indeed, $\delta_2=0$ is completely contained in the fiber along this locus. 


\subsection{$E_8$}
\label{sec:E8}

The singular Tate form for $E_8$ is
\begin{equation}
	\begin{aligned}
		y^2 - x^3 + b_1xy\zeta_0 - b_2x^2\zeta_0^2 + b_3y\zeta_0^3 - b_4x\zeta_0^4 - b_6\zeta_0^5 = 0 \,.
	\end{aligned}
\end{equation}
Again, $\zeta_0=0$ denotes the locus above which the singular $E_8$ fiber is. 
The following combination of blowups and small resolutions resolve the space in all codimensions
\begin{equation}
\ba
&		(x, y, \zeta_0; \zeta_1) \cr
&		(x, y, \zeta_1; \zeta_2) \cr
&		(y, \zeta_1; \delta_1) \cr
&		(y, \zeta_2; \delta_2) \cr
&		(\zeta_2, \delta_1; \epsilon_1) \cr
&		(\zeta_1, \delta_1;\epsilon_2)\cr
&		(\zeta_2, \delta_2; \epsilon_3) \cr
&		(\delta_1, \delta_2;\epsilon_4)\cr
&		(\delta_2, \epsilon_1;\epsilon_5)\cr
&		(\epsilon_1, \epsilon_4; \epsilon_6) \cr
&		(\delta _2,\epsilon _4;\epsilon _7)\cr
&		(\delta _2,\epsilon _5;\epsilon _8)\cr
&		(\epsilon _4,\epsilon _5;\epsilon _9)\cr
&		(\epsilon _5,\epsilon _7; \epsilon _{10})\cr
\ea
\end{equation}
Resolved geometry is then
\begin{equation}
\ba
& \delta _1 \delta _2  \epsilon _4 \epsilon _7y\left( y+    
    \zeta _0 \zeta _1 \zeta _2  \epsilon _1 \epsilon _2 \epsilon _3
    \epsilon _5 \epsilon _6 \epsilon _8 \epsilon _9 \epsilon _{10} \left(b_3 \delta _1 \zeta _1 \zeta _0^2 \epsilon _1 \epsilon _2^2
   \epsilon _4 \epsilon _5 \epsilon _6^2 \epsilon _7 \epsilon _8 \epsilon _9^2 \epsilon _{10}^2+b_1 x\right)
  \right)
 \cr
& =\zeta _1 \zeta _2 \epsilon _1 \epsilon _5\times  \cr
&\left(\delta _1 \zeta _0^2 \zeta _1 \epsilon _1 \epsilon _2^2 \epsilon _4 \epsilon _6^2 \epsilon _9 \left(b_6 \delta
   _1 \zeta _1 \zeta _0^3 \epsilon _2^2+\delta _2 \zeta _2 x \epsilon _3^2 \epsilon _5 \epsilon _7 \epsilon _8^2 \epsilon _9 \epsilon _{10}^2 \left(b_4 \delta _1
   \zeta _1 \zeta _0^2 \epsilon _1 \epsilon _2^2 \epsilon _4 \epsilon _5 \epsilon _6^2 \epsilon _7 \epsilon _8 \epsilon _9^2 \epsilon _{10}^2+b_2
   x\right)\right)
  +\delta _2 \zeta _2 x^3 \epsilon _3^2 \epsilon _8\right)
      \ea
\end{equation}
The exceptional sections are
\begin{equation}
\left\{\epsilon _3,\epsilon _7,\epsilon _8,\epsilon _{10},\epsilon _9,\epsilon
   _6,\delta _1,\epsilon _2,\zeta _0\right\} \,,
\end{equation}
and the intersection of the associated Cartan divisors is as follows
\begin{equation}
	\begin{array}{c|c|l|ccccccccc}
		\mbox{Cartan Divisor} & \mbox{Section} & \mbox{Eqution in $Y_4$} & \cr\hline
D_{-\alpha_1} &  \epsilon _3  & 0=   \delta _2-b_6 \zeta _2 \epsilon _1^2 \epsilon _5&   (-2 & 0 & 1 & 0 & 0 & 0 & 0 & 0 & 0 )\cr
D_{-\alpha_2} &  \epsilon _7  & 0= b_6 \epsilon_4 \epsilon_9 + \delta_2 \epsilon_8  &  (0 & -2 & 0 & 1 & 0 & 0 & 0 & 0 & 0 )\cr
D_{-\alpha_3} &  \epsilon _8  & 0=   \delta_2 \epsilon_7-b_6 \epsilon_5&  (1 & 0 & -2 & 1 & 0 & 0 & 0 & 0 & 0 )\cr
D_{-\alpha_4} &  \epsilon _{10}  & 0= \epsilon_7-\epsilon_5 (b_6 \epsilon_9+\epsilon_8)  &  (0 & 1 & 1 & -2 & 1 & 0 & 0 & 0 & 0 )\cr
D_{-\alpha_5} &   \epsilon _9 & 0=   \epsilon_5- \epsilon_4 \epsilon_7&  (0 & 0 & 0 & 1 & -2 & 1 & 0 & 0 & 0 )\cr
D_{-\alpha_6} &  \epsilon _6  & 0=  \delta_1 \epsilon_4-\epsilon_1 \epsilon_5 &  (0 & 0 & 0 & 0 & 1 & -2 & 1 & 0 & 0) \cr
D_{-\alpha_7} &  \delta_1  & 0=  \epsilon_1 &  (0 & 0 & 0 & 0 & 0 & 1 & -2 & 1 & 0) \cr
D_{-\alpha_8} &   \epsilon _2 & 0=  \delta_1 - \zeta_1 \epsilon_1 &  (0 & 0 & 0 & 0 & 0 & 0 & 1 & -2 & 1) \cr
D_{-\alpha_0} & \zeta_0 & 0= \delta _1 y^2-\zeta _1 x^3&  (0 & 0 & 0 & 0 & 0 & 0 & 0 & 1 & -2)
	\end{array}
\end{equation}
The intersection graph is, as expected, the affine $E_8$ Dynkin diagram.

$E_8$ has only non-minimal loci of symmetry enhancement. In particular,
along the matter locus $b_6=0$, $\delta_2$ becomes a component of the fiber. 
As in other instances, we nevertheles can consider the splitting of  the Cartan divisors, which is summarized by
\begin{equation}\ba
D_{-\alpha_{2}} &\quad \rightarrow \quad  S_{(0,-1,1,0,0,0,0,0,0)} + S_{(0,-1,-1,1,0,0,0,0,0)} \cr
D_{-\alpha_{3}} &\quad \rightarrow \quad  S_{(1,1,-1,0,0,0,0,0,0)} + S_{(0,-1,-1,1,0,0,0,0,0)} \,,
\ea\end{equation}
where the matter surfaces all carry Cartan charges corresponding to weights of  the ${\bf 248}$ representation of $E_8$. 
Along the (non-minimal) Yukawa coupling locus $b_6=Q=0$ from table \ref{BigTable}, we do not find any further splitting.


\subsection*{Acknowledgements}

\noindent
We are grateful to Joe Marsano and Dave Morrison for very helpful discussions and comments on the draft. 
We also thank Niklas Beisert, Andreas Braun, Hirotaka Hayashi, Moritz K\"untzler, Natalia Saulina and Will Walters for discussions on related matters. 
This work is supported in part by STFC. 


\newpage
\appendix


\section{Tate forms}\label{app:Tate}
\begin{table}[ht]
{
\begin{center}
\begin{tabular}{|c|c|c|c|c|c|c|c|} \hline
 Type & Group & $ a_1$ &$a_2$ & $a_3$ &$ a_4 $& $ a_6$ &$\Delta$ \\ \hline $I_0 $ & --- &$ 0 $ &$ 0
$ &$ 0 $ &$ 0 $ &$ 0$ &$0$ \\ \hline $I_1 $ & --- &$0 $ &$ 0 $ &$ 1 $ &$ 1
$ &$ 1 $ &$1$ \\ \hline $I_2 $ &$SU(2)$ &$ 0 $ &$ 0 $ &$ 1 $ &$ 1 $ &$2$ &$
2 $ \\ \hline $I_{3}^{ns} $ & $Sp(1)$ &$0$ &$0$ &$2$ &$2$ &$3$ &$3$ \\ \hline
$I_{3}^{s}$ & $SU(3)$ &$0$ &$1$ &$1$ &$2$ &$3$ &$3$ \\ \hline
$I_{2n}^{ns}$ &$ Sp(n)$ &$0$ &$0$ &$n$ &$n$ &$2n$ &$2n$ \\ \hline
$I_{2n}^{s}$ &$SU(2n)$ &$0$ &$1$ &$n$ &$n$ &$2n$ &$2n$ \\ \hline
$I_{2n+1}^{ns}$ &$Sp(n)$ & $0$ &$0$ &$n+1$ &$n+1$ &$2n+1$ &$2n+1$
\\ \hline $I_{2n+1}^s$ &$SU(2n+1)$ &$0$ &$1$ &$n$ &$n+1$ &$2n+1$ &$2n+1$
\\ \hline $II$ & --- &$1$ &$1$ &$1$ &$1$ &$1$ &$2$ \\ \hline $III$ &$SU(2)$ &$1$
&$1$ &$1$ &$1$ &$2$ &$3$ \\ \hline $IV^{ns} $ &$Sp(1)$ &$1$ &$1$ &$1$
&$2$ &$2$ &$4$ \\ \hline $IV^{s}$ &$SU(3)$ &$1$ &$1$ &$1$ &$2$ &$3$ &$4$
\\ \hline $I_0^{*\,ns} $ &$G_2$ &$1$ &$1$ &$2$ &$2$ &$3$ &$6$ \\ \hline
$I_0^{*\,ss}$ &$SO(7)$ &$1$ &$1$ &$2$ &$2$ &$4$ &$6$ \\ \hline $I_0^{*\,s}
$ &$SO(8)^*$ &$1$ &$1$ &$2$ &$2$ &$4$ & $6$ \\ \hline $I_{1}^{*\,ns}$
&$SO(9)$ &$1$ &$1$ &$2$ &$3$ &$4$ &$7$ \\ \hline $I_{1}^{*\,s}$ &$SO(10) $
&$1$ &$1$ &$2$ &$3$ &$5$ &$7$ \\ \hline $I_{2}^{*\,ns}$ &$SO(11)$ &$1$ &$1$
&$3$ &$3$ &$5$ &$8$ \\ \hline $I_{2}^{*\,s}$ &$SO(12)^*$ &$1$ &$1$ &$3$
&$3$ &$5$&$8$\\ \hline 
$I_{2n-3}^{*\,ns}$ &$SO(4n+1)$ &$1$ &$1$ &$n$ &$n+1$
&$2n$ &$2n+3$ \\ \hline $I_{2n-3}^{*\,s}$ &$SO(4n+2)$ &$1$ &$1$ &$n$ &$n+1$
&$2n+1$ &$2n+3$ \\ \hline $I_{2n-2}^{*\,ns}$ &$SO(4n+3)$ &$1$ &$1$ &$n+1$
&$n+1$ &$2n+1$ &$2n+4$ \\ \hline $I_{2n-2}^{*\,s}$ &$SO(4n+4)^*$ &$1$ &$1$
&$n+1$ &$n+1$ &$2n+1$ 
&$2n+4$ \\ \hline $IV^{*\,ns}$ &$F_4 $ &$1$ &$2$ &$2$ &$3$ &$4$
&$8$\\ \hline $IV^{*\,s} $ &$E_6$ &$1$ &$2$ &$2$ &$3$ &$5$ & $8$\\ \hline
$III^{*} $ &$E_7$ &$1$ &$2$ &$3$ &$3$ &$5$ & $9$\\ \hline $II^{*} $
&$E_8\,$ &$1$ &$2$ &$3$ &$4$ &$5$ & $10$ \\ \hline
 non-min & --- &$ 1$ &$2$ &$3$ &$4$ &$6$ &$12$ \\ \hline
\end{tabular}
\label{TateTable}
\end{center}
}\caption{\footnotesize Vanishing order of the sections $a_n$ and  $\Delta$ in the Tate forms from \cite{Bershadsky:1996nh}.} 

\end{table}\newpage


\section{Example Resolutions}
\label{app:Examples}

\subsection{$SU(5)$ Reloaded}

The resolution of the $SU(5)$ Tate form has been well understood \cite{Esole:2011sm, MS}. Nevertheless, it may be useful for the reader, in order to get familiar with our general analysis and the structures introduced, to readdress this resolution. The low $k$ cases are generally somewhat pathological, which in particular will become clear in codimension 3 for $SU(5)$. 
The Tate form of $SU(5)$ is 
\begin{equation}
y^2 +b_1 x y+b_3 \zeta _0^2 y = x^3+b_2 \zeta _0 x^2+b_4 \zeta _0^3 x+b_6 \zeta _0^5 \,,
\end{equation}
where the $SU(5)$ singular fiber is located along $\zeta_0=0$. 
The singularity in codimension 1 can be resolved by two blowups
\begin{equation}
\ba
&(x, y, \zeta_0 ; \zeta_1) \cr
& (x, y, \zeta_{1};\zeta_2) \,,
\ea
\end{equation}
using the notation defined in (\ref{BUNotation}).
After the proper transform this results in
\begin{equation}
\label{ResCodim1}
y \left(y+ b_1 x+b_3 \zeta _1 \zeta _0^2\right) =
\zeta _1 \zeta _2 \left(b_2 x^2 \zeta _0+\zeta _1 \zeta _0^3 \left(b_6 \zeta _1 \zeta _0^2+b_4 x\right)+\zeta _2 x^3\right) \,.
\end{equation}
The space (\ref{ResCodim1}) is resolved in codimension 1, however not in higher codimension. There is a network of small resolutions whose structure will be discussed in \cite{LS}. For the current purposes of resolving the space, we pick one of these, for instance
\begin{equation}
\ba
&(y, \zeta_1; \delta_1) \cr
& (y, \zeta_2; \delta_2) \,.
\ea
\end{equation}
Note that this small resolution is different from the one studied in \cite{MS}. 
The resulting geometry, after proper transformation, is smooth and given by
\begin{equation}
y \left(\delta _1 \left(b_3 \zeta _1 \zeta _0^2+\delta _2 y\right)+b_1 x\right)=
\zeta _1 \zeta _2 \left(b_2 x^2 \zeta _0+\delta _1 \zeta _1 \zeta _0^3 \left(b_6 \delta _1 \zeta _1 \zeta _0^2+b_4 x\right)+\delta _2 \zeta _2 x^3\right) \,.
\end{equation}
The projectivity relations are, as detailed in appendix \ref{app:SUn} for the general case,
\begin{equation}\label{ProjRel}
\begin{array}{ll}
& [ \delta _2 \zeta _2 x , \delta _1 \delta _2^2 \zeta _2 y , \zeta _0] \\
&[ x , \delta _1 \delta _2 y , \delta _1 \zeta _1 ]\\
 &[\delta _2 y ,\zeta _1 ]\\
 &[y , \zeta _2 ]
\end{array}
\end{equation}
The classes are
\begin{equation}
\begin{array}{l|l}
\hbox{Section} & \hbox{Class} \cr\hline
x & \sigma+2c_1 - E_1 - E_2 \cr
y& \sigma+ 3 c_1 - E_1 - E_2-E_3 -E_4 \cr
\zeta_0 & S-E_1\cr
\zeta_1 & E_1- E_2 -E_3 -E_4\cr
\zeta_2 & E_2-E_4 \cr
\delta_1 & E_3 \cr
\delta_2 & E_4 
\end{array}
\end{equation}
Each of the exceptional sections gives rise to an irreducible exceptional divisor
\begin{equation}\label{ExDivs}
\begin{array}{l|l|l}
\hbox{Divisor} &\hbox{Section} & \hbox{Equation in $Y_4$} \cr\hline
D_{-\alpha_0}&  \zeta _0 & 0 = -\zeta _1 x^3+b_1 y_4 x+y_4^2 \delta _1 \\
D_{-\alpha_1}&  \zeta _1 & 0 = x b_1+\delta _1 \\
 D_{-\alpha_2}& \zeta _2 & 0 = x b_1+\delta _1 \left(\delta _2+b_3 \zeta _1\right) \\
 D_{-\alpha_4}& \delta _1 & 0 = b_1 y_4-\zeta _1 \zeta _2 \left(\delta _2 \zeta _2+b_2 \zeta _0\right) \\
 D_{-\alpha_3}& \delta _2 & 0 = y_4 \left(x b_1+b_3 \delta _1\right)-\left(b_2 x^2+\delta _1 \left(x b_4+b_6 \delta _1\right)\right) \zeta _2\end{array}
\end{equation}
The divisors intersect in the affine $A_4$ Dynkin diagram. We compute the intersections as explained in appendix \ref{app:SUn}. To spell this out, note that the relations  in the intersection ring, contained already in projective relations (\ref{ProjRel}) are 
\begin{equation}
\ba
 \sigma (\sigma+ 2 c_1 ) (\sigma + 3 c_1) &=0 \cr
 (\sigma+ 2 c_1 - E_1 ) (\sigma + 3 c_1- E_1) (S - E_1)&=0 \cr 
 (\sigma+ 2 c_1 - E_1- E_2 ) (\sigma + 3 c_1- E_1- E_2) (E_1- E_2)&=0 \cr 
(\sigma + 3 c_1- E_1- E_2 - E_3) (E_1- E_2- E_3)&=0 \cr 
(\sigma + 3 c_1- E_1- E_2 - E_3- E_4) ( E_2- E_4)&=0 \,.
\ea
\end{equation}
From these we derive  (\ref{BURCube}).

The codimension 2 matter loci correpond to ${\bf 10}$ localized at $b_1=0$ and $\bar{\bf 5}$ matter, which is at $P=0$. From the general analysis in section \ref{sec:SUodd2}, in particular (\ref{SUoddb1Split}), we infer that along  $b_1=0$ the splittings are\footnote{The divisors $S_{v_{k+2}}$ and $S_{v_{2k}}$ would have the same labels for $k=2$, however, they are distinct, and defined as in section \ref{sec:SUodd2}.} 
 \begin{equation}
\ba
D_{-\alpha_2}\quad  &\rightarrow \quad S_{v_2}+ S_{v_{k+2}} \cr
D_{-\alpha_4} \quad &\rightarrow \quad  S_{v_{2k}}+ S_{v_{k+2}}+D_{-\alpha_1}.[b_1] \,,
\ea
\end{equation}
where $S_{v_1}=D_{-\alpha_1}.[b_1] $ is irreducible, and the weights are as in the Cartan charge table (\ref{SUoddb1Weights}) 
\begin{equation}\ba
{v_2} &= (0, 0, -1, 0, 1) \cr
{v_{2k}} &= ( 0, 1, 0, 0, -1) \cr
{v_{k+2}} &= (0, 1, -1, 1, -1)  \,.
\ea
\end{equation}

Along $P=b_2 b_3^2+b_1 \left(b_1 b_6-b_3 b_4\right)=0$, which is the locus of the ${\bf \bar{5}}$ matter, the only divisor that  becomes reducible is $\delta_{k=2}=0$. 
In particular its equation along $P=0$ factorizes
\begin{equation}
\delta_2=P=0 :\qquad  \left(b_3 \delta _3+b_1 x_2\right) \left(b_2 b_3 \delta _3 \zeta _2-b_1 \zeta _2 \left(b_4 \delta _3+b_2 x_2\right)+b_1^2 y\right) =0 \,.
\end{equation}
As of section \ref{sec:SUodd2}, or by direct computation, the split is
\begin{equation}
D_{-\alpha_3} \quad \rightarrow \quad S_{u_{4}} + S_{u_{5}} \,,
\end{equation}
where 
\begin{equation}
\ba
u_4&=(0, 0, 1, -1, 0) \cr
u_5&=(0, 0, 0, -1, 1)\,.
\ea
\end{equation}
In codimension 3 there is the $b_1=b_3=0$ SO(12) point and the $b_1=b_2=0$ or $E_6$ point, which generate bottom and top Yukawas in a GUT model. 
Along $b_1=b_3=0$ our general analysis implies
\begin{equation}
S_{v_{5}} = D_{-\alpha_4} \cdot [b_1] \quad \longrightarrow\quad  \Sigma_{v_{2}} + 2 \times  \Sigma_{u_{4}} \,.
\end{equation}
Or put differently, the  restriction of the root $\Sigma_{v_5}= S_{v_{5}} \cdot [b_3] = D_{-\alpha_4} \cdot [b_1] \cdot [b_3] $ splits to $\Sigma_{v_{2}}+ 2 \Sigma_{u_{4}}$, which all become homologous at $b_1=b_3=0$. This is consistent with a bottom type Yukawa coupling.
At $b_1=b_2=0$ the following splitting occurs, see (\ref{SUYuk1}, \ref{SUYuk2})
\begin{equation}
\ba
S_{v_{2k}} &\quad \longrightarrow \quad \Sigma_{ v_{k+2}} + \Sigma_{w'} \cr
S_{-\alpha_{k+1}} &\quad \longrightarrow \quad \Sigma_{u_{2k}} + \Sigma_{u_{2k+1}}  \,.
\ea
\end{equation}
Note that $w'$ was a weight of $\Lambda^4{V}\cong V$ for $SU(5)$, and this is therefore consistent with the top Yukawa coupling. The second splitting is an adjoint splitting into fundamental and anti-fundamental. Note that the interesection graph is that of an $E_6$ (not affine) Dynkin diagram, with multiplicities $(1,2,3,2,1,1)$. Note that this fiber does not appear in \cite{Esole:2011sm}. Likewise, along $b_1=b_3=0$ the fibers are Kodaira $D$ type
 with the correct multiplicities, which appears to be different from the findings in \cite{Esole:2011sm}, however, is in agreement with all the multiplicities obtained in \cite{MS}. Note that later on in the context of $SU(5)\times U(1)$ models \cite{Krause:2011xj}, the multiplicities in \cite{MS} were also confirmed. 
 There is an interesting network of small resolutions which gives rise to different splittings of the matter curves and will be discussed elsewhere \cite{LS}.


\subsection{$SO(10)$ }

The resolution of $SO(10)$ was done in \cite{Tatar:2012tm}, however as for $SU(5)$ we present it in the framework of the current paper, so as to have one explicit example using the general analysis for ${SO(4k+2)}$ with $k=2$. Furthermore, the low $k$ cases exhibit some pathologies, which will be explained.
The singular Tate form is
\begin{equation}
        y^2 - x^3 + b_1xy\zeta_0 -b_2x^2\zeta_0 + b_3y\zeta_0^2 
        - b_4x\zeta_0^{3} - b_6\zeta_0^{5} = 0 \,.
\end{equation}
This is resolved as in (\ref{SOoddBlowups}) with
\begin{equation}
	\begin{array}{l}
		(x, y, \zeta_0; \zeta_1) \cr 
		(x, y, \zeta_1; \zeta_2) \cr
		(y, \zeta_1, \zeta_2; \gamma_1) \cr
		(y, \zeta_1; \delta_1) \cr
		(y, \zeta_2; \delta_2) \cr
		(\zeta_1, \delta_1; \delta_0) \cr
	\end{array}
\end{equation}
The fully resolved Tate form is
\begin{equation}
\ba
&\delta _1 \left( \delta _2 y^2 +\delta _0  \zeta _0 \zeta _1 y \left(b_3 \zeta _0+b_1 \gamma _1 \delta _2 \zeta _2
   x\right) \right)
\cr
&\qquad \qquad  =\zeta _1 \zeta _2 \left(\gamma _1 \left(\delta _0^2 \delta _1 \zeta _0^3 \zeta _1 \left(b_6 \gamma _1 \delta _0^2 \delta _1 \zeta _1 \zeta _0^2+b_4
   x\right)+\delta _2 \zeta _2 x^3\right)+b_2 \zeta _0 x^2\right)
\ea   \end{equation}

The irreducible exceptional divisors are labeled by the affine roots of $SO(10)$, and reproducing the corresponding Dynkin diagram from the intersections in section \ref{sec:SOodd}
\begin{equation}
	\begin{array}{l|c|l}
                \mbox{Divisor } & \mbox{Section } & \mbox{Equation in } Y_4  \cr\hline
                D_{-\alpha_0} & \zeta_0  & 0 = y^2\delta_1 - x^3\zeta_1  \cr
		D_{-\alpha_1} & \delta_1 & 0 = \gamma_1 + b_2\zeta_0 \cr
		D_{-\alpha_2} & \delta_0 & 0 = \delta_1 - \zeta_1(\gamma_1 + b_2\zeta_0) \cr
		D_{-\alpha_3} & \gamma_1 & 0 = y\delta_1(y\delta_2 + b_3\delta_0\zeta_1) - b_2\zeta_1\zeta_2 \cr
		D_{-\alpha_4} & \zeta_2 & 0 = \delta_2 + b_3\zeta_1 \cr
		D_{-\alpha_5} & \delta_2 & 0 = b_3y - \zeta_2(b_2x^2 + \gamma_1(b_4x + b_6\gamma_1)) \cr
	\end{array}
\end{equation}
Along the codimension two locus $b_2=0$ (which is minimal for this low rank example) the Cartan divisors either remain irreducible or split, 
\begin{equation}
	\begin{array}{ll}
		D_{-\alpha_0} &\longrightarrow\quad S_{-\alpha_0} \cr
		D_{-\alpha_1} &\longrightarrow\quad S_{-\alpha_1} \cr
		D_{-\alpha_2} &\longrightarrow\quad S_{-\alpha_2} \cr
		D_{-\alpha_3} &\longrightarrow\quad S_{-\alpha_1} + S_{v_1} + S_{v_2} \cr
		D_{-\alpha_4} &\longrightarrow\quad S_{-\alpha_4} \cr
		D_{-\alpha_5} &\longrightarrow\quad S_{-\alpha_5} \,,
	\end{array}
\end{equation}
where the the restricting of a Cartan divisor that remains irreducible is denoted by the matter surface $S_{-\alpha_i} \equiv D_{-\alpha_i} \cdot [b_2]$, and the additional new components that arise from the splitting are weights of the spin representations ${\bf 16}$ and ${\bf \overline{16}}$.

Similarly, along the matter locus $b_3=0$ the only nontrivial splitting is
\begin{equation}
    D_{-\alpha_{5}} \quad \longrightarrow S_{-\alpha_4} + 2 \times S_{u_{5}} \,,
\end{equation}
where $S_{-\alpha_4} = D_{-\alpha_4} \cdot[b_3]$ and $u_5$ is a weight of the vector representation ${\bf 10}$.

Finally, consider the codimension 3 loci. Along $b_2 = b_3 = 0$ we expect the Yukawa coupling ${\bf 16}^2 {\bf 10}$. Indeed the matter surfaces restrict as in (\ref{SOoddcodim3})
\begin{equation}
	\begin{array}{ll}
		S_{-\alpha_5} &\longrightarrow\quad \Sigma_{-\alpha_4} + \Sigma_{v_3} + \Sigma_{v_4} \cr
		S_{v_2} &\longrightarrow\quad \Sigma_{v_1} + \Sigma_{v_3} \cr
	\end{array}
\end{equation}
where this is precisely the coupling ${\bf 16}^2 {\bf 10}$. Along the other loci, in particular $b_3 = b_4^2 -4 b_2b_6 = 0$ no further splitting occurs, however as in the codimension 3 locus $P=Q=0$ case for $SU(2k+1)$ the coupling generated is realizing ${\bf 10}^2 {\bf 1}$.


\subsection{Fiber Structure}

The reader may have noted that the resolution procedure used 
in this appendix differs from the procedure in the main 
body of the text\footnote{We thank our JHEP referee for the suggestion to look into this questions and making a comparison with the work that appeared later \cite{Braun:2013cb}.}. The resolution we have used here is the in 
appendix \ref{app:SOReload}. As stated there, both procedures 
give the same matter representations and 
Yukawa couplings\footnote{These two resolutions are not identical, which is reflected in the fact that the specific weights of the matter representations which appear are different, however the resolutions should be related by flop transitions.}, however the resolution used here will give 
a clearer picture of the fiber structure.

As terminology, we will refer to the resolution used in this appendix (and in appendix \ref{app:SOReload}) as the 
``$\gamma$-resolution'', and the other one in the main text, as the ``$\epsilon$-resolution''.
The $\gamma$-resolution is similar to the one in \cite{Tatar:2012tm} and is equivalent to the one in \cite{Braun:2013cb}).

The fiber structures are given in table \ref{SO10FibStruct}.

\begin{equation}\label{SO10FibStruct}
	\begin{array}{|c|c|c|}
		\hline
		\text{Higher codimension locus} & 
		\gamma\text{-resolution fiber} & 
		\epsilon\text{-resolution fiber} \cr
		\hline
		\begin{picture}(50, 40)
			\put(10, 20){$b_3 = 0$}
		\end{picture} & 
		\begin{picture}(50, 40)
			\tiny
			\put(0, 10){\circle*{5}}
			\put(0, 30){\circle*{5}}
			\put(10, 20){\circle*{5}}
			\put(20,20){\circle*{5}}
			\put(30,20){\circle*{5}}
			\put(40,10){\circle*{5}}
			\put(40,30){\circle*{5}}
			\put(10,20){\line(1,0){20}}
			\put(10,20){\line(-1,1){10}}
			\put(10,20){\line(-1,-1){10}}
			\put(30,20){\line(1,1){10}}
			\put(30,20){\line(1,-1){10}}
		\end{picture} & 
		\begin{picture}(50, 40)
			\tiny
			\put(0, 10){\circle*{5}}
			\put(0, 30){\circle*{5}}
			\put(10, 20){\circle*{5}}
			\put(20,20){\circle*{5}}
			\put(30,20){\circle*{5}}
			\put(40,10){\circle*{5}}
			\put(40,30){\circle*{5}}
			\put(10,20){\line(1,0){20}}
			\put(10,20){\line(-1,1){10}}
			\put(10,20){\line(-1,-1){10}}
			\put(30,20){\line(1,1){10}}
			\put(30,20){\line(1,-1){10}}
		\end{picture} \cr\hline
		\begin{picture}(50,40)
			\put(10, 20){$b_2 = 0$}
		\end{picture} &
		\begin{picture}(50, 60)
			\tiny
			\put(0, 20){\circle*{5}}
			\put(20, 30){\circle*{5}}
			\put(10, 20){\circle*{5}}
			\put(20,20){\circle*{5}}
			\put(30,20){\circle*{5}}
			\put(40,20){\circle*{5}}
			\put(20,40){\circle*{5}}	
			\put(0,20){\line(1,0){40}}
			\put(20,20){\line(0,1){20}}
		\end{picture} & 
		\begin{picture}(50, 60)
			\tiny
			\put(0, 20){\circle*{5}}
			\put(20, 30){\circle*{5}}
			\put(10, 20){\circle*{5}}
			\put(20,20){\circle*{5}}
			\put(30,20){\circle*{5}}
			\put(40,20){\circle*{5}}
			\put(20,40){\circle*{5}}	
			\put(0,20){\line(1,0){40}}
			\put(20,20){\line(0,1){20}}
		\end{picture} \cr\hline 
		\begin{picture}(50, 40)
			\put(-10, 20){$b_3 = b_4^2 - 4b_2b_6 = 0$}
		\end{picture} & 
		\begin{picture}(50, 40)
			\tiny
			\put(0, 10){\circle*{5}}
			\put(0, 30){\circle*{5}}
			\put(10, 20){\circle*{5}}
			\put(20,20){\circle*{5}}
			\put(30,20){\circle*{5}}
			\put(40,20){\circle*{5}}
			\put(10,20){\line(1,0){30}}
			\put(10,20){\line(-1,1){10}}
			\put(10,20){\line(-1,-1){10}}
		\end{picture} & 
		\begin{picture}(50, 40)
			\tiny
			\put(0, 10){\circle*{5}}
			\put(0, 30){\circle*{5}}
			\put(10, 20){\circle*{5}}
			\put(20,20){\circle*{5}}
			\put(30,20){\circle*{5}}
			\put(40,20){\circle*{5}}
			\put(10,20){\line(1,0){30}}
			\put(10,20){\line(-1,1){10}}
			\put(10,20){\line(-1,-1){10}}
		\end{picture} \cr\hline
		\begin{picture}(50,40)
			\put(0,20){$b_2 = b_3 = 0$}
		\end{picture} &
		\begin{picture}(70, 40)
			\tiny
			\put(60,20){\circle*{5}}
			\put(50, 20){\circle*{5}}
			\put(0, 20){\circle*{5}}
			\put(10, 20){\circle*{5}}
			\put(20,20){\circle*{5}}
			\put(30,20){\circle*{5}}
			\put(40,20){\circle*{5}}
			\put(0,20){\line(1,0){60}}
		\end{picture} & 
		\begin{picture}(50,40)
			\put(0,20){(*)}
		\end{picture} \cr\hline
	\end{array}
\end{equation}

To depict the fiber along $b_2 = b_3 = 0$ for the $\delta$-resolution it is more usefl to consider the dual graphs of the fiber.
\begin{equation}
	\begin{array}{|c|c|}
		\hline
		\gamma\text{-resolution fiber} & 
		\epsilon\text{-resolution fiber} \cr
		\hline
		\begin{picture}(80, 40)
			\put(0,10){\line(1,1){20}}
			\put(10,30){\line(1,-1){20}}
			\put(20,10){\line(1,1){20}}
			\put(25,35){\line(1,-1){25}}
			\put(40,10){\line(1,1){20}}
			\put(50,30){\line(1,-1){20}}
			\put(60,10){\line(1,1){20}}
			\put(30, 30){\circle*{5}}
		\end{picture} & 
		\begin{picture}(80, 40)
			\put(0,10){\line(1,1){20}}
			\put(10,30){\line(1,-1){20}}
			\put(20,10){\line(1,1){20}}
			\put(30,30){\line(1,-1){20}}
			\put(40,10){\line(1,1){20}}
			\put(50,30){\line(1,-1){20}}
			\put(25,25){\line(1,0){20}}
			\put(35, 25){\circle*{5}}
		\end{picture}  
		\cr\hline
	\end{array}
\end{equation}
The black dot in the dual graph represents a singular ALE fibration, as in 
\cite{Braun:2013cb}, which occurs when $b_2 = b_3 = y = \gamma_1 = 1 + b_4\delta_1 = 0$ 
in the $\gamma$-resolution, and at $b_2 = b_3 = \delta_1 = \epsilon_1 = \delta_2 = 0$ in 
the $\epsilon$-resolution. Note that fourfold is however smooth.


\section{Details of the Resolution}


\subsection{$SU(n)$ and general computational Methods}\label{app:SUn}

We consider the $SU(n)$ singularities with resolutions as explained in  (\ref{SUoddBIG}, \ref{SUoddSMALL}, \ref{SUevenBIGSMALL}). After the resolution, the sections in the smooth Tate forms (\ref{SUCodim1Smooth}) and (\ref{SUCodim1SmoothEVEN}) are in the following classes:

\begin{equation}\label{SUDivClass}
\ba
SU(2k+1): &\qquad \qquad 
	\begin{array}{c|ll}
	\hbox{Section} & \hbox{Class} &\cr \hline
		x & \sigma + 2c_1 - E_1 - \cdots - E_k & \cr
		y & \sigma + 3c_1 - E_1 - \cdots - E_{2k} & \cr
		\zeta_0 & S_2 - E_1 & \cr
		\zeta_i & E_i - E_{i+1} - E_{k+i} & i = 1,\cdots,k-1 \cr
		\zeta_k & E_k - E_{2k} & \cr
		\delta_i & E_{k+i} & i = 1,\cdots,k \cr	
	\end{array}
\cr
\cr
SU(2k): &\qquad \qquad 
	\begin{array}{c|ll}
	\hbox{Section} & \hbox{Class} &\cr \hline
		x & \sigma + 2c_1 - E_1 - \cdots - E_k & \cr
		y & \sigma + 3c_1 - E_1 - \cdots - E_{2k-1} & \cr
		\zeta_0 & S_2 - E_1 & \cr
		\zeta_i & E_i - E_{i+1} - E_{k+i} & i = 1,\cdots,k-1 \cr
		\zeta_k & E_k & \cr
		\delta_i & E_{k+i} & i = 1,\cdots,k-1 \cr	
	\end{array}
	\ea
\end{equation}
We consider $n = 2k+1$ as well as  $n = 2k$.  
After all the blowups the projective relations are
\begin{equation}
\begin{aligned}
	SU(2k+1): & \qquad 
	\begin{array}{llc}
		(i) \quad & [xG_2^k(\zeta\delta), yG_2^k(\zeta\delta)B_1^k(\delta), \zeta_0] & \cr
		(ii) \quad & [xG_{i+2}^k(\zeta\delta), yG_{i+2}^k(\zeta\delta)B_1^k(\delta), \zeta_i\delta_i] & i = 1,\cdots,k-1 \cr
		(iii) \quad & [yB_{i+1}^k(\delta), \zeta_i] & i = 1,\cdots,k-1 \cr
		(iv) \quad & [y, \zeta_k] &  \,\cr
	\end{array} \cr \cr
	SU(2k): & \qquad 
	\begin{array}{llc}
		(i) \quad & [xG_2^k(\zeta)G_2^{k-1}(\delta), yG_2^k(\zeta)G_2^{k-1}(\delta)B_1^{k-1}(\delta), \zeta_0] & \cr
		(ii) \quad & [xG_{i+2}^k(\zeta)G_{i+2}^{k-1}(\delta), yG_{i+2}^k(\zeta)G_{i+2}^{k-1}(\delta)B_1^{k-1}(\delta), \zeta_i\delta_i] & i = 1,\cdots,k-1 \cr
		(iii) \quad & [yB_{i+1}^{k-1}(\delta), \zeta_i] & i = 1,\cdots,k-1 \cr
	\end{array} \cr \cr
\end{aligned}
\end{equation}
where, in the usual way, $[a, b, c]$ for instances means, that not all three entries can vanish at the same time. 
Recall the notation
\be
\ba
	A(z) &= \prod_{i=1}^{k-1}z_{i}^{i-1} \cr
	C(z) &= \prod_{i=1}^{k-1}z_{i}^{k-(i+1)} \cr
	 B(z) &= \prod_{i=1}^{k-1}z_i \,, 
\ea
\ee
where it is understood that e.g.  $A(\zeta \delta)= \prod_{i=1}^{k-1}(\zeta_{i}\delta_i)^{i-1}$.
Futhermore define
\begin{equation}
\ba
		G_i^n(\zeta)& = \prod_{j=i}^n \zeta_j^{j-i+1} \cr	
	B_i^n(\zeta) &= \prod_{j=i}^n\zeta_j \,,
\ea
\end{equation}
where it is understood that $B_i^n(x) = G_i^n(x) = 1$ when $i > n$.
For each singularity type we compute two types of intersections: 
\begin{itemize}

\item 
Intersections of the Cartan divisors 
\begin{equation}\label{CartanInts}
 D_{-\alpha_i} \cdot_{Y_4} C_{-\alpha_j} = 
 D_{-\alpha_i} \cdot_{X_5} D_{-\alpha_j} \cdot_{X_5} [Y_4] \cdot_{X_5} D_1 \cdot_{X_5}  D_2
\end{equation}
whith  $C_{-\alpha_j}$ being the component of the fiber associated to the root $-\alpha_j$, and $D_{-\alpha_i}$ is the Cartan divisor obtained by fibering $C_{-\alpha_i}$ over $S_2$. In the last equality we have chosen $D_1, D_2$ such that $D_1 \cdot_{B_3}  D_2 \cdot_{B_3} S_2 = 1$.
The only non-trivial intersection in the fiber is $\sigma^2$, so that from the above intersection we need to extract the $\sigma^2 S_2$ coefficient in the fivefold intersection. 

\item 
Intersection of Cartan divisors with Matter surfaces. The matter surfaces are of the type, e.g. for $b_1=0$ matter 
\begin{equation}
S_{v} = D_{-\alpha_i} \cdot [b_1]
\end{equation}
To determine the weights (``Cartan charges") $v$ we need to intersect with the original Cartans:

\end{itemize}

To compute these intersections, we need to take into account the relations that result from the blowups.
These allow us to replace certain combinations of intersections and rewrite everything in terms of independent intersections. 
The blowup relations are 
\begin{equation}
	\begin{array}{lll}
		(i)\  &0=  (\sigma + 2c_1 - E_1)(\sigma + 3c_1 - E_1)(S_2 - E_1)  & \cr
		(ii)\  & 0=(\sigma + 2c_1 - E_1 - \cdots - E_{i+1})(\sigma + 3c_1 - E_1 - \cdots - E_{i+1})(E_i - E_{i+1}) & i=1,\cdots,k-1 \cr
		(iii)\  & 0=(\sigma + 3c_1 - E_1 - \cdots - E_{k+i})(E_i - E_{i+1} - E_{k+i})  & i = 1,\cdots,k-1 \cr
		(iv) \ &0= (\sigma + 3c_1 - E_1 - \cdots - E_{2k})(E_k - E_{2k})  & \cr
	\end{array}
\end{equation}
Where $(iv)$ only holds in the $SU(2k+1)$ case. The relations $(i)$ and $(ii)$, which arise from the blowups, can be solved for $E_i^3$, whereas the ones from the small resolutions $(iii)$ and $(iv)$
are solved for $E_{k+i}^2$. 
We also have the blowup relation coming from the original $\mathbb{P}^2$ bundle, i.e. the projective relation $[x, y, w]$,
\begin{equation}
	\sigma^3 = -5c_1\sigma^2 - 6c_1^2\sigma  \,.
\end{equation}
Futhermore, the blowups were all in the patch $w=1$, which means that the divisor $w=0$ does not intersect the exceptional divisors, i.e.
\begin{equation}\label{sigmaE}
\sigma \cdot E_i = \sigma \cdot E_{k+i}= 0  \,.
\end{equation}

The computational strategy is now as follows: as explained above, the only non-trivial intersection in the fiber is $\sigma^2$. In order to obtain a non-trivial intersection for Cartan divisors, we thus need to determine the coefficient of $\sigma^2 S_2$ in (\ref{CartanInts}). So, given a fivefold intersection in $X_5$, we first use the the blowup relations to reduce the expressions to independent intersection products, as well as (\ref{sigmaE}), to extract the coefficient of $\sigma^2 S_2$, which intersects once with $D_1 D_2$. We will now first use the blowup relations to determine an independent set of triple intersections.

From the blowup relations we can determine the following triple intersections, which allow us to replace all $E_i^3$ and $E_{k+i}^2$ in all intersection computations
\begin{equation}\label{BURCube}
	\begin{array}{llr}
		E_i^3 & = \sigma^2S_2 + \cdots & i = 1,\cdots,k \cr
		E_jE_{k+i}^2 & = (\delta_{ij} - \delta_{ji+1})\sigma^2S_2 + \cdots & j = 1,\cdots,k; i = 1,\cdots,k-1 \cr
		E_{k+j}E_{k+i}^2 &= ((3 - \delta_{i1})\delta_{ij} - \delta_{ij+1})\sigma^2S_2 + \cdots &
					i,j = 1,\cdots,k-1 \cr
	\end{array}
\end{equation}
For $SU(2k+1)$ we have in addition the following relations
\begin{equation}
	\begin{array}{llr}
		E_jE_{2k}^2 & = \delta_{ik}\sigma^2S_2 + \cdots & j = 1,\cdots,k \cr
		E_{k+j}E_{2k}^2 & = (\delta_{jk} - \delta_{kj+1})\sigma^2S_2 + \cdots & j = 1,\cdots,k \,. \cr
	\end{array}
\end{equation}
Note that the $\cdots$ indicate terms not involving $\sigma^2 S_2$. 
For instance we can compute the intersections of the Cartan divisors which have  classes (\ref{SUDivClass}) with this information. Consider the product 
(\ref{CartanInts}) and expand it in terms of $E_i$, and apply the blowup relations (\ref{BURCube}), exctracting the coefficient of $\sigma^2 S_2$ and making use of (\ref{sigmaE}).


\subsection{$SO(4k+2)$}
\label{app:SOodd}

After performing the blowups (\ref{SOoddBlowups}) on the geometry 
(\ref{SOoddTate}) then the classes of the sections in the new, 
resolved geometry are
\begin{equation}
	\begin{array}{c|l}
	\hbox{Section} & \hbox{Class} \cr \hline
		x & \sigma + 2c_1 - E_1 - \cdots - E_k \cr
		y & \sigma + 3c_1 - E_1 - \cdots - E_{2k} \cr
		\zeta_0 & S_2 - E_1 \cr
		\zeta_1 & E_1 - E_2 - E_{k+1} - E_{2k+1} \cr
		\zeta_i & E_i - E_{i+1} - E_{k+i} - E_{2k+i} - E_{3k+i-1} \cr
		\zeta_k & E_k - E_{2k} - E_{4k-1} \cr
		\delta_i & E_{k+i} - E_{2k+i} - E_{3k+i} \cr
		\delta_k & E_{2k} \cr
		\kappa_i & E_{2k+i} \cr
		\epsilon_i & E_{3k+i} \cr
		Y_4 & 3\sigma + 6c_1 - 2E_1 - \cdots - 2E_k - E_{k+1} - \cdots - E_{3k-1} - E_{3k+1} - \cdots - E_{4k-1} \cr
	\end{array}
\end{equation}
The projective relations in the resolved geometry are
{\footnotesize
\begin{equation}
	\begin{array}{c|l}
		(i) & [xG_2^k(\delta\zeta)G_2^{k-1}(\kappa^2\epsilon)G_1^{k-1}(\epsilon)B_1^{k-1}(\epsilon),
		yB_1^{k-1}(\delta\kappa\epsilon^2)\delta_kG_2^k(\zeta\delta)G_2^{k-1}(\kappa^2\epsilon)G_1^{k-1}(\epsilon), \zeta_0] \cr
		(ii) & [xG_3^k(\zeta\delta)G_3^{k-1}(\kappa^2\epsilon)G_2^{k-1}(\epsilon)B_2^{k-1}(\epsilon),
		yB_1^{k-1}(\delta\kappa\epsilon)\delta_kxG_3^k(\zeta\delta)G_3^{k-1}(\kappa^2\epsilon)G_2^{k-1}(\epsilon)B_2^{k-1}(\epsilon),
		\zeta_1\delta_1\kappa_1^2\epsilon_1] \cr
		(iii) & [xG_{i+2}^k(\zeta\delta)G_{i+2}^{k-1}(\kappa^2\epsilon)G_{i+1}^{k-1}(\epsilon)B_{i+1}^{k-1}(\epsilon),
		yB_1^{k-1}(\delta\kappa\epsilon)\delta_kxG_{i+2}^k(\zeta\delta)G_{i+2}^{k-1}(\kappa^2\epsilon)
		G_{i+1}^{k-1}(\epsilon)B_{i+1}^{k-1}(\epsilon), \zeta_i\delta_i\kappa_i^2\epsilon_i\epsilon_{i-1}] \cr
		&\hfill  i= 2, \cdots, k-1 \cr
		(iv) & [yB_2^{k-1}(\delta\kappa\epsilon)\delta_k, \zeta_1\kappa_1] \cr
		(v) & [yB_{i+1}^{k-1}(\delta\kappa\epsilon)\delta_k, \zeta_i\kappa_i\epsilon_{i-1}] \hfill i=2, \cdots, k-1\cr
		(vi) & [y, \zeta_k\epsilon_{k-1}] \cr
		(vii) & [\zeta_1, \delta_1\epsilon_1] \cr
		(viii) & [\zeta_i\epsilon_{i-1}, \delta_i\epsilon_i] \hfill i=2, \cdots, k-1  \cr
		(ix) & [\delta_i, \zeta_{i+1}] \hfill i=1,\cdots, k-1 \cr
		(x) & [\delta_{k-1}, \zeta_k] \cr
	\end{array}
\end{equation}
}

The projective relations give rise to the blowup relations 
{\footnotesize
\begin{equation}
	\begin{array}{c|l}
		(i) & (\sigma + 2c_1 - E_1)(\sigma + 3c_1 - E_1 )(S_2 - E_1) = 0 \cr
		(ii) & (\sigma + 2c_1 - E_1 - E_2)(\sigma + 3c_1 - E_1 - E_2)(E_1 - E_2) = 0 \cr
		(iii) & (\sigma + 2c_1 - E_1 - \cdots - E_{i+1})(\sigma + 3c_1 - E_1 - \cdots - E_{i+1})(E_i - E_{i+1}) = 0 \cr
		(iv) & (\sigma + 3c_1 - E_1 - \cdots - E_{k+1})(E_1 - E_2 - E_{k+1}) = 0 \cr
		(v) &  (\sigma + 3c_1 - E_1 - \cdots - E_{k+i})(E_i - E_{i+1} - E_{k+i}) = 0 \cr
		(vi) &  (\sigma + 3c_1 - E_1 - \cdots - E_{2k})(E_k - E_{2k}) = 0 \cr
		(vii) & (E_1 - E_2 - E_{k+1} - E_{2k+1})(E_{k+1} - E_{2k+1}) = 0 \cr
		(viii) & (E_i - E_{i+1} - E_{k+i} - E_{2k+i})(E_{k+i} - E_{2k+i}) = 0 \cr
		(ix) & (E_{k+i} - E_{2k+i} - E_{3k+i})(E_{i+1} - E_{i+2} - E_{k+i+1} - E_{2k+i+1} - E_{3k+i}) = 0 \cr
		(x) & (E_{2k-1} - E_{3k-1} - E_{4k-1})(E_{1} - E_{2k} - E_{4k-1}) = 0 \cr
	\end{array}
\end{equation}
}
Again we solve these in terms of an independent set of triple intersections, noting only 
the $\sigma^2S_2$ component, where $\cdots$ represent terms that are not of the 
form $A\sigma^2S_2$.
\begin{equation}
	\begin{array}{rl}
		E_i^3 &= \sigma^2S_2 + \cdots \cr
		E_jE_{k+i}^2 &= (\delta_{i,j} - \delta_{j,i+1})\sigma^2S_2 + \cdots \cr
		E_{k+j}E_{k+i}^2 &= ((3 - \delta_{i,1})\delta_{i,j} - \delta_{i,j+1})\sigma^2S_2 + \cdots \cr
		E_jE_{2k}^2 &= \delta_{j,k}\sigma^2S_2 + \cdots \cr
		E_{k+j}E_{2k}^2 &= -\delta_{k,j+1}\sigma^2S_2 + \cdots \cr
		E_{2k}^3 &= \sigma^2S_2 + \cdots \cr
		E_jE_{2k+i}^2 &= (\delta_{i,j} - \delta_{j,i+1})\sigma^2S_2 + \cdots \cr
		E_{k+j}E_{2k+i}^2 &= ((1 - \delta_{i,1})\delta_{i,j} - \delta_{i,j+1})\sigma^2S_2 + \cdots \cr
		E_{2k}E_{2k+i}^2 &= 0 + \cdots \cr
		E_{2k+j}E_{2k+i}^2 &= 2\delta_{i,j}\sigma^2S_2 + \cdots \cr
		E_jE_{3k+i}^2 &= \cdots \cr
		E_{k+j}E_{3k+i}^2 &= (\delta_{i,j} - \delta_{j,i+1})\sigma^2S_2 + \cdots \cr
		E_{2k}E_{3k+i}^2 &= -\delta_{k,i+1}\sigma^2S_2 + \cdots \cr
		E_{2k+j}E_{3k+i}^2 &= -(\delta_{i,j} + \delta_{j,i+1})\sigma^2S_2 + \cdots \cr
		E_{3k+j}E_{3k+i}^2 &= 4\delta_{i,j}\sigma^2S_2 + \cdots \cr
		E_jE_{4k-1}^2 &= 0 + \cdots \cr
		E_{k+j}E_{4k-1}^2 &= \delta_{k,j+1}\sigma^2S_2 + \cdots \cr
		E_{2k}E_{4k-1}^2 &= -\sigma^2S_2 + \cdots \cr
		E_{2k+j}E_{4k-1}^2 &= -\delta_{k,j+1}\sigma^2S_2 + \cdots \cr
		E_{3k+j}E_{4k-1}^2 &= 3\delta_{k,j+1}\sigma^2S_2 + \cdots \cr
	\end{array}
\end{equation}

Using the blowup relations and the classes of the exceptional sections we find the non-zero intersections 
of the Cartans with each other and $Y_4$ to be
\begin{equation}
	\begin{array}{rl}
		[\zeta_0]._{X_5}[\zeta_0]._{X_5}[Y_4] &= -2\sigma^2S_2 + \cdots \cr
		[\zeta_0]._{X_5}[\kappa_i]._{X_5}[Y_4] &= \delta_{i,1}\sigma^2S_2 + \cdots \cr
		[\delta_1]._{X_5}[\delta_1]._{X_5}[Y_4] &= -2\sigma^2S_2 + \cdots \cr
		[\delta_1]._{X_5}[\kappa_i]._{X_5}[Y_4] &= \delta_{i,1}\sigma^2S_2 + \cdots \cr
		[\kappa_i]._{X_5}[\kappa_j]._{X_5}[Y_4] &= -2\delta_{i,j}\sigma^2S_2 + \cdots \cr
		[\kappa_i]._{X_5}[\epsilon_j]._{X_5}[Y_4] &= (\delta_{i,j} + \delta_{i,j+1}\sigma^2S_2 + \cdots \cr
		[\epsilon_i]._{X_5}[\epsilon_j]._{X_5}[Y_4] &= -2\delta_{i,j}\sigma^2S_2 + \cdots \cr
		[\epsilon_i]._{X_5}[\zeta_k]._{X_5}[Y_4] &= \delta_{k,i+1}\sigma^2S_2 + \cdots \cr
		[\epsilon_i]._{X_5}[\delta_k]._{X_5}[Y_4] &= \delta_{k,i+1}\sigma^2S_2 + \cdots \cr
		[\zeta_k]._{X_5}[\zeta_k]._{X_5}[Y_4] &= -2\sigma^2S_2 + \cdots \cr
		[\delta_k]._{X_5}[\delta_k]._{X_5}[Y_4] &= -2\sigma^2S_2 + \cdots \cr
	\end{array}
\end{equation}
With all other intersections of this form between any 2 exceptional sections 
and $Y_4$ having no factor like $A\sigma^2S_2$.
Note that the coefficients of $\sigma^2S_2$ here reproduce the negative of the Cartan matrix for  \
$SO(4k+2)$. 

When we go to the codimension 2 loci certain divisors become 
reducible and we need to calculate the Cartan charge 
vector associated to each irreducible component. To 
calculate these we need to know the following intersections (where $\cdots$ are terms not of the type $A \sigma^2 S_2$)
\begin{equation}
	\begin{array}{rll}
		[\kappa_i]._{X_5}[\delta_i]._{X_5}[\zeta_0] & = \cdots & \cr
		[\kappa_i]._{X_5}[\delta_i]._{X_5}[\delta_1] & = \delta_{i,2}\sigma^2S_2 + \cdots & \cr
		[\kappa_i]._{X_5}[\delta_i]._{X_5}[\kappa_j] & = -\delta_{i,j}\sigma^2S_2 + \cdots & \cr
		[\kappa_i]._{X_5}[\delta_i]._{X_5}[\epsilon_j] & = \delta_{i,j}\sigma^2S_2 + \cdots & \cr
		[\kappa_i]._{X_5}[\delta_i]._{X_5}[\zeta_{k}] & = \cdots & \cr
		[\kappa_i]._{X_5}[\delta_i]._{X_5}[\delta_k] & = \cdots & \cr
		[\epsilon_{k-1}]._{X_5}[\delta_{k-1}]._{X_5}[\zeta_0] & = \cdots & \cr
		[\epsilon_{k-1}]._{X_5}[\delta_{k-1}]._{X_5}[\delta_1] & = \cdots & \cr
		[\epsilon_{k-1}]._{X_5}[\delta_{k-1}]._{X_5}[\kappa_i] & = \delta_{k,i+1}\sigma^2S_2 + \cdots & \cr
		[\epsilon_{k-1}]._{X_5}[\delta_{k-1}]._{X_5}[\epsilon_i] & = -\delta_{k,i+1}\sigma^2S_2 + \cdots & \cr
		[\epsilon_{k-1}]._{X_5}[\delta_{k-1}]._{X_5}[\zeta_{k}] & = \cdots & \cr
		[\epsilon_{k-1}]._{X_5}[\delta_{k-1}]._{X_5}[\delta_k] & = \sigma^2S_2 + \cdots & \cr
		[\delta_k]._{X_5}[\delta_{k-1}]._{X_5}[\zeta_0] & = \cdots & \cr
		[\delta_k]._{X_5}[\delta_{k-1}]._{X_5}[\delta_1] & = \cdots & \cr
		[\delta_k]._{X_5}[\delta_{k-1}]._{X_5}[\kappa_i] & = \cdots & \cr
		[\delta_k]._{X_5}[\delta_{k-1}]._{X_5}[\epsilon_i] & = \delta_{k,i+1}\sigma^2S_2 + \cdots & \cr
		[\delta_k]._{X_5}[\delta_{k-1}]._{X_5}[\zeta_k] & = \cdots & \cr
		[\delta_k]._{X_5}[\delta_{k-1}]._{X_5}[\delta_k] & = -\sigma^2S_2 + \cdots & \cr
	\end{array}
\end{equation}


\subsection{$SO(4k+4)$}\label{app:SOeven}

The classes of the various sections in the resolved geometry are
\begin{equation}
	\begin{array}{c|ll}
		\text{Section} & \text{Class} \cr\hline
		x & \sigma + 2c_1 - E_1 - \cdots - E_{k+1} & \cr
		y & \sigma + 3c_1 - E_1 - \cdots - E_{2k+1} & \cr
		\zeta_0 & S_2 - E_1 & \cr
		\zeta_1 & E_1 - E_2 - E_{k+2} - E_{2k+2} & \cr
		\zeta_i & E_i - E_{i+1} - E_{k+1+i} - E_{2k+1+i} - E_{3k+i} & 
			i = 2,\cdots,k \cr
		\zeta_{k+1} & E_{k+1} & \cr
		\delta_i & E_{k+1+i} - E_{2k+1+i} - E_{3k+1+i} & i = 1,\cdots,k-1 \cr
		\delta_k & E_{2k+1} - E_{3k+1} & \cr
		\kappa_i & E_{2k+1+i} & i = 1,\cdots,k \cr
		\epsilon_i & E_{3k+1+i} & i = 1,\cdots,k-1 \cr
		Y_4 & 3\sigma + 6c_1 - 2E_1 - \cdots - 2E_{k+1} - E_{k+2} - \cdots 
			- E_{4k} & \cr
	\end{array}
\end{equation}
And the projective relations are
{\footnotesize
\begin{equation} 
	\begin{array}{c|ll}
		(i) & [xG_2^{k+1}(\zeta)G_2^k(\delta)G_2^{k-1}(\kappa^2\epsilon)G_1^{k-1}(\epsilon)B_1^{k-1}(\epsilon), 
			yB_1^{k-1}(\delta\kappa\epsilon)\delta_kG_2^{k+1}(\zeta)G_2^k(\delta)
			G_2^{k-1}(\kappa^2\epsilon)G_1^{k-1}(\epsilon)B_1^{k-1}(\epsilon), \zeta_0] & \cr
		(ii) & [xG_3^{k+1}(\zeta)G_3^k(\delta)G_3^{k-1}(\kappa^2\epsilon)G_2^{k-1}(\epsilon)B_2^{k-1}(\epsilon),
			yB_1^{k-1}(\delta\kappa\epsilon)\delta_kG_3^{k+1}(\zeta)G_3^k(\delta)
			G_3^{k-1}(\kappa^2\epsilon)G_2^{k-1}(\epsilon)B_2^{k-1}(\epsilon), \zeta_1\delta_1\kappa_1^2\epsilon_1] & \cr
		(iii) & [xG_{i+2}^{k+1}(\zeta)G_{i+2}^k(\delta)G_{i+2}^{k-1}(\kappa^2\epsilon)G_{i+1}^{k-1}(\epsilon)B_{i+1}^{k-1}(\epsilon), 
			yB_1^{k-1}(\delta\kappa\epsilon)\delta_kG_{i+2}^{k+1}(\zeta)G_{i+2}^k(\delta)
			G_{i+2}^{k-1}(\kappa^2\epsilon)G_{i+1}^{k-1}(\epsilon)B_{i+1}^{k-1}(\epsilon), \cr
			& \hfill \zeta_i\delta_i\kappa_i^2\epsilon_{i-1}\epsilon_i] \,,\quad i=2, \cdots, k-1\cr
		(iv) & [x, yB_1^{k-1}(\delta\kappa\epsilon)\delta_k, \zeta_k\delta_k\kappa_k^2\epsilon_{k-1}] & \cr
		(v) & [yB_2^{k-1}(\delta\kappa\epsilon)\delta_k, \zeta_1\kappa_1] & \cr
		(vi) & [yB_{i+1}^{k-1}(\delta\kappa\epsilon)\delta_k, \zeta_i\kappa_i\epsilon_{i-1}] \hfill  i=2, \cdots, k-1 \cr
		(vii) & [\zeta_1, \delta_1\epsilon_1] & \cr
		(viii) & [\zeta_i\epsilon_{i-1}, \delta_i\epsilon_i] \hfill  i=2, \cdots, k-1& \cr
		(ix) & [\zeta_k\epsilon_{k-1}, \delta_k] & \cr
		(x) & [\zeta_{i+1}, \delta_i] \hfill i=1, \cdots, k-2 & \cr
		(xi) & [\zeta_k, \delta_{k-1}] & \cr
	\end{array}
\end{equation}
}
The projective relations imply blowup relations
\begin{equation}
	\begin{array}{ll}
		(\sigma + 2c_1 - E_1)(\sigma + 3c_1 - E_1)(S_2 - E_1) = 0 & \cr
		(\sigma + 2c_1 - E_1 - \cdots - E_{i+1})(\sigma + 3c_1 - E_1 - \cdots - E_{i+1})(E_i - E_{i+1}) = 0 & i = 1,\cdots,k \cr
		(\sigma + 3c_1 - E_1 - \cdots - E_{k+1+i})(E_i - E_{i+1} - E_{k+1+i}) = 0 & i = 1,\cdots,k \cr
		(E_i - E_{i+1} - E_{k+1+i} - E_{2k+1+i})(E_{k+1+i} - E_{2k+1+i}) = 0 & i = 1,\cdots,k \cr
		(E_{i+1} - E_{i+2} - E_{k+2+i} - E_{2k+2+i} - E_{3k+1+i})(E_{k+1+i} - E_{2k+1+i} - E_{3k+1+i}) = 0 & i = 1,\cdots,k-1 \cr
	\end{array}
\end{equation}
We eliminate the non-independent  3-way intersections. 
keeping terms of the type  $\sigma^2 S_2$ only 
\begin{equation}
	\begin{array}{rll}
		E_i^3 &= \sigma^2S_2 + \cdots & \cr
		E_jE_{k+1+i}^2 &= (\delta_{i,j} - \delta_{j,i+1})\sigma^2S_2 + \cdots & \cr
		E_{k+1+j}E_{k+1+i}^2 &= ((3 - \delta_{i,1})\delta_{i,j} - \delta_{i,j+1})\sigma^2S_2 + \cdots & \cr
		E_jE_{2k+1+i}^2 &= (\delta_{i,j} - \delta_{j,i+1})\sigma^2S_2 + \cdots & \cr
		E_{k+1+j}E_{2k+1+i}^2 &= ((1 - \delta_{i,1})\delta_{i,j} - \delta_{i,j+1})\sigma^2S_2 + \cdots & \cr
		E_{2k+1+j}E_{2k+1+i}^2 &= 2\delta_{i,j}\sigma^2S_2 + \cdots \cr
		E_jE_{3k+1+i}^2 &= \cdots & \cr
		E_{k+1+j}E_{3k+1+i}^2 &= (\delta_{i,j} - \delta_{j,i+1})\sigma^2S_2 + \cdots & \cr
		E_{2k+1+j}E_{3k+1+i}^2 &= -(\delta_{i,j} + \delta_{j,i+1})\sigma^2S_2 + \cdots & \cr
		E_{3k+1+j}E_{3k+1+i}^2 &= 4\delta_{i,j}\sigma^2S_2 + \cdots & \cr
	\end{array}
\end{equation}

The non-zero (in the coefficient of $\sigma^2S_2$ sense) intersections 
involving two of the exceptional sections and $Y_4$ are
\begin{equation}
	\begin{array}{rll}
		[\zeta_0]._{X_5}[\zeta_0]._{X_5}[Y_4] &= -2\sigma^2S_2 + \cdots & \cr
		[\zeta_0]._{X_5}[\kappa_i]._{X_5}[Y_4] &= \delta_{i,1}\sigma^2S_2 + \cdots & \cr
		[\delta_1]._{X_5}[\delta_1]._{X_5}[Y_4] &= -2\sigma^2S_2 + \cdots & \cr
		[\delta_1]._{X_5}[\kappa_i]._{X_5}[Y_4] &= \delta_{i,1}\sigma^2S_2 + \cdots & \cr
		[\kappa_i]._{X_5}[\kappa_j]._{X_5}[Y_4] &= -2\delta_{i,j}\sigma^2S_2 + \cdots & \cr
		[\kappa_i]._{X_5}[\epsilon_j]._{X_5}[Y_4] &= (\delta_{i,j} - \delta_{i,j+1})\sigma^2S_2 + \cdots & \cr
		[\kappa_i]._{X_5}[\zeta_{k+1}]._{X_5}[Y_4] &= \delta_{i,k}\sigma^2S_2 + \cdots & \cr
		[\kappa_i]._{X_5}[\delta_{k}]._{X_5}[Y_4] &= \delta_{i,k}\sigma^2S_2 + \cdots & \cr
		[\epsilon_i]._{X_5}[\epsilon_j]._{X_5}[Y_4] &= -2\delta_{i,j}\sigma^2S_2 + \cdots & \cr
		[\zeta_{k+1}]._{X_5}[\zeta_{k+1}]._{X_5}[Y_4] &= -2\sigma^2S_2 + \cdots & \cr
		[\delta_k]._{X_5}[\delta_k]._{X_5}[Y_4] &= -2\sigma^2S_2 + \cdots & \cr
	\end{array}
\end{equation}

Additionally for higher codimension it will be necessary to know 
the following intersections to work out the Cartan charge vectors 
of the irreducible matter surfaces
\begin{equation}
	\begin{array}{rll}
		[\kappa_i]._{X_5}[\delta_i]._{X_5}[\zeta_0] & = \cdots & \cr
		[\kappa_i]._{X_5}[\delta_i]._{X_5}[\delta_1] & = \delta_{i,2}\sigma^2S_2 + \cdots & \cr
		[\kappa_i]._{X_5}[\delta_i]._{X_5}[\kappa_j] & = -\delta_{i,j}\sigma^2S_2 + \cdots & \cr
		[\kappa_i]._{X_5}[\delta_i]._{X_5}[\epsilon_j] & = \delta_{i,j}\sigma^2S_2 + \cdots & \cr
		[\kappa_i]._{X_5}[\delta_i]._{X_5}[\zeta_{k+1}] & = \cdots & \cr
		[\kappa_i]._{X_5}[\delta_i]._{X_5}[\delta_k] & = \cdots & \cr
		[\kappa_k]._{X_5}[\delta_{k-1}]._{X_5}[\zeta_0] & = \cdots & \cr
		[\kappa_k]._{X_5}[\delta_{k-1}]._{X_5}[\delta_1] & = \cdots & \cr
		[\kappa_k]._{X_5}[\delta_{k-1}]._{X_5}[\kappa_i] & = -\delta_{i,k}\sigma^2S_2 + \cdots & \cr
		[\kappa_k]._{X_5}[\delta_{k-1}]._{X_5}[\epsilon_i] & = \delta_{k,i+1}\sigma^2S_2 + \cdots & \cr
		[\kappa_k]._{X_5}[\delta_{k-1}]._{X_5}[\zeta_{k+1}] & = \cdots & \cr
		[\kappa_k]._{X_5}[\delta_{k-1}]._{X_5}[\delta_k] & = \sigma^2S_2 + \cdots & \cr
	\end{array}
\end{equation}



\subsection{$E_7$}
\label{app:E7}

In this section we provide a few details on the resolution of the $E_7$ singularity.
The coordinates have to satisfy the following projective relations
\begin{equation}
        \begin{array}{l}
                [x\delta_2\varepsilon_1\varepsilon_3^3\varepsilon_4\varepsilon_5^2\varepsilon_6^2\zeta_2, 
                y\delta_1\delta_2^2\varepsilon_1^2\varepsilon_2\varepsilon_3^4\varepsilon_4^3\varepsilon_5^4\varepsilon_6^5\zeta_2,
                \zeta_0] \cr
                [x, y\delta_1\delta_1\varepsilon_1\varepsilon_2\varepsilon_3\varepsilon_4^2\varepsilon_5^2\varepsilon_6^3, 
                \delta_1\varepsilon_1\varepsilon_2^2\varepsilon_4\varepsilon_5\varepsilon_6^2\zeta_1] \cr
                [y\delta_1\varepsilon_3\varepsilon_4\varepsilon_5\varepsilon_6, \varepsilon_2\zeta_1] \cr
                [y, \varepsilon_1\varepsilon_3\varepsilon_5\varepsilon_6\zeta_2] \cr
                [\varepsilon_3\zeta_2, \delta_1\varepsilon_2\varepsilon_4\varepsilon_6] \cr
                [\zeta_1, \delta_1\varepsilon_4\varepsilon_6] \cr
                [\zeta_2, \delta_2\varepsilon_4\varepsilon_5\varepsilon_6] \cr
                [\delta_1, \delta_2\varepsilon_5] \cr
                [\delta_2, \varepsilon_1\varepsilon_6] \cr
                [\varepsilon_1, \varepsilon_4] \cr
        \end{array}
\end{equation}
The various sections have the following classes
\begin{equation}
        \begin{array}{l|l}
                \hbox{Section} &\hbox{Class}\cr\hline
                x & \sigma + 2c_1 - E_1 - E_2 \cr
                y & \sigma + 3c_1 - E_1 - E_2 - E_3 - E_4 \cr
                \zeta_0 & S_2 - E_1 \cr
                \zeta_1 & E_1 - E_2 - E_3 - E_6 \cr
                \zeta_2 & E_2 - E_4 - E_5 - E_7 \cr
                \delta_1 & E_3 - E_5 - E_6 - E_8 \cr
                \delta_2 & E_4 - E_7 - E_8 - E_9 \cr
                \varepsilon_1 & E_5 - E_9 - E_{10} \cr
                \varepsilon_2 & E_6 \cr
                \varepsilon_3 & E_7 \cr
                \varepsilon_4 & E_8 - E_{10} \cr
                \varepsilon_5 & E_9 \cr
                \varepsilon_6 & E_{10} \cr
                Y_4 & 3\sigma + 6c_1 - 2E_1 - 2E_2 - E_3 - E_4 - E_5 - E_6 - E_7 - E_8 - E_9 - E_{10} \cr
        \end{array}
\end{equation}


\subsection{$E_8$}
\label{app:E8}

In section \ref{sec:E8} we resolved the $E_8$ singularity. 
The projectivity relations for the coordinates that were used to resolve the geometry are
\begin{equation}
        \begin{array}{l}
                [x\delta_2\varepsilon_1\varepsilon_{10}^4\varepsilon_3^2\varepsilon_4\varepsilon_5^2
                \varepsilon_6^2\varepsilon_7^2\varepsilon_8^3\varepsilon_9^3\zeta_2,
                y\delta_1\delta_2^2\varepsilon_1^2\varepsilon_{10}^9\varepsilon_2\varepsilon_3^3\varepsilon_4^3
                \varepsilon_5^4\varepsilon_6^5\varepsilon_7^5\varepsilon_8^6\varepsilon_9^7\zeta_2, \zeta_0] \cr
                [x, y\delta_1\delta_2\varepsilon_1\varepsilon_{10}^5\varepsilon_2\varepsilon_3\varepsilon_4^2
                \varepsilon_5^4\varepsilon_6^3\varepsilon_7^3\varepsilon_8^3\varepsilon_9^4, 
                \delta_1\varepsilon_1\varepsilon_{10}^2\varepsilon_2^2\varepsilon_4\varepsilon_5\varepsilon_6^2
                \varepsilon_7\varepsilon_8\varepsilon_9^2\zeta_1] \cr
                [y, \delta_2\varepsilon_{10}^3\varepsilon_3\varepsilon_4\varepsilon_5\varepsilon_6
                \varepsilon_7^2\varepsilon_8^2\varepsilon_9^2, \varepsilon_2\zeta_1] \cr
                [y, \varepsilon_1\varepsilon_{10}\varepsilon_3\varepsilon_5\varepsilon_6\varepsilon_8\varepsilon_9\zeta_2] \cr
                [\varepsilon_3\zeta_2, \delta_1\varepsilon_{10}\varepsilon_2\varepsilon_4\varepsilon_6\varepsilon_7\varepsilon_9] \cr
                [\zeta_1, \delta_1\varepsilon_{10}\varepsilon_4\varepsilon_6\varepsilon_7\varepsilon_9] \cr
                [\zeta_2, \delta_2\varepsilon_{10}^3\varepsilon_4\varepsilon_5\varepsilon_6\varepsilon_7^2
                \varepsilon_8^2\varepsilon_9^2] \cr
                [\delta_1, \delta_2\varepsilon_{10}^2\varepsilon_5\varepsilon_7\varepsilon_8^2\varepsilon_9] \cr
                [\delta_2\varepsilon_{10}\varepsilon_7\varepsilon_8, \varepsilon_1\varepsilon_6] \cr
                [\varepsilon_1, \varepsilon_4\varepsilon_7\varepsilon_9\varepsilon_{10}] \cr
                [\delta_2\varepsilon_8, \varepsilon_4\varepsilon_9] \cr
                [\delta_2, \varepsilon_5\varepsilon_9\varepsilon_{10}] \cr
                [\varepsilon_4, \varepsilon_5\varepsilon_{10}] \cr
                [\varepsilon_5, \varepsilon_7] \cr
        \end{array}
\end{equation}
The various sections have the following classes
\begin{equation}
        \begin{array}{l|l}
                \hbox{Section} &\hbox{Class}\cr\hline
                x & \sigma + 2c_1 - E_1 - E_2 \cr
                y & \sigma + 3c_1 - E_1 - E_2 - E_3 - E_4 \cr
                \zeta_0 & S_2 - E_1 \cr
                \zeta_1 & E_1 - E_2 - E_3 - E_6 \cr
                \zeta_2 & E_2 - E_4 - E_5 - E_7 \cr
                \delta_1 & E_3 - E_5 - E_6 - E_8 \cr
                \delta_2 & E_4 - E_7 - E_8 - E_9 - E_{10} - E_{11} \cr
                \varepsilon_1 & E_5 - E_9 - E_{11} \cr
                \varepsilon_2 & E_6 \cr
                \varepsilon_3 & E_7 \cr
                \varepsilon_4 & E_8 - E_{10} - E_{11} - E_{13} \cr
                \varepsilon_5 & E_9 - E_{12} - E_{13} - E_{14} \cr
                \varepsilon_6 & E_{10} \cr
                \varepsilon_7 & E_{11} - E_{14} \cr
                \varepsilon_8 & E_{12} \cr
                \varepsilon_9 & E_{13} \cr
                \varepsilon_{10} & E_{14} \cr
        \end{array}
\end{equation}
Finally, the class of the fourfold after the resolutions is given by
\begin{equation}
        [Y_4] = 3\sigma + 6c_1 - 2E_1 - 2E_2 - E_3 - E_4 - E_5 - E_6 - E_7 - E_8 - E_9 - E_{10} - E_{11} - E_{12} - E_{13} - E_{14} \,. 
\end{equation}

\section{Alternative Resolution of $SO(4k+2)$ and Non-Minimality}
\label{app:SOReload}

In this appendix we discuss an alternative resolution  of $SO(4k+2)$, which we believe is related by flops to the one in the main text. The structure of the minimal singular loci is unchanged, however, something interesting happens along the non-minimal matter locus $b_2=0$, which we will discuss here. 

Again, our starting point is the Tate form for  $SO(4k+2)$ 
\begin{equation}\label{SOoddTateALT}
        y^2 - x^3 + b_1xy\zeta_0 -b_2x^2\zeta_0 + b_3y\zeta_0^k 
        - b_4x\zeta_0^{k+1} - b_6\zeta_0^{2k+1} = 0 \,.
\end{equation}
We consider here the resolution, using the notation of 
(\ref{BUNotation}), where we first apply as many blowups as possible, and then additional small resolutions to make the exceptional divisors irreducible
\begin{equation}\label{SOoddBlowupsALT}
        \begin{array}{rl}
                (x, y, \zeta_i; \zeta_{i+1}) & \quad i = 0,\cdots,k-1 \cr
                (y, \zeta_i, \zeta_{i+1}; \gamma_i) & \quad i = 1,\cdots,k-1 \cr
                (y, \zeta_i; \delta_i) & \quad i = 1,\cdots,k \cr
                (\delta_1, \zeta_1; \delta_0)& 
        \end{array}
\end{equation}
The last blowup is not needed to resolve the space, merely to make
each of our exceptional sections irreducible. The resolved geometry is 
\begin{equation}
        \begin{aligned}
                &y^2B(\delta)\delta_k 
                - x^3B(\zeta\delta)A(\zeta\delta\gamma^2)
                        \zeta_k^k\delta_k^{k-1}\delta_0 
                + b_1xy\zeta_0B(\zeta\delta\gamma)\zeta_k\delta_k\delta_0 
                \cr &- b_2x^2\zeta_0B(\zeta)\zeta_k 
                + b_3y\zeta_0^kB(\zeta\delta\gamma)
                        C(\zeta\delta\gamma^2)\delta_0^{2k-3} 
                \cr &- b_4x\zeta_0^{k+1}B(\zeta^2\delta\gamma)
                        C(\zeta\delta\gamma^2)\zeta_k\delta_0^{2k-2} 
                - b_6\zeta_0^{2k+1}B(\zeta^3\delta^2\gamma^2)
                        C(\zeta^2\delta^2\gamma^4)\zeta_k\delta_0^{4k-4} \,,
        \end{aligned}
\end{equation}
along with the projective relations which are readily obtained from the set of resolutions (\ref{SOoddBlowupsALT}). 

In codimension 1, the Cartan divisors are now given in terms of 
\begin{equation}
        \begin{array}{l|c|l}\label{SOoddDivsALT} 
                \mbox{Divisor } & \mbox{Section } & \mbox{Equation in } Y_4  \cr\hline
                D_{-\alpha_0} & \zeta_0 & 0 = y^2\delta_1 - x^3\zeta_1  \cr
                D_{-\alpha_1} & \delta_1  & 0 = \gamma_1 + b_2\zeta_0  \cr
                D_{-\alpha_2} & \delta_0  & 0 = \delta_1 - (\gamma_1 + b_2\zeta_0)\zeta_1 \cr
                D_{-\alpha_{2i+1}} \, \quad i = 1,\cdots,k-2 & \gamma_i  & 0 = y^2\delta_i\delta_{i+1} - b_2\zeta_i\zeta_{i+1}  \cr
                D_{-\alpha_{2i}} \qquad i = 2,\cdots,k-1 & \zeta_i & 0 = \delta_i  \cr 
                D_{-\alpha_{2k-1}} & \gamma_{k-1} & 0 = y^2\delta_{k-1}\delta_k + b_3y\zeta_{k-1}\delta_{k-1} - b_2\zeta_{k-1}\zeta_k  \cr
                D_{-\alpha_{2k}} & \zeta_k & 0 = \delta_k + b_3\zeta_{k-1} \cr
                D_{-\alpha_{2k+1}} & \delta_k & 0 = b_3y - b_2x^2\zeta_k - b_4x\gamma_{k-1}\zeta_k - b_6\gamma_{k-1}^2\zeta_k 
        \end{array}
\end{equation}
Let us consider the codimension 2 loci. 
Along $b_3=0$ the splitting into matter surfaces is completely analogous to the one in the main text, and the same Cartan charge for $u_{2k+1}^{(i)}$ appears.

\begin{figure}
\begin{center}
\begin{picture}(280, 120)
\color{green}

	
	\put(0,40){\color{blue}\line(-2,-3){20}}
	\put(0,40){\color{blue}\line(-2,3){20}}
	\put(-20,70){\color{blue}\line(2,-1){60}}
	\put(-23,70){\line(2,-1){60}}

	\put(40,10){\color{blue}\line(0,1){30}}
	\put(39,10){\line(0,1){30}}
	
	\put(40,10){\color{blue}\line(4,3){80}}

	\put(120,10){\color{blue}\line(0,1){60}}
	\put(119,10){\line(0,1){60}}	
	
	\put(120,10){\color{blue}\line(4,3){20}}	
	\put(160,40){\color{blue}\line(4,3){40}}	
	\put(160,40){\color{blue}\line(-4,-3){10}}

	\put(200,10){\color{blue}\line(0,1){60}}
	\put(199,10){\line(0,1){60}}

	\put(200,10){\color{blue}\line(4,3){80}}

	\put(280,10){\color{blue}\line(0,1){60}}
	\put(280,70){\color{blue}\line(2,-3){20}}	
	\put(279,70){\line(2,-3){20}}
	\put(280,10){\line(2,3){20}}
		
	\put(280,10){\color{blue}\line(0,1){60}}

	\put(300,70){\color{blue}\line(0,-1){30}}	
	\put(280,10){\color{blue}\line(1,0){20}}		


	\put(-20,70){\color{blue}\circle*{10}}
	\put(-20,70){\circle*{5}}
	\put(-45,65){$\delta_1$}
	
	\put(-20,10){\color{blue}\circle*{10}}
	\put(-20,10){\circle*{5}}
	\put(-45,5){$\zeta_0$}

	\put(300,70){\color{blue}\circle*{10}}
	\put(300,70){\circle*{5}}
	\put(315,65){$\zeta_k$}
	
	\put(300,10){\color{blue}\circle*{10}}
	\put(300,10){\circle*{5}}
	\put(315,5){$\delta_k$}
	
	\put(315,37){$\gamma_{k-1}$}


	\put(0,40){\color{blue}\circle*{10}}
	\put(0,40){\circle*{5}}

	\put(40,40){\color{blue}\circle*{10}}
	\put(40,40){\circle*{5}}
	\put(80,40){\color{blue}\circle*{10}}
	\put(80,40){\circle*{5}}
	\put(120,40){\color{blue}\circle*{10}}	
	\put(120,40){\circle*{5}}

	\put(133,40){\color{blue}$\ldots$}

	\put(160,40){\color{blue}\circle*{10}}
	\put(160,40){\circle*{5}}
	\put(200,40){\color{blue}\circle*{10}}
	\put(200,40){\circle*{5}}	
	\put(240,40){\color{blue}\circle*{10}}
	\put(240,40){\circle*{5}}
	\put(300,40){\color{blue}\circle*{10}}
	\put(300,40){\circle*{5}}

	\put(-2,20){$\delta_0$}
	\put(28, 20){$\gamma_1$}
	\put(75,20){$\zeta_2$}
	\put(105,20){$\gamma_2$}
	\put(150,20){$\zeta_{k-2}$}
	\put(178,20){$\gamma_{k-2}$}
	\put(235,20){$\zeta_{k-1}$}

	\put(40,10){\color{blue}\circle*{10}}
	\put(120,70){\color{blue}\circle*{10}}
	\put(120,10){\color{blue}\circle*{10}}
	\put(200,70){\color{blue}\circle*{10}}
	\put(200,10){\color{blue}\circle*{10}}
	\put(280,70){\color{blue}\circle*{10}}
	\put(280,10){\color{blue}\circle*{10}}

	\put(35,-10){\color{blue}$v_{4}$}

	\put(115,85){\color{blue}$v_{5}$}
	\put(115,-10){\color{blue}$v_{7}$}

	\put(195,85){\color{blue}$v_{3k-7}$}
	\put(195,-10){\color{blue}$v_{3k-5}$}

	\put(265,85){\color{blue}$v_{3k-4}$}
	\put(265,-10){\color{blue}$v_{3k-2}$}

\end{picture}
\end{center}
\caption{Intersection graph of the fibers in codimension 2 along $b_2=0$ for $SO(4k+2)$ for the alternative resolution (\ref{SOoddBlowupsALT}). 
  $\color{blue}\bullet$ are the irreducible components in codimension 2, bicolored nodes either remain irreducible when passing to the codimension 2, or have a remaining component, e.g. $\gamma_i$ split off into three components, generically, which is indicated by two new blue nodes and one bicolored one. Green lines indicate the splitting, blue lines the intersections. }\label{fig:DynkinSOodd2ALT} 
\end{figure}
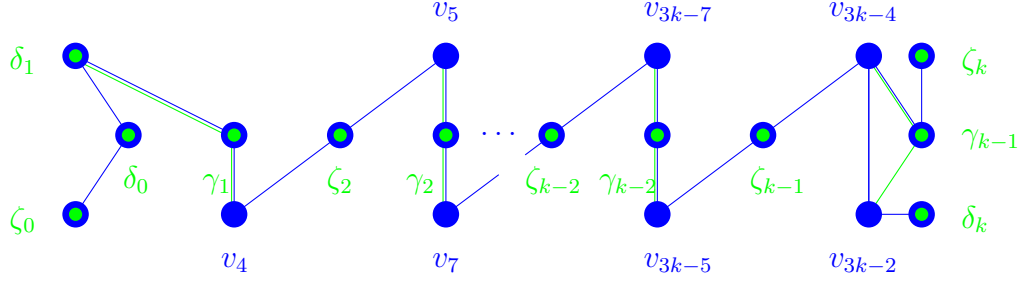


The interesting things happen along the $b_2 = 0$ non-minimal locus. Restricting the Cartan divisors to $b_2=0$, the irreducible components are now as follows --  in particular, the last column in the table, which indicates the Cartan charges, makes it clear that the weights that appear are different from the ones in the resolution in the main text:
\begin{equation}
        \begin{array}{l|c|l|l}
                \text{Matter surface } & \text{Section} & \hbox{Equation in $Y|_{b_2=0}$} & j\text{th Cartan charge} \cr\hline
                S_{-\alpha_0} & \zeta_0 & 0 = y^2\delta_1 - x^3\zeta_1  & \delta_{j,2} - 2\delta_{j,0} \cr
                S_{-\alpha_1} & \delta_1 & 0 = \gamma_1  & \delta_{j,2} - 2\delta_{j,1} \cr
                S_{-\alpha_2} & \delta_0 & 0 = \delta_1 - \gamma_1\zeta_1  & \delta_{j,0} + \delta_{j,1} + \delta_{j,3} - 2\delta_{j,2} \cr
                S_{v_3} & \gamma_1 & 0 =y  & \delta_{j,1} - \delta_{j,3} \cr
                S_{v_{3i}} \,,\ \ \quad  i = 2,\cdots,k-2& \gamma_i & 0 = y  & -\delta_{j,2i+1} \cr
                S_{v_{3i-1}} \,, \quad i = 2,\cdots,k-1 & \gamma_i & 0 = \delta_i  & \delta_{j,2i} \cr
                S_{v_{3i+1}}  \,,\quad  i = 1,\cdots,k-2 & \gamma_i & 0 = \delta_{i+1} & \delta_{j,2i+2} \cr
                S_{v_{3k-3}} & \gamma_{k-1} & 0 = y  & \delta_{j,2k+1} - \delta_{j,2k-1} \cr
                S_{v_{3k-2}} & \gamma_{k-1} & 0 = y\delta_k + b_3\zeta_{k-1}  & \delta_{j,2k} - \delta_{j,2k-1} \cr
                S_{-\alpha_{2k}} & \zeta_k & 0 = \delta_k + b_3\zeta_{k-1}  & \delta_{j,2k-1} - 2\delta_{j,2k} \cr
		S_{-\alpha_{2k+1}} & \delta_k & 0 = b_3y - \gamma_{k-1}\zeta_k(b_4x + b_6\gamma_{k-1}) & \delta_{j,2k-1} - 2\delta_{j,2k+1} \cr
		S_{-\alpha_{2i}}\,,\quad i = 2,\cdots,k-1  & \zeta_i & 0 = \delta_i & \delta_{j,2i-1} + \delta_{j,2i+1} - 2\delta_{j,2i} \cr
        \end{array} 
\end{equation}
In addition to the spin representations, in this case we find weights of the other fundamental representations $\Lambda^iV$. 
The splitting of the Cartan divisors can be summarized as follows
\begin{equation}
  \ba
                D_{-\alpha_3} &\quad\longrightarrow\quad S_{-\alpha_1} + 2\times S_{v_3} + S_{v_4} \cr
                D_{-\alpha_{2i+1}} &\quad\longrightarrow\quad  S_{v_{3i-1}} + 2 \times S_{v_{3i}} + S_{v_{3i+1}} \cr
                D_{-\alpha_{2k-1}} &\quad\longrightarrow\quad  S_{v_{3k-4}} + S_{v_{3k-3}} + S_{v_{3k-2}} \,,
\ea
\end{equation}

Denote the highest weights of the fundamental representations by
$ (\mu_{\Lambda^iV})_j = \delta_{i,j}$, then the weights appering in the fiber along $b_2=0$ are summarized as follows
\begin{equation}
        \begin{array}{l|l}
                \text{Matter surface } & \text{Weight } \cr\hline
                S_{v_3} & \mu_{\Lambda^2V} - \left(\alpha_1 + 3\alpha_2 + \sum_{j=3}^{2k-1}4\alpha_j + 2\alpha_{2k} + 2\alpha_{2k+1}\right) \cr
                S_{v_{3i-1}} & \mu_{\Lambda^{2i}V} \cr
                S_{v_{3i+1}} & \mu_{\Lambda^{2i+2}V} \cr
                S_{v_{3i}} & \mu_{\Lambda^{2i+1}} - \left(\sum_{j=1}^{2i } 2 j\alpha_j   + 2 (2i+1)\sum_{j=2i+1}^{2k-1} \alpha_j  + (2i+1) \alpha_{2k} + (2i+1) \alpha_{2k+1}\right) \cr
                S_{v_{3k-3}} & \mu_{S^+} - \left(\sum_{i=1}^{2k-1} i \alpha_i + k \alpha_{2k} + (k-1) \alpha_{2k+1} \right) \cr
                S_{v_{3k-2}} & \mu_{S^+} - \left(\sum_{i=1}^{2k-1} i \alpha_i + 2k \alpha_{2k} + 2k \alpha_{2k+1} \right)  \cr
                S_{u_{2k+1}} & \mu_{V} - \left(\sum_{{i=1}}^{2k-1}\alpha_i + \alpha_{2k+1}\right) \cr
        \end{array}
\end{equation}
It would be interesting to understand the mathematics behind which representations appear when passing to a non-minimal 
locus, and how this is encoded in the choice of small resolutions.


\newpage

\bibliographystyle{JHEP}

\providecommand{\href}[2]{#2}\begingroup\raggedright\endgroup


\end{document}